\documentclass[twocolumn,eqsecnum,floatfix,aps,prc]{revtex4}
\usepackage{epsfig}
\def\rhovec{\mbox{\boldmath $\rho$}}
\begin{document}

%\parindent=10pt
%%%%%%%%%%%%%%%%%%%%%%%%%%%%%%%%%%%%%%%%%%%%%%%%%%%%%%%%%%%%%%%%

\title {Study of various few-body systems using Gaussian % \\
%\vskip 0.3cm
expansion method (GEM)
}

\author{Emiko Hiyama}
\email{hiyama@phys.kyushu-u.ac.jp}
\affiliation{Department of Physics, Kyushu University,
Fukuoka 819-0395, Japan, \\
RIKEN Nishina Center, RIKEN, Wako 351-0198, Japan}

\author{Masayasu Kamimura}
\email{mkamimura@riken.jp}
\vskip 0.3cm
\affiliation{RIKEN Nishina Center, RIKEN, Wako 351-0198, Japan}

\date{\today}

%%%%%%%%%%%%%%%%%%%%%%%%%%%%
\begin{abstract}
We review our calculation method,  
Gaussian expansion method (GEM), and its applications to various 
few-body (3- to 5-body) systems
such as \mbox{1) few-nucleon} systems, 
2)~few-body structure of hypernuclei, 
3)~clustering structure of light nuclei and unstable nuclei, 
4)~exotic atoms/molecules,
5)~cold atoms,
6)~nuclear astrophysics and 
7)~structure of exotic hadrons. 
Showing examples in our published papers, we explain 
\mbox{i)  high} accuracy of GEM calculations and its reason,
\mbox{ii) wide} applicability  of  GEM,
\mbox{iii) successful} predictions by GEM calculations before measurements.
GEM was proposed 30 years ago and
has been applied to a variety of subjects.
To solve  few-body Schr\"{o}dinger 
equations accurately, use is made of the Rayleigh-Ritz variational method 
for bound states,
the complex-scaling method for resonant states and
the Kohn-type variational principle to  $S$-matrix for scattering
states.  The total wave function is 
expanded in terms of few-body Gaussian 
basis functions spanned over all the sets of rearrangement  
Jacobi coordinates.  Gaussians with ranges in {\it geometric progression}
work very well both for short-range and long-range behavior 
of the few-body wave functions.  Use of Gaussians with complex ranges
gives much more accurate solution 
especially when the wave function has many oscillations.
\end{abstract}

\maketitle

%\keywords{few-body problems, Gaussian expansion method, Gaussian ranges 
%in geometric progression}

%\pacsnumbers{03.65.Ge, 21.45.+v, 21.60.-n, 21.80.+a, 36.10.-k, 67.85.-d}

\vskip 3.3cm
\begin{center}
\vspace{3cm} 
Contents
\end{center}

\parindent=0pt
{\bf I.}  {\bf Introduction}  \hfill 2
\vskip 0.3cm

{\bf II.} {\bf Gaussian expansion method (GEM) for \\
\hspace*{4mm} few-body systems}  \hfill 3
\vskip 0.1cm
$\quad$ A.  Use of all the sets of Jacobi coordinates \hfill 3

$\quad$ B. Gaussian basis functions
 with ranges in \\
\hspace*{8mm} geometric progression  \hfill 4

$\quad$ C. Easy optimization of nonlinear variational \\
\hspace*{8mm} parameters  \hfill 4

$\quad$ D. Complex-range Gaussian basis function \hfill 5

$\quad$ E. Infinitesimally-shifted Gaussian lobe basis \\
\hspace*{8mm} functions \hfill 5
\vskip 0.3cm
{\bf III.} {\bf Accuracy of GEM calculations}  \hfill 6

\vskip 0.1cm
$\quad$ A. Muonic molecule in muon-catalyzed fusion cycle \hfill 6

$\quad$ B. 3-nucleon bound state ($^3$H and $^3$He) \hfill 7

$\quad$ C. Benchmark test calculation of 4-nucleon 
       \\ \hspace*{8mm} ground and second $0^+$ states  \hfill 8

$\quad$ D. Determination of antiproton mass  by GEM \hfill 10

$\quad$ E.  Calculation of $^4$He-atom tetramer in 
       \\ \hspace*{8mm} cold-atom physics (Efimov physics)  \hfill 11

%\vskip 2.8cm
\vskip 0.3cm
{\bf IV.} {\bf Successful predictions by GEM}  \hfill 13
\vskip 0.1cm

$\quad$ A. Prediction of energy level of antiprotonic \\
\hspace*{8mm} He atom \hfill 13

$\quad$ B. Prediction of shrinkage of hypernuclei \hfill 13

\vfill
\pagebreak
$\quad$ C.  Prediction of spin-orbit splitting in hypernuclei  \hfill 15

$\quad$ D. Prediction of neutron-rich hypernuclei \hfill 16

$\quad$ E.  Prediction of hypernuclear states with \\
\hspace*{8mm} strangeness $S=-2$        \hfill 17

$\quad$ F. Strategy of studying hypernuclei and $YN$ \\
\hspace*{8mm} and $YY$ interactions  \hfill 18

\vskip 0.3cm
{\bf V.} {\bf Extension of GEM}  \hfill 18
\vskip 0.1cm

$\quad$ A.  Few-body resonances with 
complex-scaling \\
\hspace*{8mm} method   \hfill   18

$\quad$ $\;$ $\;\;\;$ A1. Tetraneutron$(^4n)$ resonances   \hfill 19

$\quad$ $\;$ $\;\;\;$ A2. 3-body resonances in $^{12}$C
studied with \\ \hspace*{17mm} complex-range Gaussians    \hfill 20

$\quad$ B. Few-body reactions with 
Kohn-type variational \\ \hspace*{8mm} principle to $S$-matrix   \hfill 21

$\quad$ $\;$ $\;\;\;$ B1. Muon transfer reaction  in $\mu$CF cycle \hfill 21

$\quad$ $\;$ $\;\;\;$ B2. Catalyzed big-bang nucleosynthesis \\ 
\hspace*{17mm} (CBBN) reactions    \hfill 22

$\quad$ $\;$ $\;\;\;$ B3. Scattering calculation of 5-quark  $(uudd{\bar s})$
\\ \hspace*{17mm} systems   \hfill 22

\vskip 0.3cm
{\bf VI.} {\bf Summary}    \hfill  23
\vskip 0.3cm
$\quad$ {\bf Acknowledgements}   \hfill  24

\vskip 0.2cm
$\quad$ {\bf Appendix}   \hfill  24

$\quad$ $\;\;$  Examples of accurate 2-body GEM calculations 

\vskip 0.2cm
$\quad$ {\bf References}   \hfill 28

%\vspace{3cm} 

%\vskip 3.3cm
\vfill

\pagebreak
\parindent=10pt
\section{INTRODUCTION}

There are many examples of 
{\it precision} numerical 
calculations that contributed to the study of 
fundamental laws and constants in physics. 
One of the recent examples may be 
a contribution of our calculation method, Gaussian expansion 
method (GEM)~\cite{Kamimura88,Kameyama89,Hiyama03,Hiyama12FEW,Hiyama12PTEP}
for few-body systems, to the determination
of antiproton mass. 
In Particle Listings 2000~\cite{Listing2000}, 
the Particle Data Group provided, for the first time, 
the recommended value of \mbox{antiproton mass} ($m_{\bar p}$),
compared with proton mass ($m_p$), in the form of  
$|m_{\bar p}-m_p|/m_p < 5 \times 10^{-7}$, and commented that
this could be used for a test of $CPT$ invariance.
This value was derived by a collaboration of 
experimental and theoretical
studies of  highly-excited metastable states in
the antiprotonic helium atom (He$^{2+} + {\bar p} + e^-$),    
namely, by a high-resolution laser spectroscopy experiment 
at CERN~\cite{Torii99} 
% and by Hori {\it et al.} \cite{Hori01}
and a precision Coulomb 3-body GEM calculation~\cite{Kino98,Kino99}
with the accuracy of 10 significant figures in the level energies
(cf. Sec.~\ref{sec:pbar-mass}).

%%%%%%%%%%%%%%%%%%%%%%%%%%%%%%%%%%%%%%%%%%%%%%
\begin{figure}[b!]
\begin{center}
\epsfig{file=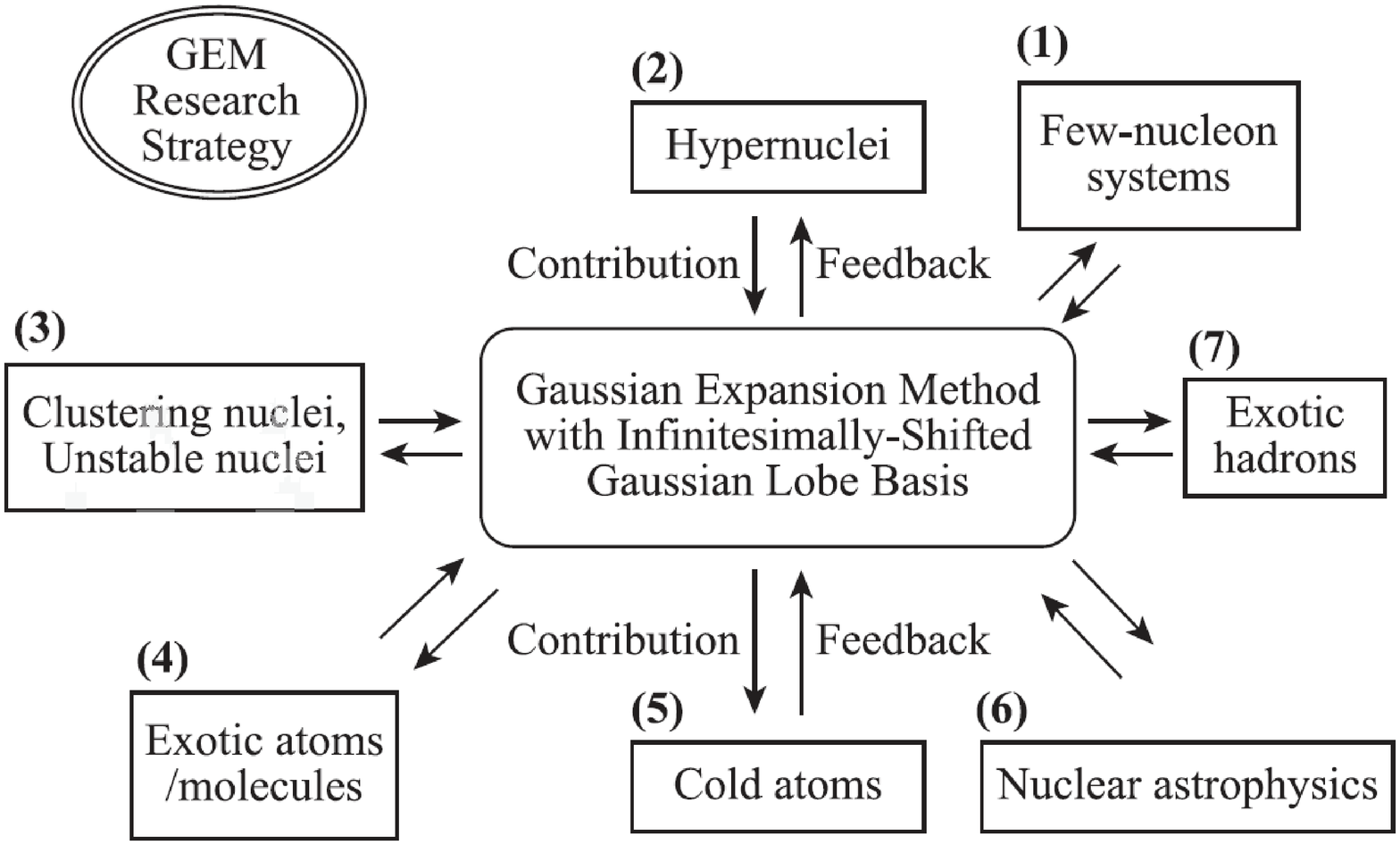,width=8.5cm,height=5.3cm}
\caption{
Research strategy for few-body physics with GEM.
}  
\label{fig:strategy}
\end{center}	
\end{figure}
%%%%%%%%%%%%%%%%%%%%%%%%%%%%%%%%%%%%%%%%%%%%%%

\vskip 0.1cm
Many  important problems in physics can be addressed by 
{\it accurately} solving the  Schr\"{o}dinger equations 
for bound state, resonances and reaction processes in
few-body (especially, 3- and 4-body) systems. 
It is of  particular importance to
develop various numerical 
methods for high-precision calculations of such problems.
For this purpose, 
the present authors and collaborators proposed and have been
developing the Gaussian expansion method for few-body 
systems~\cite{Kamimura88,Kameyama89,Hiyama03,Hiyama12FEW,Hiyama12PTEP}.
        
\vskip 0.1cm
Using the GEM, the present authors and collaborators 
have been studying many subjects in various research
fields of physics. 
Our strategy for such studies is as follows: 
As shown in Fig.~\ref{fig:strategy}, we have our own calculation method GEM 
in the center  and have been applying it 
to a variety of systems, such as \mbox{(1) few-nucleon} systems, 
\mbox{(2) hypernuclei}, 
\mbox{(3) clustering} nuclei and unstable nuclei, 
\mbox{(4) exotic} atoms/molecules, 
\mbox{(5) cold} atoms, 
\mbox{(6) nuclear} astrophysics and
\mbox{(7) exotic} hadrons. 

\vskip 0.1cm
As indicated in Fig.~\ref{fig:strategy} by arrows back to the center,
we often obtained useful {\it feedback} from the calculation effort 
in each field, so that we further  developed the GEM itself. 
We then applied the so-improved GEM to a  new field 
where the present authors and collaborators had not enter before. 
We have been repeating this research cycle under this strategy.

\vskip 0.1cm
The purpose of the present review paper is to explain 
\mbox{i)  high} accuracy of GEM calculations and its reason,
\mbox{ii) wide} applicability  of  GEM to various few-body systems, and
\mbox{iii) predictive} power of  GEM calculations.

\vskip 0.1cm

In the case of bound states, 
the  few-body Schr\"{o}dinger equation is solved on the basis of
the Rayleigh-Ritz variational principle;  
the total wave function  is 
expanded in terms of the $L^2$-integrable basis functions by which 
Hamiltonian is diagonalized.

\vskip 0.1cm
We employ few-body Gaussian basis functions that are 
spanned over all the sets of 
rearrangement Jacobi coordinates (for example,
Eq.~(\ref{eq:3body-wf-disting}) for 3-body and
Eq.~(\ref{eq:Psi-expansion}) for 4-body systems). 
This construction of few-body basis functions using
all the Jacobi coordinates 
makes the function space significantly larger
than that spanned by the basis functions of 
single set of Jacobi coordinates.
 
\vskip 1.1cm
In the authors' opinion, a very useful set of basis functions
%% (with angular-momentum $l$ and its \mbox{$z$-component $m$)}
along any Jacobi coordinate ${\bf r}$ is  
\begin{eqnarray}
e^{- \nu_n r^2} r^l \, Y_{lm}(\widehat{\bf r}), \;\;
r_n=\nu_n^\frac{1}{2} =r_1 a^{n-1} \;\;  (n=1,...,N) , \nonumber
\end{eqnarray}
where the ranges are taken 
in {\it geometric progression}~\cite{Kamimura88}; and similarly for the other
Jacobi coordinates.
We refer to them as Gaussian basis functions.
  
\vskip 0.1cm
The geometric progression $\{ r_n\}$ is 
dense at short distances so that the description of the dynamics 
mediated by short range potentials can be properly treated. 
Moreover, 
though single Gaussian decays quickly, appropriate
superposition of many Gaussians can decay accurately (exponentially)
up to a sufficiently large distance. 
We~show many example figures 
for 2-, 3- and 4-body cases in this paper  
(a  reason why the `geometric progression' works well
is mentioned in Sec.~\ref{sec:geomet}). 

\vskip 0.1cm
Use of Gaussians with complex ranges~\cite{Hiyama03}, 
\begin{eqnarray}
%\begin{equation}
e^{- \eta_n r^2 } r^l \, Y_{lm}(\widehat{\bf r}), \;\;
 \;\; \eta_n=(1 \pm i \, \omega)\,\nu_n, \nonumber
%\end{equation}
\end{eqnarray}
makes the function space much wider than that of Gaussians with real ranges
mentioned above since the former has oscillating part explicitly
(cf.~Sec.~\ref{sec:complex-range}). 
The new  basis functions
are especially suitable for describing 
wave functions having many oscillating nodes 
\mbox{(cf. Figs.~\ref{fig:howf-complex} and \ref{fig:sincos-coul}}
in Appendix).

\vskip 0.1cm
Therefore, in the study of few-body resonances using the complex-scaling 
method (for example, \cite{Aoyama2006} and cf.~Sec.~\ref{sec:CSM}),
the complex-range Gaussian basis functions are specially useful 
since the resonance wave function in the method is highly oscillating
when the rotation angles $\theta$  is large 
in the \mbox{scaling $r \to r\, e^{i \theta}$.}

\vskip 0.1cm
Another important advantage of using real- and complex-range
Gaussians  is that calculation of
the Hamiltonian matrix elements among the few-body basis functions
can easily be performed~\cite{Hiyama03}. This advantage is much more
enhanced if one uses the infinitesimally-shifted Gaussian basis
functions~\cite{Hiyama94,Hiyama95u,Hiyama03} introduced by the R.H.S. of   
%\begin{equation}
\begin{eqnarray}
e^{-\nu_n r^2} r^l \,Y_{lm}({\widehat {\bf r}}) 
=\!\lim_{\varepsilon \rightarrow 0} 
  \frac{1}{(\nu_n \varepsilon)^l} 
\sum_{k=1}^{k_{\rm max}} \,  C_{lm,k}\; 
e^{-\nu_n ({\bf r}\,-\,\varepsilon {\bf D}_{lm,k})^2} \nonumber
%\end{equation}
\end{eqnarray}
because 
the tedious angular-momentum algebra (Racah algebra) 
does not appear
when calculating the few-body matrix elements (cf. Sec.~\ref{sec:ISGL}).

\vskip 0.1cm
In the study of few-body scattering and reaction processes,
we emply the Kohn-type variational principle to 
$S$-matrix~\cite{Kamimura77}.
The wave-function amplitude in the interaction region
is expanded in terms of the few-body real- (complex-)range Gaussian basis 
functions constructed on all the sets of Jacobi coordinates.
We consider the basis functions are nearly complete in the 
restricted region; examples will be discussed 
in Sec.~\ref{sec:reaction}.

\vskip 0.1cm
As long as the employed interactions among constituent particles (clusters)
of the few-body system concerned are all well-established ones,
accurate results by the GEM calculations are so reliable
that we can use them to make a prediction before measurements % (if any)
 about that system. If some members of such interactions 
are ambiguous (or not established),
we first try to improve them phenomenologically in order to reproduce
the existing experimental data for all the subsystems
(possible combinations of the constituent members).
Then, it is possible for the GEM calculation to make 
a {\it prediction}  about the full system (cf. a strategy in our study of
hypernuclear physics, Fig.~\ref{fig:strategy-hyper},
in Sec.~\ref{sec:hyper-strategy}).  
Examples of successful predictions by the GEM calculations
will be presented in Sec.~\ref{sec:predict}.

\vskip 0.1cm
This article is organized as follows:
Outline of the GEM framework  is capitulated in Sec.~\ref{sec:GEM}. 
Examples of high-precision  GEM calculations are
demonstrated in Sec.~\ref{sec:AccuracyGEM}. 
We review, in Sec.~\ref{sec:predict},
examples of successful GEM predictions before measurements.
Extension of GEM to few-body resonances and few-body reactions are
presented in Sec.~\ref{sec:extension}. 
Summary is given in Sec.~\ref{sec:summary}. 
In Appendix, we present several examples of  \mbox{2-body} GEM calculations
in order to show the high accuracy of the real- and complex-range
Gaussian basis functions, taking visible cases.

%\end{document}   %arxiv

%%%%%%%%%%%%%%%%%%%%%%%%%%%%%%%%%%%%%%%%%%%%%%%%%%%%%%%%%%%%%%%%%%%
%\section{Gaussian expansion method (GEM) for few-body systems}
\section{GAUSSIAN EXPANSION METHOD (GEM) FOR FEW-BODY SYSTEMS}
\label{sec:GEM}
%%%%%%%%%%%%%%%%%%%%%%%%%%%%%%%%%%%%%%%%%%%%%%%%%%%%%%%%%%%%%%%%%%%

GEM has already been applied to various 3-, 4- and 5-body systems.
In this section, we briefly  explain the method taking
the case of 3-body {\it bound} states for simplicity.

Applications to complex-scaling calculations 
for 3- and 4-body {\it resonant} states 
are shown Sec.~\ref{sec:CSM} and 
those to  {\it reactions} 
are presented in Sec.~\ref{sec:reaction}.

%%%%%%%%%%%%%%%%%%%%%%%%%%%%%%%%%%%%%%%%%%%%%%%%%%%%%
\subsection{Use of all the Jacobi-coordinate sets}
\label{sec:3body-jacobi}
%%%%%%%%%%%%%%%%%%%%%%%%%%%%%%%%%%%%%%%%%%%%%%%%%%%%%

\begin{figure}[b!]
\begin{center}
\epsfig{file=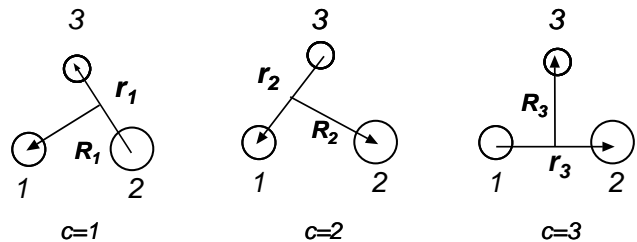,scale=0.45}
\end{center}
\caption{
Three sets of Jacobi coordinates of 3-body system.
All of them are used in GEM calculations
}
\label{fig:Jacobi-3body}
\end{figure}

In GEM, solution to the 
Schr\"{o}dinger equation for the bound-state wave function $\Psi_{JM}$
with the total angular momentum $J$
and its $z$-component $M$,
\begin{equation}
( H - E ) \Psi_{JM}=0 \, , 
\end{equation}  
%for 3-body bound states
is obtained by diagonalizing the Hamiltonian
in a space spanned by a finite number
of $L^2$-integrable  3-body basis functions which are
constructed on all the sets of Jacobi coordinates 
(Fig.~\ref{fig:Jacobi-3body}). 
%\vskip 0.2cm

%%%%%%%%%%%%%%%%%%  Fig. 2 (Jacobi) %%%%%%%%%%%
%%%%%%%%%%%%%%%%%%%%%%%%%%%%%%%%%%%%%%%%%%%%%%%%%
%
%Keypoints of GEM may be stated as follows
%(degrees of freedom other than spatial variables omitted):

The total wave function $ \Psi_{JM}$
is written as a sum of component functions
of {\it  all the 3 rearrangement channels}
\begin{eqnarray}
 \Psi_{JM}& =&
  \sum_{\alpha=1}^{\alpha_{\rm max}} 
 A_\alpha \,\Phi_\alpha^{(1)} ({\bf r}_1, {\bf R}_1) 
+  \sum_{\beta=1}^{\beta_{\rm max}}
 B_\beta  \, \Phi_\beta^{(2)} ({\bf r}_2, {\bf R}_2)  \nonumber \\
&+&  \sum_{\gamma=1}^{\gamma_{\rm max}} 
C_\gamma\, \Phi_\gamma^{(3)} ({\bf r}_3, {\bf R}_3) , \qquad
\label{eq:3body-wf-disting}
\end{eqnarray}
where spins and isospins are omitted for simplicity.
The 3-body basis functions are taken as 
\begin{eqnarray}
&& \!\!\!\! \!\!\!\! \Phi_\alpha^{(1)} ({\bf r}_1, {\bf R}_1)\! =\!
% \Phi_\alpha^{(1)} ({\bf r}_1, {\bf R}_1)\! =\!
\phi_{n_1 l_1}^{(1)}(r_1)\, \psi_{N_1L_1}^{(1)}(R_1)  
\Big[ Y_{l_1}(\widehat{{\bf r}}_1)\,
Y_{L_1}(\widehat{{\bf R}}_1) \Big]_{J M},\nonumber \\ 
%\label{eq:3-boson-amp1} 
%
&&\!\!\!\!\!\!\!\!  \Phi_\beta^{(2)} ({\bf r}_2, {\bf R}_2)\! =\!
%% \Phi_\beta^{(2)} ({\bf r}_2, {\bf R}_2) =
\phi_{n_2 l_2}^{(2)}(r_2)\, \psi_{N_2L_2}^{(2)}(R_2)  
\Big[ Y_{l_2}(\widehat{{\bf r}}_2)\,
Y_{L_2}(\widehat{{\bf R}}_2) \Big]_{J M}, \nonumber \\
%\label{eq:3-boson-amp2} 
%
&&\!\!\!\!\!\!\!\!  \Phi_\gamma^{(3)} ({\bf r}_3, {\bf R}_3)\! =\!
%% \Phi_\gamma^{(3)} ({\bf r}_3, {\bf R}_3) =
\phi_{n_3 l_3}^{(3)}(r_3)\, \psi_{N_3L_3}^{(3)}(R_3)  
\Big[ Y_{l_3}(\widehat{{\bf r}}_3)\,
Y_{L_3}(\widehat{{\bf R}}_3) \Big]_{J M},  \nonumber\\
\label{eq:3-particles} 
\end{eqnarray}
\vskip -0.2cm
\noindent
where $\alpha$, $\beta$ and $\gamma$ specify    
\begin{eqnarray}
&&\alpha \equiv \{n_1, l_1, N_1, L_1\}\:,\quad
\beta \equiv \{n_2, l_2, N_2, L_2\}\:,\quad  \nonumber \\
&&\gamma \equiv \{n_3, l_3, N_3, L_3\}\:,
\end{eqnarray}
with $l, L$ denoting angular momenta and
$n, N$ specifying  radial dependence (namely, Gaussian ranges; see below).
%\vskip 0.2cm
Energies $E$ and  wave-function coefficients $A_\alpha$, 
$B_\beta$ and $C_\gamma$  are determined 
simultaneously by using the Rayleigh-Ritz variational principle,
namely by diagonalizing the Hamiltonian using the basis functions.

%\vskip 0.2cm
If the three particles are identical particles, 
Eq.~(\ref{eq:3body-wf-disting}) is to be replaced by
\begin{eqnarray}
\!\!\!\Psi_{JM}\! =\!
  \sum_{\alpha=1}^{\alpha_{\rm max}} \! A_\alpha  
    \Big[ \Phi_\alpha ({\bf r}_1, {\bf R}_1) 
               + \Phi_\alpha ({\bf r}_2, {\bf R}_2)
               + \Phi_\alpha ({\bf r}_3, {\bf R}_3) \Big]. 
\nonumber 
\label{eq:3-identical}
\end{eqnarray}
%
%\vskip 0.2cm
This construction of 3-body basis functions on
all the sets of Jacobi coordinates 
makes the function space significantly larger
%(even if $l$ and $L$ are strongly restricted) 
than the case using the basis functions of  single channel alone.
Also it makes the non-orthogonality between the basis functions
much less troublesome than in the single-channel case.
These types of 3-body 
basis functions are particularly suitable  for describing
compact clustering between two particles along any $r_c~(c=1-3)$
and  for a weakly coupling of the third particle  
along any $R_c$.
We also emphasize that the 3-channel basis functions 
are particularly appropriate for systems composed of 
mass-different (distinguishable) particles.

%%%%%%%%%%%%%%%%%%%%%%%%%%%%%%%%%%%%%%%%%%%%%%%%%%%%%%%%%%%%%%%%%%%
\subsection{Gaussians with ranges in geometric progression}
\label{sec:geomet}
%%%%%%%%%%%%%%%%%%%%%%%%%%%%%%%%%%%%%%%%%%%%%%%%%%%%%%%%%%%%%%%%%%%

%\vskip 0.2cm
Radial dependence of the basis functions 
$\phi_{n l}(r)$ and $\psi_{N L}(R)$ is taken as
Gaussians (multiplied by $r^l\,$ and $R^L$) 
with ranges in {\bf geometric progression}
\cite{Kamimura88,Kameyama89,Hiyama03}: 
\begin{eqnarray}
\phi_{n l}(r) &\!\!=\!\!&  N_{nl}\, 
r^{l}\:e^{- \nu_n r^2},  \nonumber \\
\nu_n &\!\!=\!\!&1/r_n^2   \;, \quad \nonumber\\
r_{n}&\!\!=\!\!& r_1\, a^{n-1} 
 \quad \:(n=1,...,n_{\rm max})  
\label{eq:gauss-r} 
\end{eqnarray}
and
\begin{eqnarray}
 \psi_{N L}(R)&\!\!=\!\!&  N_{NL}\, R^{L}\:e^{- \lambda_N R^2}, 
  \nonumber \\
 \lambda_N &\!\!=\!\!& 1/R_N^2 \;, \nonumber\\
 R_{N}&\!\!=\!\!& R_1\, A^{N-1}
\quad \:(N=1,...,N_{\rm max}) 
\label{eq:gauss-R} 
\end{eqnarray}
%%%%%%%%%%%%%%%%%%%%%%%%%%%%%%%%
with the normalization constants $N_{nl}$ and $N_{NL}$.

\vskip 0.1cm
The geometric progression is 
dense at short distances so that the description of the dynamics 
mediated by short range potentials can be properly treated. 
Moreover, 
though single Gaussian decays quickly, appropriate
superposition of many Gaussians can decay accurately (exponentially)
up to a sufficiently large distance.  
Good examples in 2-body systems are 
demonstrated in Figs.~\ref{fig:deuteron-av8-long} and \ref{fig:dimer-long}
in Appendix. 

\vskip 0.1cm 
Even for 3- and 4-body systems, the Gaussian basis functions 
so chosen can describe accurately both short range correlations and
long range asymptotic behavior simultaneously. 
Here, we emphasize that it is not \mbox{necessary}
to introduce {\it a priori} the Jastrow  
correlation factor in the total wave function
so as to describe the strong short-range correlations;
it is enough for the purpose to use the Gaussian basis 
functions~(\ref{eq:gauss-r}) and (\ref{eq:gauss-R})  
as will be shown in successful results of \mbox{Figs.~10 and  
\ref{fig:tetra-den-short}} in 4-body systems.
%\ref{fig:deuteron-av8-short} and \ref{fig:dimerpot}).

\begin{figure} [h]
\begin{center}
\epsfig{file=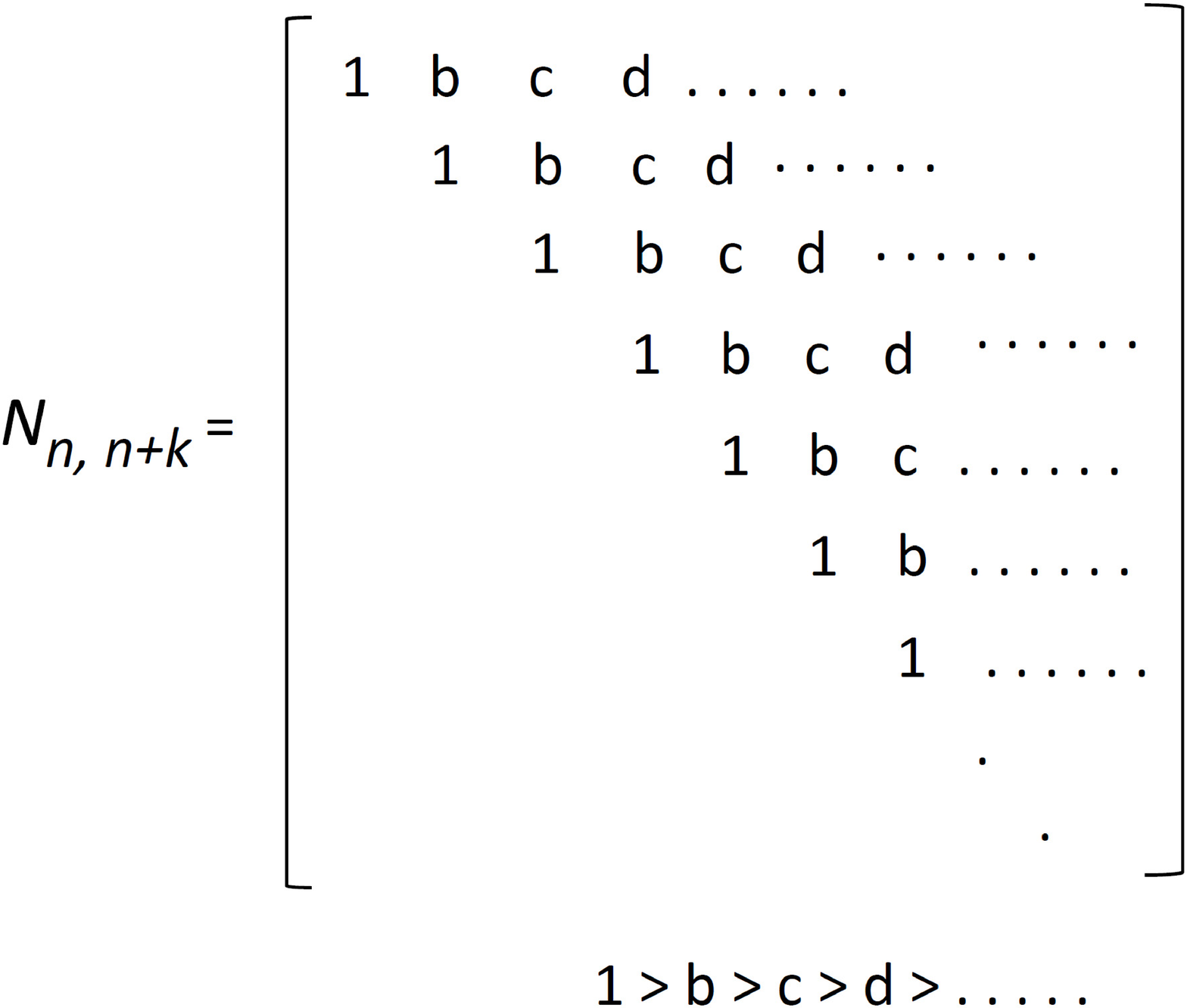,width=6.0cm,height=4.1cm}
\caption{The norm-overlap matrix (\ref{eq:norm-overlap})
in the case where the
Gaussian ranges are given in geometric progression 
(\ref{eq:gauss-r}). 
}
\label{fig:norm-mat}
\end{center}	
\end{figure}

\vskip 0.1cm
A reason why the Gaussians with ranges in geometric 
progression work well may be stated as follows~\cite{Hiyama12COLD-1}:
The norm-overlap matrix elements, $N_{n, n+k}\, (k=0,...,n_{\rm max})$,
between the basis functions is given as

%\pagebreak
%%%%%%%%%%%%%%%%%%%%%%%%%%%%%%%%
\begin{equation}
N_{n, n+k} = \langle \phi_{n \, l} \,|\, \phi_{n+k \, l} \rangle =
%[\frac{ 2a^k}{(1+a^{2k})}]^{l+3/2} ,
\left( \frac{2a^k}{1+a^{2k}} \right)^{l+3/2},  
\label{eq:norm-overlap}
\end{equation}
%%%%%%%%%%%%%%%%%%%%%%%%%%%%%%%%%
which shows that the overlap with the $k$-th neighbor
is \mbox{{\it independent}} of $n$ and  decreases gradually with increasing $k$
as illustrated in Fig.~\ref{fig:norm-mat}.
We then expect that the coupling among the whole basis functions
take place smoothly and coherently so as to
describe properly both the short-range structure and
long-range asymptotic \mbox{behavior} simultaneously.

%%%%%%%%%%%%%%%%%%   Fig. 3  %%%%%%%%%%%%%%%%%%%%%%%%%%%%
%%%%%%%%%%%%%%%%%%%%%%%%%%%%%%%%%%%%%%%%%%%%%%

\vskip 0.1cm
The Gaussian shape of basis functions makes the
calculation of the Hamiltonian matrix elements easy
even between different rearrangement channels.
On the other hand, according to the experience by the authors,
eigenfunctions of a harmonic-oscillator potential
(namely, Gaussian times Laguerre polynomials)
is not suitable for describing  three- and more-body systems 
because of the tediousness in the coordinate transformation 
and in the  many-dimensional integration when calculating the matrix elements. 
Also, it is difficult to describe a very weakly bound state 
that has a long-range tail since the long-range 
harmonic-oscillator eigenfunctions
inevitably oscillate many times up to the tail region.
%Use of the Gaussians with ranges in geometric progression
%does not suffer from such  difficulties.

%%%%%%%%%%%%%%%%%%%%%%%%%%%%%%%%%%%%%%%%%%%%%%%%
\subsection{Easy optimization of nonlinear variational parameters}
\label{sec:easy-opt}
%%%%%%%%%%%%%%%%%%%%%%%%%%%%%%%%%%%%%%%%%%%%%%%%

The setting of Gaussians with ranges in geometric progression
as in Eqs.~(\ref{eq:gauss-r}) and (\ref{eq:gauss-R})
enables us to optimize the ranges using a small number 
of free parameters; we recommend to take the sets 
\{$n_{\rm max}, r_1, r_{n_{\rm max}}$\} and 
\{$N_{\rm max}, R_1, R_{N_{\rm max}}$\} 
without using the ratios $a$ and $A$ 
which are given by 
$a=(r_{n_{\rm max}}/r_1)^{1/(n_{\rm max}-1)}$ and  
$A=(R_{N_{\rm max}}/R_1)^{1/(N_{\rm max}-1)}$.
%Note that, in principle, choice of the sets depends on
%the angular momenta \{$l_c, L_c$\} $(c=1,2,3)$. 

\vskip 0.1cm
Since the computation time by the use of the Gaussian
basis functions is very short, we can take rather large number for
${n_{\rm max}}$ and ${N_{\rm max}}$, even {\it more than enough}.
\mbox{It is} therefore satisfactory to optimize the Gaussian ranges
\{$r_1, r_{n_{\rm max}}, R_1, R_{N_{\rm max}}$\}  using
{\it round numbers}  
(cf. the \mbox{2-body} examples in Appendix); this is 
due to the fact that small change of
the ranges does not significantly change the function space 
since the space is already sufficiently wide by taking 
{\it more-than-enough} large numbers for
${n_{\rm max}}$ and ${N_{\rm max}}$.

\vskip 0.1cm
In the calculation of the 3-nucleon bound states ($^3$H and $^3$He)
using a realistic $NN$ potential (AV14),
the well-converged GEM calculation~\cite{Kameyama89}
took totally 3600 basis functions, but
only the 3 cases of  {\it round-number} sets
\vskip 0.24cm
$r_1=0.05, \:r_{n_{\rm max}}=15.0,\: R_1=0.3, \:R_{N_{\rm max}}=9.0$ fm,

$r_1=0.1\:\:, \:r_{n_{\rm max}}=15.0,\: R_1=0.3, \:R_{N_{\rm max}}=9.0$ fm,

$r_1=0.1\:\:, \:r_{n_{\rm max}}=10.0,\: R_1=0.3, \:R_{N_{\rm max}}=6.0$ fm

\vskip 0.24cm
\noindent
 (depending on $l, L$ and spins; cf.
Table I of  Ref.~\cite{Kameyama89}) 

\noindent
were so satisfactory that the binding energy 
converges with the 1-keV accuracy of four significant figures;
as will be explained in Fig.~\ref{fig:econverge} 
in Sec.~\ref{sec:3H-bound},
this convergence with respect to the increasing number of
angular-momentum channels is more rapid than that of the
Faddeev-method calculations of the same problem.

\vskip 0.1cm
Our method is quite transparent in the sense that all the nonlinear
variational parameter employed can explicitly be listed in a small table.
Therefore, one can examine the GEM results
by making a check calculation with the {\it same parameters}. 
For example, even in a well-convered 
{\it 4-body} calculation in the cold-atom physics
in Ref.~\cite{Hiyama12COLD-1} by the present authors, 
all the nonlinear variational parameters for totally 23504 basis functions
were listed in a small table of only 14 lines (Table V of that paper).
This calculation will be introduced in Sec.~\ref{cold-atom}.

\vskip 0.1cm
Good choice of the Gaussian ranges %$(r_1, r_{n_{\rm max}})$
depends mostly on  size and  shape of the interaction 
and spatial extension of the system.
But, to the authors' opinion,
slight experience is enough to master how to find such a choice
thanks to the properties of the Gaussian basis functions mentioned above.

%%%%%%%%%%%%%%%%%%%%%%%%%%%%%%%%%%%%%%%%%%%%%%%%
\subsection{Complex-range Gaussian basis functions}
\label{sec:complex-range}
%%%%%%%%%%%%%%%%%%%%%%%%%%%%%%%%%%%%%%%%%%%%%%%%

\vskip 0.1cm
In spite of many successful examples of the 
use of the Gaussian basis functions in the few-body calculations, 
it was hard to describe accurately  highly-oscillatory 
wave functions  having more than 
several nodes since the Gaussians themselves had
no radial nodes. 

\vskip 0.1cm
To overcome this difficulty, the present authors proposed~\cite{Hiyama03}  
new types of basis functions which 
have radial oscillations but tractable as easily as Gaussians;
namely, Gaussians with complex ranges
$\eta_n$ and $\eta^*_n$ instead of  real range 
$\nu_n \,(n=1,..., n_{\rm max}) $:
\begin{eqnarray}
&& r^l e^{-\eta_n r^2}\:,  \quad   
\;\; \eta_n=(1+i\,\omega)\,\nu_n\:, \nonumber \\
&& r^l e^{-\eta_n^* r^2}\:,  \quad 
\;\; \eta_n^*=(1-i\,\omega)\,\nu_n\:,
\label{eq:complex-range}
\end{eqnarray}
with $\nu_n$ in geometric progression as in (\ref{eq:gauss-r}).
They are equivalent to the set 
\begin{eqnarray}
\!\!\!\!\!\!&&\quad r^l\:e^{-\nu_nr^2} {\rm cos} \: \omega \nu_n r^2 \; 
=r^l (e^{-\eta_n r^2} + e^{-\eta_n^* r^2})/2\:, \quad   
\nonumber \\
\!\!\!\!\!\!&&\quad r^l\:e^{-\nu_nr^2}  {\rm sin} \: \omega \nu_n r^2 \; 
=r^l (e^{-\eta_n r^2} - e^{-\eta_n^* r^2})/2i.  \quad   
\label{eq:sin-cos}
\end{eqnarray}

%%%%%%%%%%%%%%%%%%%%%%%%%%%  Fig. 4   %%%%%%%%%%%%%%%%%%
\begin{figure}[t!]
\begin{center}
\epsfig{file=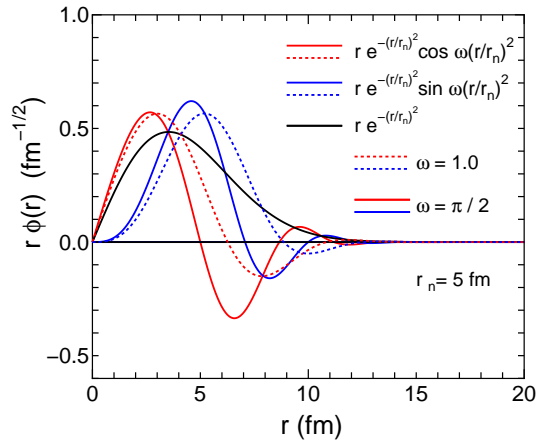,scale=0.41}
\caption{
An example of the $l=0$ real-range and complex-range Gaussian basis functions  
(multiplied by $r$) of Eqs.~(\ref{eq:gauss-r}) and (\ref{eq:sin-cos})
with $r_n=1/\sqrt{\nu_n}=5$ fm and $\omega=1.0$ and $\pi/2$.
They are normalized to unity. 
}
\end{center}
\label{fig:sindou}
\end{figure}
%%%%%%%%%%%%%%%%%%%%%%%%%%%%%%%%%%%%%%%%%%%%%%%%%
%

We refer to these oscillating functions~(\ref{eq:complex-range}) and
(\ref{eq:sin-cos}) 
as {\it complex-range} Gaussians.
From our experiences, we recommend to take 
simply $\omega= 1$ or $\pi/2$ 
as well as adopting  geometric progression for $\nu_n$.
In order to compare visually the real-range and complex-range Gaussians,
we plot an example of them in Fig.~4.  %?? Fig.~\ref{fig:sindou}.  

\vskip 0.1cm

In Appendix A.6 for 2-body examples with a harmonic 
oscillator potential and a Coulomb potential, we show that use of 
the complex-range Gaussian basis functions
makes it possible to represent oscillating functions 
having more than 20 radial nodes accurately 
\mbox{(cf. Figs.~\ref{fig:howf-complex} and \ref{fig:sincos-coul})}.

\vskip 0.1cm
Hamiltonian matrix elements
between the complex-range Gaussians
can be calculated with  essentially the
same computation program for the real-range Gaussians
with some real variables replaced by complex ones;
this is another advantage of the complex-range Gaussians.

\vskip 0.1cm
Since the complex-range Gaussian basis functions makes 
the function space of few-body systems much wider than
that with real-range Gaussians, 
applicability of  GEM becomes much extended,
for example, in Refs.~\cite{Ohtsubo2013,Hiyama12COLD-1,Hiyama12COLD-2,
Hiyama14COLD-3,Matsumoto03PS,Matsumoto03CDCC,Kamimura09}
by the authors and collaborators.

%%%%%%%%%%%%%%%%%%%%%%%%%%%%%%%%%%%%%%%%%%%%%%%%
\subsection{Infinitesimally-shifted 
Gaussian-lobe (ISGL) basis functions}
\label{sec:ISGL}
%%%%%%%%%%%%%%%%%%%%%%%%%%%%%%%%%%%%%%%%%%%%%%%%

\vskip 0.1cm
When we proceed to 4-body systems,
calculation of the Hamiltonian matrix elements  
becomes much laborious especially when treating
many spherical harmonic functions  $Y_{lm}({\widehat {\bf r}})$
in the matrix element calculation.
In order to make the 4-body calculation 
tractable even for complicated interactions, 
one of the present authors (E.H.) 
proposed the infinitesimally-shifted Gaussian-lobe (ISGL)
basis functions~\cite{Hiyama94,Hiyama95u,Hiyama03}. 
The Gaussian function 
$r^l e^{-\nu_n r^2} Y_{lm}(\widehat {\bf r})$
is replaced by a superposition of 
infinitesimally-shifted Gaussians as
\begin{eqnarray}
&& N_{nl}\, r^l e^{-\nu_n r^2} Y_{lm}({\widehat {\bf r}}) 
=N_{nl}\, \lim_{\varepsilon \rightarrow 0}\; 
  \frac{1}{(\nu_n \varepsilon)^l}  \nonumber \\
&& \times
\sum_{k=1}^{k_{\rm max}}  \; C_{lm,k}\; 
e^{-\nu_n ({\bf r}\,-\,\varepsilon {\bf D}_{lm,k})^2}.
\label{eq:lobe0}
\end{eqnarray}
whose shift parameters $\{C_{lm,k}, {\bf D}_{lm,k};\,
k=1-k_{\rm max}\}$ are so determined that RHS  
is equivalent to  LHS (see \mbox{Appendix} A.1 in Ref.~\cite{Hiyama03}).

We make similar replacement of the basis functions
in all the other Jacobian coordinates.
Thanks to the absence of the spherical harmonics,
use of the ISGL basis functions makes the few-body Hamiltonian 
matrix-element calculation much easier 
with no tedious angular-momentum algebra (Racah \mbox{algebra}).
When and how to take lim$_{\varepsilon \to 0}$ is 
important (see \mbox{Appendix} A.1 in Ref.~\cite{Hiyama03}).
The Gaussian range $\nu_n$ can be taken to be complex $\eta_n$ 
as in Eq.~(\ref{eq:complex-range}) of the previous 
Sec.~\ref{sec:complex-range}.

\vskip 0.1cm
Owing to this advantage, applicability of GEM becomes
very wide in various research fields (cf. Fig.~\ref{fig:strategy}).
Furthermore, use of ISGL basis functions
make  it easier to calculate few-body resonance 
states (cf. Sec.~\ref{sec:CSM}) 
with the use of the complex-scaling 
method \mbox{(cf. Ref.~\cite{Aoyama2006}} for a review)
and to calculate few-body scattering 
states (cf.~\ref{sec:reaction}) with the use of the
Kohn-type variational principle to $S$-matrix~\cite{Kamimura77}.

\vskip 0.1cm
Here, we note  a history 
about 'Gaussian-lobe basis functions' 
(those {\bf not} taking lim$_{\varepsilon \to 0}$ but using 
a small $\varepsilon$ in Eq.~(\ref{eq:lobe0})).
Such basis functions (whose shift parameters were 
different from ours) were advocated in 1960's by several 
authors~\cite{old-lobe} on the basis of their simplicity
to mimic $Y_{lm}({\widehat {\bf r}})$ with \mbox{$l > 0$.}  
But, the functions have \mbox{serious weakpoints}; namely,
computation with very small $\varepsilon$ makes 
the result easily suffer from heavy round-off error, whereas
use of a \mbox{not-very-small $\varepsilon$} 
meets an inevitable admixture of higher-order 
$Y_{l'm}({\widehat {\bf r}})$ with $l'>l$.
Therefore, the functions  were not utilized 
in actual research calculations
and seemed soon forgotten when big computers came to real use.

But, some \mbox{30 years} after,  this difficulty was solved by 
one of the authors (E.H.)~\cite{Hiyama94,Hiyama95u,Hiyama03}
by introducing the ISGL basis functions
with properly \mbox{taking} lim$_{\varepsilon \to 0}$ 
{\it after} performing the analytical integration of the Hamiltonian 
matrix elements (see Appendix A3 and A.4 of Ref.~\cite{Hiyama03});
therefore, $\varepsilon$ does not appear in the computation program.

%\pagebreak
%%%%%%%%%%%%%%%%%%%%%%%%%%%%%%%%%%%%%%%%%%%%%
%%%%%%%%%%%%%%%%%%%%%%%%%%%%%%%%%%%%%%%%%%%%%
\section{Accuracy of GEM calculations}
\label{sec:AccuracyGEM}
%%%%%%%%%%%%%%%%%%%%%%%%%%%%%%%%%%%%%%%%%%%%%
%%%%%%%%%%%%%%%%%%%%%%%%%%%%%%%%%%%%%%%%%%%%%

%%%%%%%%%%%%%%%%%%%%%%%%%%%%%%%%%%%%%%%%%%
\subsection{Muonic molecule in muon-catalyzed fusion cycle}
\label{sec:MCF}
%%%%%%%%%%%%%%%%%%%%%%%%%%%%%%%%%%%%%%%%%%%%%

\begin{figure} [b!]
\begin{center}
\vskip -0.0cm
\epsfig{file=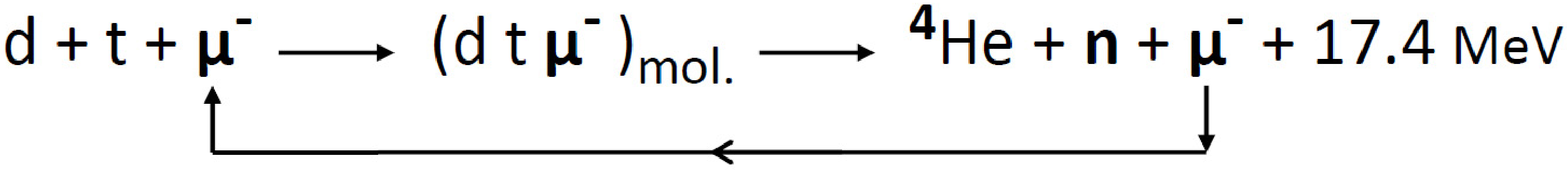,width=8.5cm}
\caption{Essence of muon-catalyzed fusion cycle 
in which the 3-body $dt\mu^-$
molecule (at $J=v=1$ state) plays a key role
to cause the fusion reaction $d+t \to~\!^4{\rm He}+n+17.4$MeV.
}
\label{fig:cycle}
\end{center}	
\end{figure}

The Gaussian expansion method was first proposed~\cite{Kamimura88}
in 1988 in the 3-body study of muonic molecule $dt\mu^-$
that appears in the cycle of 
muon-catalyzed $d$-$t$ fusion (for example, see 
Secs.~5 and 8 of Ref.~\cite{Hiyama03}
for a short survey, and Ref.~\cite{Nagamine1998} 
for a precise review).
The $d+t+\mu^-$ system is known to be a key to the possible
energy production by the muon-catalyzed 
fusion ($\mu$CF) as shown in Fig.~\ref{fig:cycle} for the
essential part of the catalyzed cycle.  

%%%%%%%%%%%%%%%%%%%%  Fig.5  %%%%%%%%%%%%%%%%%%%%%%%%%%

%%%%%%%%%%%%%%%%  Fig. 6%%%%%%%%%%%%%%%%%%%%%%%%%%%%%%
\begin{figure} [b!]
\begin{center}
\epsfig{file=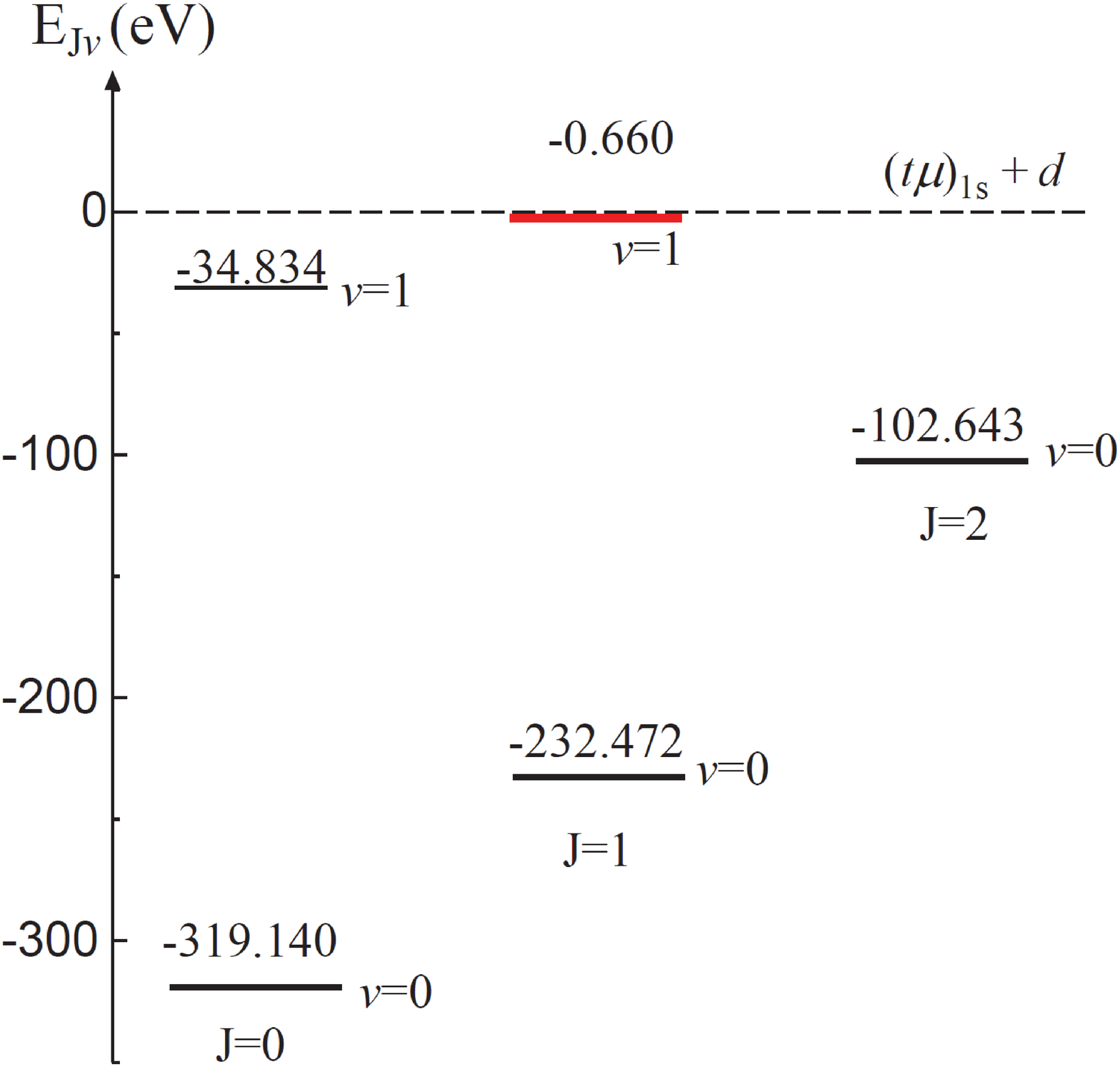,scale=0.21}
\caption{
Theoretically predicted energy levels of the $dt\mu^-$ molecule.
The near-threshold 
$J=v=1$ state (red) is important as the doorway to
the muon catalyzed fusion (Fig.~\ref{fig:cycle}).
This figure is reproduced from  Ref.~\cite{Kamimura88}.
}
\label{fig:mulevel}
\end{center}
\end{figure}
%%%%%%%%%%%%%%%%%%%%%%%%%%%%%%%%%%%%%%%%%% 
%%%%%%%%%%%%%%%%%  Fig. 7 (Jacobi dtmu) %%%%%%%%%%%
\begin{figure}[t!]
\begin{center}
\epsfig{file=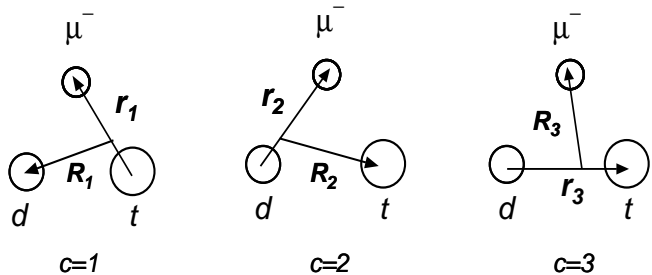,width=8.7cm,height=3.8cm}
\end{center}
\caption{
Three Jacobi coordinates of the $dt\mu^-$ molecule in 
muon-catalyzed fusion cycle. Use of them all is suitable for
describing the key $J=v=1$ state that is very weekly bound
from the $(t\mu)_{1s}+d$  threshold~\cite{Kamimura88}.
}
\label{fig:Jacobi-dtmu}
\end{figure}
%%%%%%%%%%%%%%%%%%%%%%%%%%%%%%%%%%%%%%%%%%%%%%%%%
%

\vskip 0.1cm
When negative muons $\mu^-$ are injected into the D$_2$/T$_2$ mixture,
muonic molecules $dt\mu^-$ are resonantly formed in its $J=v=1$ state 
(Fig.~\ref{fig:mulevel}) which is very loosely 
bound below the $(t\mu)_{1s}+d$ threshold
and is the key to $\mu$CF.
In~order~to analyze the 
observed data of the $dt\mu^-$ molecular formation rate,
accuracy of 0.001 eV is required in 
the calculated energy of the \mbox{$J=v=1$} state with respect to the
$(t\mu)_{1s}+d$ threshold. Since the threshold energy is $-2711.242$ eV
from the $d+t+\mu$ 3-body breakup threshold,
the accuracy of 7 significant figures is required 
in the Coulomb 3-body calculation.

\vskip 0.1cm
This difficult Coulomb 3-body problem 
was challenged during 1980's by many theoreticians 
from chemistry, atomic/molecular physics and nuclear physics.
The problem was finally solved in 1988 with the accuracy of
\mbox{7 significant} figures by three 
groups from USSR, USA and Japan giving  the
same energy of $-0.660$ eV 
from the $(t\mu)_{1s}+d$ threshold
using different calculation methods;
namely  
using a variational methods, respectively,
with elliptic basis~\cite{Korobov87},
with Slater geminal basis~\cite{Alexander88}
and with the GEM basis~\cite{Kamimura88}
(cf. Fig.~\ref{fig:Jacobi-dtmu} and Secs.~\ref{sec:3body-jacobi}
and \ref{sec:geomet}).

\vskip 0.1cm
An interesting point is the computation time
to solve the 3-body Sch\"{o}dinger equation
for \mbox{single} set of nonlinear variational parameters.
In the two methods~\cite{Korobov87,Alexander88} from
chemistry and atomic/molecular physics,
main difficulty comes from the severe non-orthogonality between 
their basis functions;  
diagonalization of the energy and overlap matrices 
%($\sim$~2000$\times$2000)
required {\it quadruple}-precision computation 
($\sim$30 decimal-digit arithmetics)
and the computation time of $\sim\!$~10~hours on
the computers at that time. 

\vskip 0.1cm
On the other hand, GEM~\cite{Kamimura88}  %%%%%%%%%%%
needed only 3 minutes. 
This rapid computation is owing to the
use of Gaussian basis functions, which are spanned over
the 3~rearrangement channels and 
have the ranges in geometrical progressions. Use of them 
suffers little from the trouble of severe non-orthogonality  
between large-scale ($\sim$2000) basis functions.
Therefore the method works entirely 
in double-precision ($\sim$14 decimal-digit arithmetics)
on supercomputers at that time.
Another reason was that the function form of the basis functions 
is particularly suitable for  {\it vector}-type supercomputers.

%%%%%%%%%%%%%%%%%%%%%%%%%%%%%%%%%%%%%%%%%%%%%%%%%%%%%%%%%%%
\subsection{3-nucleon bound states ($^3$H and $^3$He)}
\label{sec:3H-bound}
%%%%%%%%%%%%%%%%%%%%%%%%%%%%%%%%%%%%%%%%%%%%%%%%%
%
%

One of the best tests of three-body calculation method
is to solve three-nucleon bound states ($^3$H and $^3$He)
using a realistic $NN$ force.
This test was done for GEM  
in Ref.~\cite{Kameyama89} using the AV14 force~\cite{Wiringa84} and  in
Ref.~\cite{Kamimura90}  using the AV14 force plus the Tucson-Melborne
(TM) \mbox{3-body} force~\cite{Coon79}.  %%%%%%%%%%.
We shortly review them here.

\vskip 0.1cm

In practical calculations, we have to truncate the 
angular-momentum space  of the basis functions. 
It is to be stressed, however,
that the {\it interaction} is {\it not} truncated
in the angular-momentum space in the GEM calculations.
In the calculation described below 
we restrict the orbital
angular momenta $(l, L)$ of the spatial part of the basis functions
in Eqs.~(\ref{eq:gauss-r}) and (\ref{eq:gauss-R})
to $l +L \leq 6$, which results in
26 types of the $LS$-coupling configurations.
We  refer to such configurations as 
3-body angular-momentum channels.
% similarly to the terminology the Faddeev calculations. 
The 26 channels employed in our calculation are listed in 
Table I of Ref.~\cite{Kameyama89}
together with the Gaussian parameters. It is to be emphasized that 
all the nonlinear variational parameters of the GEM 
\mbox{calculation}
are explicitly listed in such a small table; in principle,
one can examine
the calculated results by using the {\it same} parameters.
%\vskip 0.4cm

%%%%%%%%%%%%%%%%%%%%%%%%  Fig.8  Convergence  %%%%%%%%%%%%%%%
\begin{figure}[t!]
\begin{center}
%\epsfig{file=h3-conv.eps,scale=0.43}
%\epsfig{file=h3-conv-new.eps,scale=0.35}
%##\epsfig{file=fig8-h3-conv-new.eps,scale=0.35}
\epsfig{file=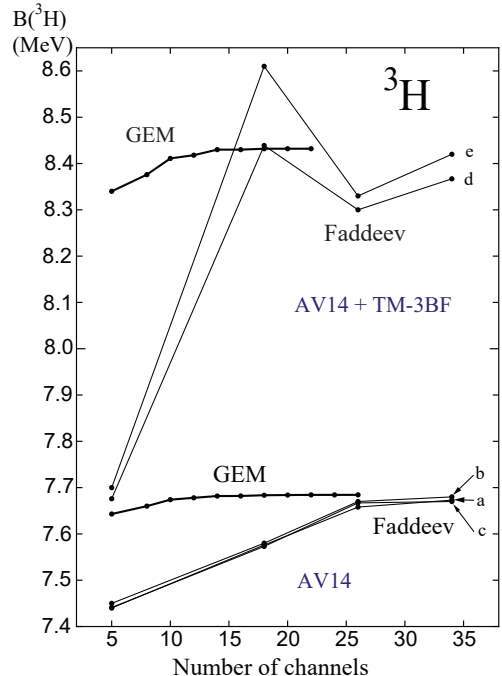,scale=0.35}
\caption{
Convergence of the binding energies of $^3$H 
calculated by the present method 
\cite{Kameyama89,Kamimura90} and by the Faddeev method
 with respect to the number of the three-body channels.  
Interactions used are AV14 (lower lines) and AV14+TM-3BF 
(upper lines). 
Ref.~\cite{Chen85} for line c, Ref.~\cite{Ishikawa86}
for b and e,  and Ref.~\cite{Sasakawa88} for a, d, f, g.
This figure is taken from 
Ref.~\cite{Kameyama89,Kamimura90}, where
a similar figure for $^3$He is given.
}
\label{fig:econverge}
\end{center}
\end{figure}
%%%%%%%%%%%%%%%%%%%%%%%%%%%%%%%%%%%%%%%%%%%%%%%%%
%

\vskip 0.1cm
Convergence of the binding energy of $^3$H 
with respect to the number of the 3-body 
angular-momentum channels
is illustrated in Fig.~\ref{fig:econverge}.
The results shown are those given by GEM in 
Refs.~\cite{Kameyama89,Kamimura90} some $\sim$ 30 years ago
\mbox{together} with those given by the  
Faddeev calculations at that time.
The convergence is very rapid in GEM.

\vskip 0.1cm
We note that one of the reasons for such a rapid convergence 
in the GEM framework comes from the fact that 
the interaction is treated {\it without} partial-wave decomposition
(namely, no truncation in the angular-momentum space).
%though the wave function is truncated in the space.
This is a difference from the Faddeev-method calculations 
and is also pointed out in \S 2.2 of Ref.~\cite{Gibson93}
by Payne and Gibson.

%%%%%%%%%%%%%%%%%%%%%%%%%%%%%%%%%%%%%%%%%%%%%
\subsection{Benchmark test calculations of 
4-nucleon ($^4$He) ground and second $0^+$ states}
\label{sec:bench-mark}
%%%%%%%%%%%%%%%%%%%%%%%%%%%%%%%%%%%%%%%%%%%%%

%\vskip 0.1cm
%\noindent
%i) {\it $^4${\rm He} ground state}
\subsubsection{$^4${\rm He} ground state}
%\vskip 0.2cm

Calculation of the 4-nucleon bound state ($^4$He)
using realistic $NN$ force
is  useful for
testing  methods and  schemes for few-body calculations.
In 2001, a very severe benchmark test calculation of 
the 4-body bound state was performed in Ref.~\cite{Kamada01}
by 18 authors, including the present \mbox{authors},
from seven research groups with the use of
their own efficient calculation methods, namely,
the Faddeev-Yakubovsky equation method (FY), 
the \mbox{Gaussian} expansion method (GEM),
the stochastic variational method (SVM),
the hyperspherical harmonic variational method (HH), 
the Green's function Monte Carlo (GFMC) method,
the no-core shell model (NCSM) and
effective interaction hyperspherical harmonic method (EIHH).
Those different calculation methods were
explained briefly in the paper~\cite{Kamada01}. 

\vskip 0.1cm
They used the  $NN$ realistic force, AV8$'$ interaction~\cite{AV8P97}
(consisting of central, spin-orbit and tensor forces),
and compared the calculated energy eigenvalues
and some wave function properties of the $^4$He ground state.

\vskip 0.1cm
The present authors (GEM) employed 4-body Gaussian basis functions
spanned over the full 18 sets of Jacobi coordinates
(composed of the K-type and H-type ones) as shown 
in Fig.~9.
%
%%%%%%%%%%%%%%%%%%%%%%%%  Fig. 9  %%%%%%%%%%%%%
\begin{figure}[b!]
\begin{center}
\epsfig{file=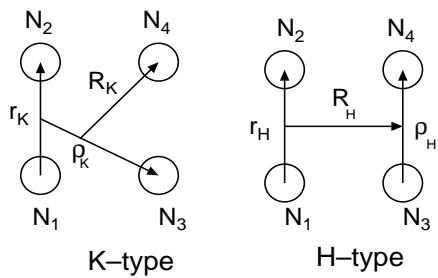,width=5.8 cm,height=3.6 cm}
\caption{K-type and H-type Jacobi coordinates for the
\mbox{4-nucleon} systems. Antisymmetrization
of the 4 particles 
generates the Jacobi coordinate sets $c=1, ..., 12$ (K-type) and 
$c=13, ...,  18$ (H-type).  See Fig.~18 of Ref.~\cite{Hiyama03}
for explicit figures of the 18 sets.
}
\end{center}
\label{fig:jacobi-4body}
\end{figure}
%\end{wrapfigure}
%%%%%%%%%%%%%%%%%%%%%%%%%%%%%%%%%%%%%

In the GEM approach, the most general 4-nucleon wave
function (with $J$ for the total angular momentum and
$T$ for the isospin)
is written as a sum of the component functions in the K-
and H-type Jacobi coordinates % (Fig.~\ref{fig:jacobi-4body}),
employing the $LS$ coupling scheme:

%\end{document}   %%% arxiv
   
%\pagebreak
\begin{eqnarray}
\!\!\! \Psi_{JM, TT_z} =
\!\sum_{\alpha}  C^{({\rm K})}_{\alpha}
\Phi^{({\rm K})}_\alpha +
\!\sum_{\alpha}  C^{({\bf H})}_{\alpha}
\Phi^{({\rm H})}_\alpha  ,% \nonumber
\label{eq:Psi-expansion}
\end{eqnarray}
where the antisymmetrized 4-body basis functions $\Phi^{({\rm
K})}_\alpha$ and $\Phi^{({\rm H})}_\alpha$ (whose suffix $JM,TT_z$
are dropped for simplicity) are described by
\begin{eqnarray}
 \Phi^{({\rm K})}_\alpha\!\! &=& \!\!       {\cal A}
 \left\{
\Big[ \big[ [\phi_{nl}^{({\rm K})}({\bf r}_{\rm K})
          \varphi_{\nu\lambda}^{({\rm K})}
          (\rhovec_{\rm K})]_\Lambda \;
       \psi_{NL}^{({\rm K})}({\bf R}_{\rm K}) \big]_{I}
     \right.    \nonumber \\
&& \quad \times      \big[ [\chi_s(12)
        \chi_{1/2}(3)]_{s'}
        \chi_{1/2}(4)
          \big]_{S}    \Big]_{JM}   \nonumber \\
&&\quad \times   \left.      \big[ [\eta_t(12)
        \eta_{1/2}(3)]_{t'}
        \eta_{1/2}(4)
          \big]_{TT_z} \right\} ,
\label{eq:Psi-K}  \\
\Phi^{({\rm H})}_\alpha \!\!& =& \!\!     {\cal A}
 \left\{
\Big[  \big[[\phi_{nl}^{({\rm H})}({\bf r}_{\rm H})
         \varphi_{\nu\lambda}^{({\rm H})}
           (\rhovec_{\rm H})]_\Lambda \;
        \psi_{NL}^{({\rm H})}({\bf R}_{\rm H}) \big]_{I}
     \right.     \nonumber \\
&&\quad \times     \big[ \chi_s(12)\chi_{s'}(34)
          \big]_{S}    \Big]_{JM}    \nonumber \\
&&\quad \times   \left.
          \big[ \eta_t(12)\eta_{t'}(34)
          \big]_{TT_z}  \right\}  ,
          \label{eq:amp}
\end{eqnarray}
with $\alpha \equiv \{nl,\nu \lambda,\Lambda,NL,I,s,s',S,t,t' \}$.
$\cal{A}$ is the 4-nucleon antisymmetrizer.
Parity of the wave function is given by $\pi=(-)^{l+\lambda+L}$.
The $\chi$'s and $\eta$'s are the spin and isospin functions,
respectively.  The spatial basis functions $\phi_{nlm}({\bf r})$,
 $\varphi_{\nu\lambda\mu} (\rhovec)$ and $\psi_{NLM}({\bf R})$
are taken to be Gaussians multiplied by spherical harmonics:
\begin{eqnarray}
&&\phi_{nlm}({\bf r}) =
N_{nl}\,r^l\:e^{-(r/r_n)^2}\:
Y_{lm}({\widehat {\bf r}})   \;,\nonumber \\
%\label{eq:3gaussa}
&&\varphi_{\nu \lambda \mu}(\rhovec)=
N_{\nu \lambda}\,\rho^\lambda\:e^{-(\rho/\rho_\nu)^2}\:
Y_{\lambda \mu}({\widehat {\rhovec}})
\;, \\ % \nonumber
\label{eq:4gauss}
&&\psi_{NLM}({\bf R}) =
N_{NL}\,R^L\:e^{-(R/R_N)^2}\:Y_{LM}({\widehat {\bf R}})
\;.  \nonumber
%\label{eq:4gauss}
\end{eqnarray}
\noindent
It is important to postulate that the Gaussian ranges lie in
geometric progression as in Eqs.~(\ref{eq:gauss-r}) and 
(\ref{eq:gauss-R}).

%%%%%%%%%%%%%%%%%%%%%%%%  Table 1  %%%%%%%%%%%%%%
\begin{table}[b!]
\begin{center}
\caption{Calculated results for some of 
$^4$He properties (binding energy, r.m.s radius and $D$-state
probability) by  seven methods of calculation. 
Reproduced from Ref.~\cite{Kamada01}.
%(note that GEM was referred to as CRC-GV in Ref.~\cite{Kamada01}).
}
\label{table:bench}
%\small
\begin{tabular}{llll}
\noalign{\vskip 0.05 true cm} \cr
\hline 
\hline 
\noalign{\vskip 0.2 true cm} 
%\vspace{0.5 mm} \\
Method    &B.E. (MeV)  &$\sqrt{\langle r^2 \rangle}$ (fm) 
&$D$ (\%)   \\
\noalign{\vskip 0.2 true cm} 
\hline 
%\vspace{-4 mm} \\
\vspace{-2 mm} \\
FY  &$25.94(9)$       &1.485(3)     &13.91  \\
GEM  &$25.90$       &1.482       &13.90  \\
SVM  &$25.92$       &1.486      &13.91  \\
HH  &$25.90(1)$       &1.483     &13.91  \\
GFMC  &$25.93(2)$       &1.490(5)  & $\;\; -$ \\
NCSM  &$25.80(20)$       &1.485    &12.98  \\
EIHH  &$25.944(10)$       &1.486   &13.89(1)  \\
\noalign{\vskip 0.1 true cm} 
\hline 
\hline\\
\end{tabular}
\end{center}
\end{table}
%%%%%%%%%%%%%%%%%%%%%%
%
%%%%%%%%%%%%%%%%%%%%%    Fig. 10   %%%%%%%%%%%%%%%%
\begin{figure}[t!]
\epsfig{file=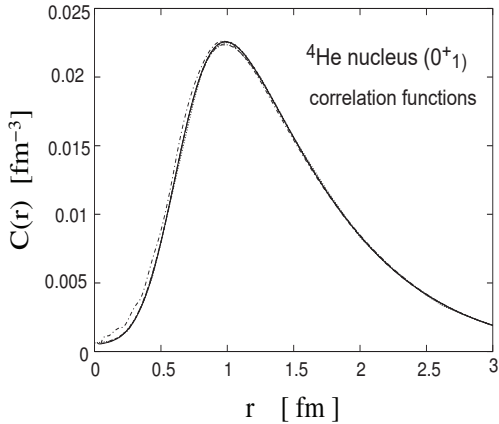,width=7.0 cm,height=6.0 cm}
\begin{center}
\vskip -0.3cm
\caption{Correlation functions (two-body density) of 
$^4$He, 
$C(r)\!=\!\langle \Psi | \delta({\bf r}{\rm-}{\bf r}_{12}) | \Psi \rangle $,
in the different calculational schemes:
FY, GEM, SVM, HH, and NCSM (overlapping curves)
and EIHH(dashed-dotted curve), 
except GFMC. 
Taken from Ref.~\cite{Kamada01}.
}
\end{center}
\label{fig:bench1}
\end{figure}
%%%%%%%%%%%%%%%%%%%%%%%%%%%%%%%%%%%%%

\vskip 0.1cm
The work of benchmark test~\cite{Kamada01} demonstrated 
that the Schr\"{o}dinger equation for the 
4-nucleon ground state can be handled very reliably by 
the different seven methods, leading to very good agreement between them in
the calculated results (some examples are shown
in Table~\ref{table:bench} and Fig.~10).     % \ref{fig:bench1}). 
This fact is quite remarkable in view of the very different techniques
of calculation and the complexity of the 
nuclear force chosen.

%\end{document}   %%% arxiv

%\vskip 0.3cm
%\noindent
%ii) {\it $^4${\rm He} second $0^+$ state}
\subsubsection{$^4${\rm He} second $0^+$ state}
%\vskip 0.1cm

Soon after the benchmark test, the present authors  
succeeded~\cite{Hiya04SECOND}, using the same GEM framework,  
in extending the $^4$He ground-state calculation
to the second $0^+$ state that has a
very loose spatial distribution compared with the
compact ground state. It can be a severe test for 
few-body calculation methods
to describe simultaneously the two $0^+$ states
that have very different properties.

%%%%%%%%%%%%%%%%%%%%%%  Table 2    %%%%%%%
%
\begin{table} [h!]  
\begin{center}
\caption{ (Upper) Calculated and observed binding energies of $^3$H, $^3$He, 
$^4$He$(0^+_1)$ and $^4$He$(0^+_2)$. The 4-body GEM 
calculation~\cite{Hiya04SECOND} takes the AV8$'$ and Coulomb potential
plus a phenomenological 3-body force.
(Lower) Calculated probability percentages 
of the $S,  P$ and $D$ states, which are
nearly the same between $^3$H ($^3$He) and $^4$He$(0^+_2)$.
This table is reproduced from Ref.~\cite{Hiya04SECOND}.
}
% to reproduce the binding energies
% of simultaneously}
\begin{tabular}{cccccc}
\noalign{\vskip 0.3 true cm} 
\hline 
\hline
\noalign{\vskip 0.1 true cm} 
  B.E. (MeV)  &   $\;\;^3$H$\;\;$  &  $^3$He &$\;$ & 
 $^4$He $(0^+_1)$ & $^4$He $(0^+_2)$ \\  
\noalign{\vskip 0.0 true cm} 
\hline 
\noalign{\vskip 0.1 true cm} 
GEM    &  8.41 &  7.74 && 28.44 & 8.19   \\
\noalign{\vskip 0.1 true cm} 
EXP    &  8.48 &  7.72 && 28.30 & 8.09  \\
\noalign{\vskip 0.1 true cm} 
\hline
\noalign{\vskip 0.3 true cm} 
  $P_S$ (\%)    & 90.96 & 90.99  &&  85.54   &   91.18 \\
  $P_P$ (\%)    & 0.08  & 0.08   &&  0.38    &    0.08 \\
  $P_D$ (\%)    & 8.97  & 8.93   && 14.08    &    8.74 \\
\noalign{\vskip 0.1 true cm} 
\hline
\label{table:3n-4n-energy}
\end{tabular}
\end{center}
\end{table}
%%%%%%%%%%%%%%%%%%%%%%%%%%%%%%%%%%%%%%

First, in order to reproduce simultaneously
the observed binding energies of $^3$H, $^3$He and $^4$He$(0^+_1)$  
before entering the $^4$He$(0^+_2)$  state,
we introduced a phenomenological 3-body force (Eq.~(3.1) of
Ref.~\cite{Hiya04SECOND}) in addition to
the AV8$'$ and  Coulomb forces. A good agreement 
for the former three states was
obtained as shown Table~\ref{table:3n-4n-energy} (upper part).
At the same time, the calculated  
binding energy of the $^4$He$(0^+_2)$ state was found to reproduce
the observed one well.

%%%%%%%%%%%%%%%%%%  Fig. (density)  %%%%%%%%%%%%%
\begin{figure}[b!]
\begin{center}
\epsfig{file=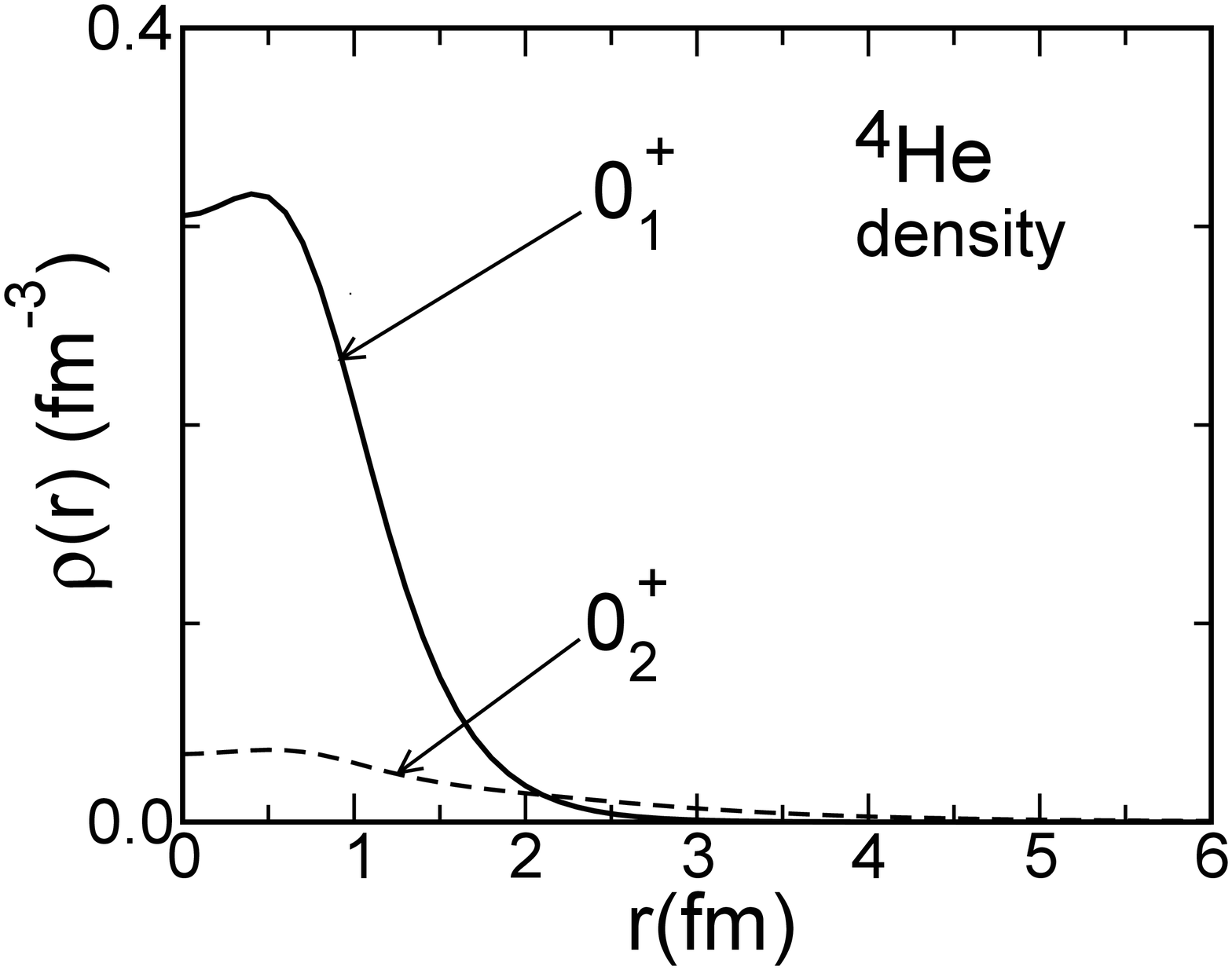,width=6.0cm,height=4.5cm}
\end{center}
\end{figure}
\begin{figure}[b!]
\begin{center}
\vskip -1.2cm
\epsfig{file=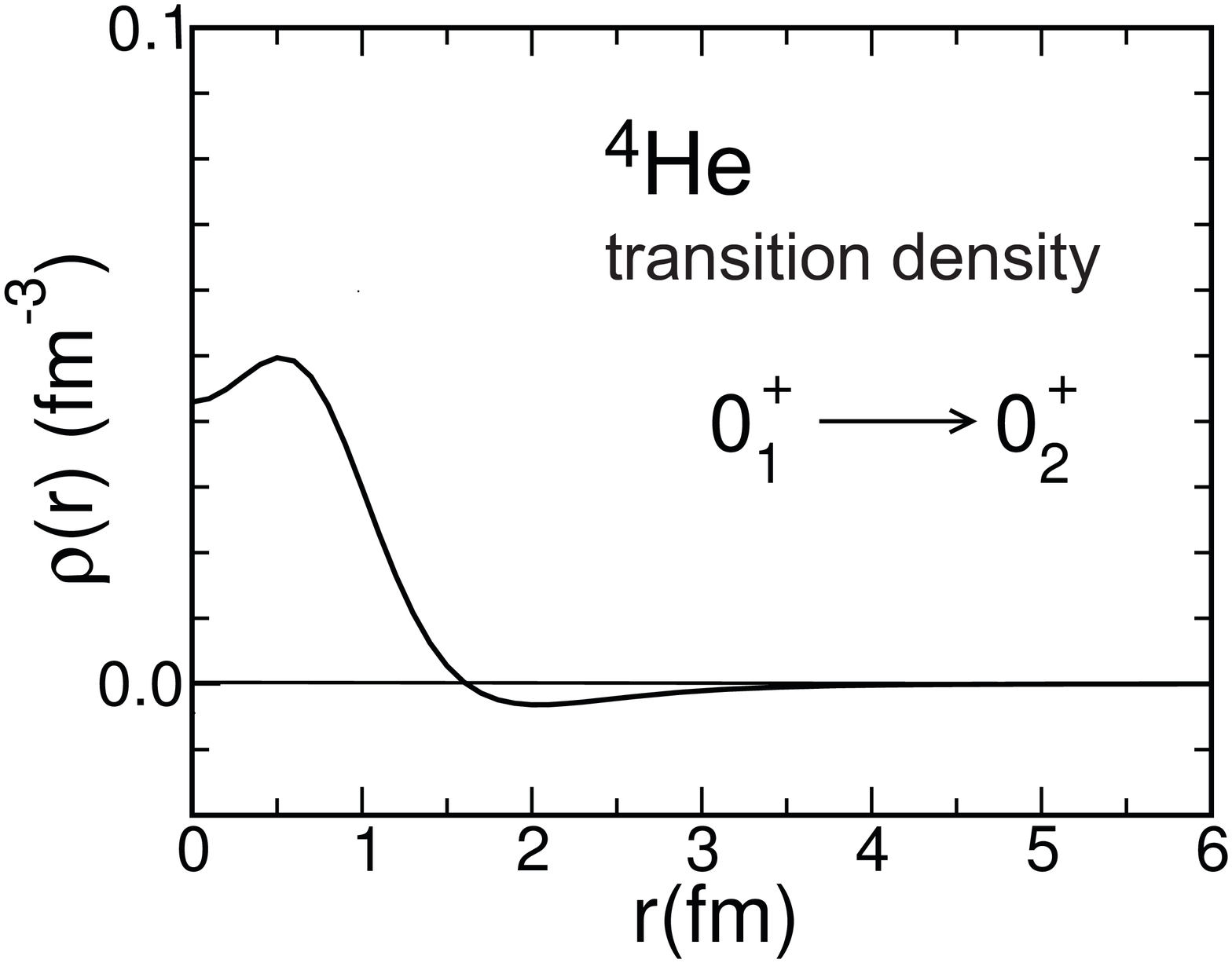,width=6.0cm,height=4.6cm}
\end{center}
\vskip -0.4cm
\caption{
Mass densities of the $0^+_1$ and $0^+_2$ states of
$^4$He (upper) and the transition density between them (lower).
Taken from Ref.~\cite{Hiya04SECOND}. }
\label{fig:density}
\end{figure}
%%%%%%%%%%%%%%%%%%%%%%%%%%%%%%%%%%%%%%%%%%%%%%
%\vskip -0.5cm

%%%%%%%%%%%%%%%%%%%%% Fig. 12  %%%%%%%%%%%%%%
\begin{figure}[b!]
\begin{center}
\epsfig{file=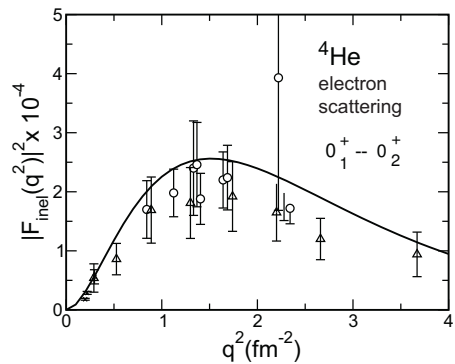,scale=0.26}
\end{center}    
\vskip -0.3cm
\caption{A GEM 4-body calculation~\cite{Hiya04SECOND} (solid line)
of the electron-scattering form factor
for $^4{\rm He}(e,e')^4{\rm He}(0^+_2)$, which
is compared with the available experimental data (for references, 
see Ref.~\cite{Hiya04SECOND}) in good agreement.
Taken from Ref.~\cite{Hiya04SECOND}.
} 
%\cite{Fro68,Wal70,Koe83}, $ \triangle $, $ \times $, 
%and O, respectively. }
\label{fig:he4inela}
\end{figure}
%%%%%%%%%%%%%%%%%%%%%%%%%%%%%%%%%%%%%%%%%%%
%

The lower part of Table~\ref{table:3n-4n-energy}
gives calculated probability percentages of the $S, P$ and $D$
components. Interestingly, they are almost the same 
between $^3$H ($^3$He) and $^4$He$(0^+_2)$.
This means that the loosely coupled
$^3{\rm H}+p$ ($^3{\rm He}+n$) configuration 
is dominant in the second $0^+$ state.

\vskip 0.1cm
As shown in Fig.~\ref{fig:density} (upper), 
distribution of the calculated mass densities
are quite different between the $0^+_1$ and $0^+_2$ 
states as expected. The transition density \mbox{between} the two states
in Fig.~\ref{fig:density} (lower) 
provides, via a Fourier transformation, 
the inelastic electron-scattering 
form factor of $^4$He($e,e')^4$He($0^+_2$).  
Therefore, comparison of the form factor with the observed one
can be another severe test of the GEM calculation.
We reproduced, for the first time using  realistic $NN$ interaction,
the observed $^4$He($e,e')^4$He($0^+_2$) data as shown in
Fig.~\ref{fig:he4inela}.

\vskip 0.1cm
We note that our results for the second $0^+$ state can be used 
in another new benchmark test calculation
(the result of Table~\ref{table:3n-4n-energy} was 
reproduced by Ref.~\cite{WHoriuchi}).
 
%\end{document}   %%% arxiv

%%%%%%%%%%%%%%%%%%%%%%%%%%%%%%%%%%%%%%%%%%%%%%%%%%%
\subsection{Determination of antiproton mass by GEM}
\label{sec:pbar-mass}
%%%%%%%%%%%%%%%%%%%%%%%%%%%%%%%%%%%%%%%%%%%%%%%%%%%

The mass of antiproton has been believed to be the same 
as the mass of proton, but there was no precise experimental 
information on it before 2000.
In the 1998 edition of Particle Listings
\cite{Listing1998}, %%%%%%%%%%%%%%%%
the Particle Data Group gave no recommended value of the
antiproton mass. 

%Instead,  they only cited several scattered values obtained until that time.
%The reason why it is difficult to determine antiproton mass
%is as follows.
%The charge-to-mass ratio was determined very precisely, 
%with $9 \times 10^{-11}$ uncertainty
%\cite{Gabrielse99}, %%%%%%%%
%from the periodic motion of an antiproton in a magnetic field.
%Another relation between the charge and the mass 
%is given by the energy of the X-ray from 
%${\bar p}$ atoms, 
%but the experimental error is as large as
%$10^{-5}$ to $10^{-4}$.  

\vskip 0.1cm
In the Particle Listings 2000~\cite{Listing2000}, %%%%%%%%%%%%%%%
a recommended value was given for the first time;
the relative deviation of the antiproton mass from 
the proton mass ($|m_{\bar p}-m_p|/m_p)$ is within 
$5 \times 10^{-7}$.
% The 2002 edition~\cite{Hagiwara02} %%%%%%%%%%%%
% reported an order of magnitude smaller value of upper limit,  
% $6\times10^{-8}$.

\vskip 0.1cm
This value was derived by a collaboration of 
experimental and theoretical
studies of % highly-excited metastable states in
the antiprotonic helium atom (${\bar p}$He$^+$)
composed of He$^{2+} + {\bar p} + e^-$,    
namely, by the high-resolution laser spectroscopy experiment 
at CERN by Torii {\it et al.}~\cite{Torii99} 
% and by Hori {\it et al.} \cite{Hori01}
and the precision 3-body calculations 
by Kino, Kudo and one of the authors (M.K.)~\cite{Kino98,Kino99}
(summarized in Sec.~6 of Ref.~\cite{Hiyama03} 
together with Refs.~\cite{Kino00,Kino01}).

\begin{figure} [b!]
\begin{center}
\epsfig{file=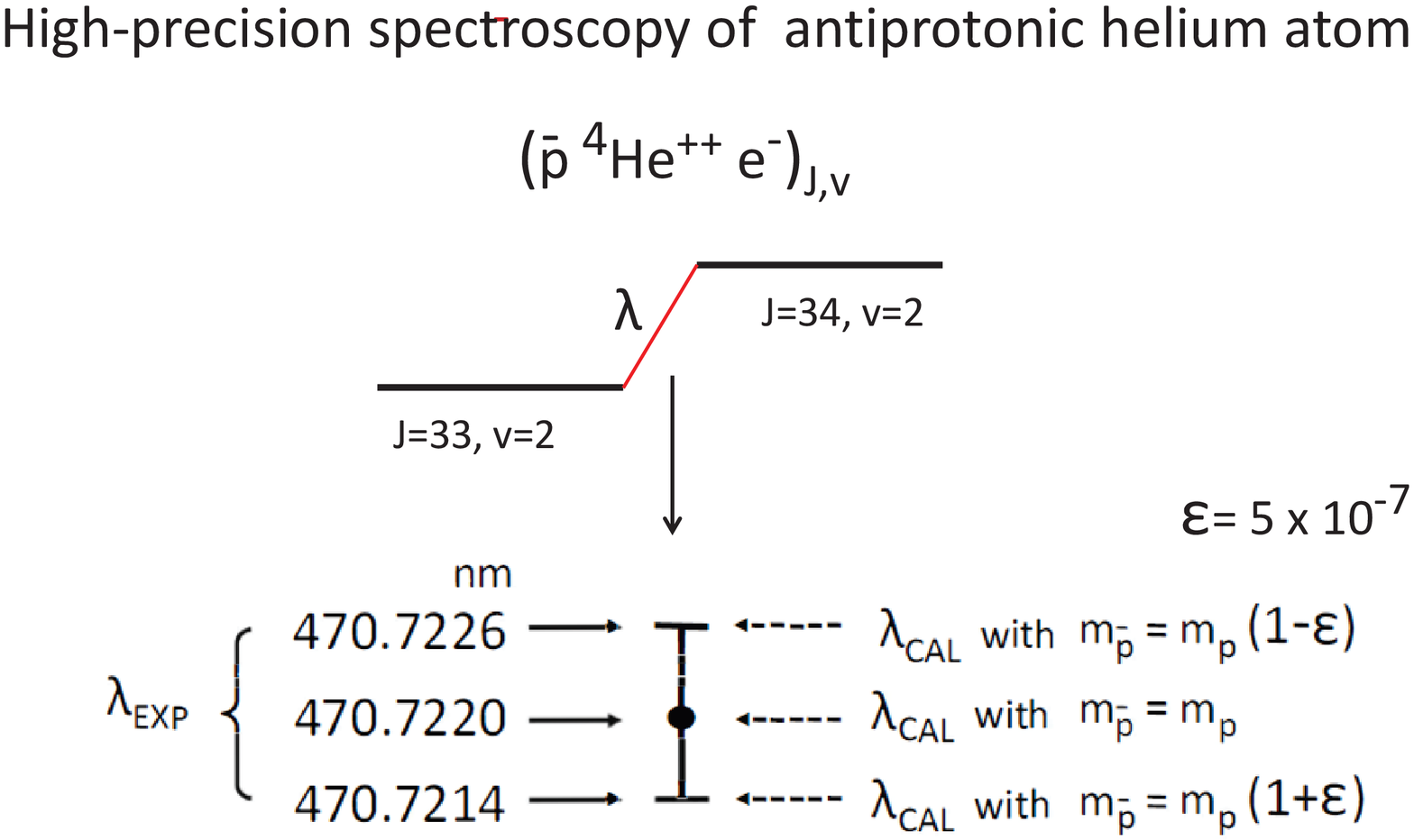,width=8.5cm}
\vskip -1.0cm
\caption{$\;$ 
Relative difference of the antiproton mass $(m_{\bar p})$ from
the proton mass ($m_p$), $\varepsilon = |m_{\bar p}-m_p|/m_p$,
was determined by the comparison between the
spectroscopic experimental data ($\lambda_{\rm EXP}$)~\cite{Torii99}
and the 3-body GEM calculation ($\lambda_{\rm CAL}$)~\cite{Kino99}
on the antiprotonic He atom (${\bar p}$He$^+$).
This gave $\varepsilon = 5 \times 10^{-7}$.
}
\label{fig:pbar-mass-error}
\end{center}	
\end{figure}

\vskip 0.1cm
The experiment for the transition between the
highly-excited metastable states with
$(J,v)=(34,2)$ and $(J,v)=(33,2)$ 
gave the wave length $\lambda_{\rm EXP}=470.7220 (6)$ nm
(Fig.~\ref{fig:pbar-mass-error}).
But, it is to be noted that this value of $\lambda_{\rm EXP}$ itself
does not directly give any information on the antiproton mass.
In Ref.~\cite{Kino99}, the data were analyzed so that  
the mass of antiproton could be derived.

\begin{figure}[b!]
\begin{center}
\epsfig{file=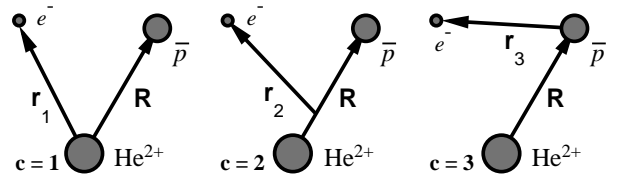,width=8cm}
\end{center}
\caption[]{
Three rearrangement channels for 
the antiprotonic helium atom (He$^{2+}+e^- +{\bar p}$).
Channels $c=1$ and $c=2$ are  suitable for 
describing the atomic picture and the molecular picture,
respectively, of this system. 
Channel $c=3$ is good at treating the correlation 
between the electron and the antiproton explicitly.
The mass-polarization term in the kinetic-energy operator
due to this choice of the coordinates
is exactly treated.
}
\label{fig:kinojacobi}
\end{figure}

\vskip 0.1cm
In the following, we briefly explain the GEM calculation
of the antiprotonic helium atom that is called 
\mbox{{\it atomcule}} since it
has two different facets, 
\mbox{i) atomic} picture of a positive-charge nucleus (He$^{2+}$)
plus two negative-charge particles
and \mbox{ii) molecular} picture of two heavy particles 
(He$^{2+}$ and ${\bar p}$) plus an electron 
(Fig.~\ref{fig:kinojacobi}).

%The data were precisely analyzed  
%in Ref.~\cite{Kino98,Kino99,Kino00,Kino01}
%and derived the antiproton mass mentioned above.
%\vskip 0.2cm
%%%%%%%%%%%%%%%%%%%%%%%  Fig. 13  %%%%%%%%%%%%%%%%%%%%%%%
%%%%%%%%%%%%%%%%%%%%%%%%%%%%%%%%%%%%%%%%%%%%%%

%\clearpage

%%%%%%%%%%%%%%%%%%   Fig. 14  %%%%%%%%%%%%%%%%%%%%%%%%%
%%%%%%%%%%%%%%%%%%%%%%%%%%%%%%%%%%%%%%%%%

This complicated system has 
difficult but important issues as follows:
\vskip -0.2cm

\begin{description}
\item{1)} The two different facets mentioned above should be well
described simultaneously (GEM takes the channels $c=1$ and 2 
in Fig.~\ref{fig:kinojacobi}).

\item{2)}
The excited states concerned are not true bound states
but so-called Feshbach resonances (GEM takes the complex-scaling method
of Sec.~\ref{sec:CSM}).

\vskip -0.2cm
\item{3)} Quantum number of the total angular momentum  concerned is 
as high as $J \sim 30 - 40$.

\item{4)} The inter-nuclear motion between the helium nucleus 
($Z=+2$) and the antiproton ($Z=-1$)  can 
not be treated adiabatically when they are close to each other
(GEM is a non-adiabatic method).

\item{5)} The correlation between  the electron and the
antiproton must be accurately taken into account
(GEM takes the channel $c=3$ explicitly).

\item{6)} Accuracy of 8 significant figures in the
transition energy (10 figures in eigenenergies before subtraction) 
is required 
to compare with the laser experiment of the transition frequency.
\end{description}

All of the issues 1) through 6) are difficult, but  
the GEM calculation in Refs.~\cite{Kino98,Kino99}
cleared them all and made it possible to  determine 
the antiproton mass recommended in Particle Listings 2000.
We explain how to determine the antiproton mass
using the eigenenergies given by the 3-body GEM calculation.

\vskip 0.1cm
The authors of Ref.~\cite{Kino99} showed 
that the central value of $\lambda_{\rm EXP}$ was reproduced 
by $\lambda_{\rm CAL}$ when taking \mbox{$m_{\bar p}=m_p$}
and that the upper and lower bounds 
of $\lambda_{\rm EXP}$
were respectively reproduced by assuming  
(cf. Fig.~\ref{fig:pbar-mass-error})
%%%%%%%%
\begin{eqnarray}
m_{\bar p}= (1 \mp \varepsilon) m_p \quad
\mbox{with} \quad \varepsilon = 5 \times 10^{-7}.  
\end{eqnarray}
%%%%%%%%%%%%%%%%%%%%%%%
Here, the relativistic and QED corrections 
were taken into account; the corrections 
are $\sim 10^{-5}$ times smaller than the
non-relativistic result.

\vskip 0.1cm
The authors of Ref.~\cite{Kino99}
then considered -- even if the antiproton mass $m_{\bar p}$ 
is deviated from $m_p$, the calculated wavelength $\lambda_{\rm CAL}$
using the $m_{\bar p}$ should be 
within the experimental error (namely, the experimental error is
fully attributed to the ambiguity of the antiproton mass). 
Then, they  reached the conclusion 
%%%%%%%%%%%%%%%%
\begin{eqnarray}
(1 - \varepsilon) m_p < m_{\bar p} <(1 + \varepsilon) m_p
\end{eqnarray}
%%%%%%%%%%%%%%%%
namely, 
\begin{eqnarray}
\frac{|m_{\bar p}-m_p|}{m_p} < \varepsilon = 5 \times 10^{-7},
\end{eqnarray}
%%%%%%%%%%%%%%%%%
which was cited in Particle Listings 2000~\cite{Listing2000};
it was commented that this can be a test of $CPT$ invariance.
GEM is so accurate as to contribute to such a
fundamental issue.
More about the ${\bar p}$He$^+$ atom and $m_{\bar p}$ is
given in Sec.~\ref{antiprotonic}. 

%%%%%%%%%%%%%%%%%%%%%%%%%%%%%%%%%%%%%%%%%%%%%
\subsection{Calculation of $^4$He-atom tetramer 
in cold-atom physics (Efimov physics)}
\label{cold-atom}
%%%%%%%%%%%%%%%%%%%%%%%%%%%%%%%%%%%%%%%%%%%%%

%\vskip 0.2cm
%\noindent
\subsubsection{Universality in few-body systems}
%\vskip 0.1cm

An essential issue in the cold-atom physics (Efimov physics) 
may be stated as that
%{\it universality} in few-body systems with large scattering length;
particles with short-range interactions 
and a large scattering length 
have \mbox{universal} low-energy properties that do not depend on
the details of their structure or their interactions at short 
distances (see, for example, Ref.~\cite{Braaten2006} for a review).
Such an pair-interaction is often called \mbox{`resonant interaction'}
since the interacting pair has a resonance or a bound state
that is located very closely to the 2-body breakup threshold.
Typical examples are the interaction between $\alpha$ particles
($^4$He nucleus) and   that between $^4$He atoms.

\begin{figure}[t!]
\begin{center}
%\vskip -1.3cm
%\epsfig{file=hiyama-fig-6.eps,width=7.8cm,height=4.7cm}
%\epsfig{file=level-He-clusters4.eps,width=8.5cm} 
\epsfig{file=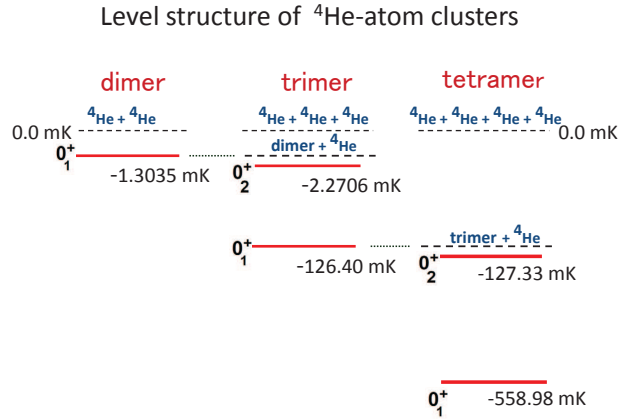,width=8.5cm} 
\end{center}
\caption{Level structure of $^4$He-atom clusters calculated by
the present authors~\cite{Hiyama12COLD-1}
using the realistic interaction between $^4$He atoms. 
Note that this structure is quite resemble to that of
the  2-, 3- and 4-$\alpha$ clusters ($^8$Be, $^{12}$C and $^{16}$O nuclei)
due to the universality in Efimov physics.
}
\label{fig:level-cluster}
\end{figure}

The level structure of $^4$He-atom dimer, trimer and tetramer
is illustrated in Fig.~\ref{fig:level-cluster};
calculation of all of the levels using the realistic 
interactions between $^4$He atoms   %~\cite{LM2M2}
was performed, for the first time,  by
the present \mbox{authors~\cite{Hiyama12COLD-1}}
as discussed below.
It is interesting to note that this level structure is
very similar to that of the lowest-lying $0^+$ states
in 2$\alpha$ ($^8$Be), 3$\alpha$ ($^{12}$C) and 4$\alpha$ ($^{16}$O)
nuclei though the scale of the two interactions is quite different to
each other; this is due to the universality mentioned above.

%%%%%%%%%%%%%%%%%%%%%%  Fig. 15   %%%%%%%%%%%%%
%%%%%%%%%%%%%%%%%%%%%%%%%%%%%%%%%%%%%%%%%%%

\vskip 0.1cm
Theoretical study of energies and wave functions of
the 3- and 4-body $^4$He-atom clusters
is one of the fundamental subjects in
the cold-atom physics 
since the realistic interaction between $^4$He atoms
is a prototype and well-studied 
interaction in the Efimov physics.
The interaction has an extremely strong short-range 
repulsive core due to the Pauli principle between
electrons \mbox{($\sim 10^6$ K} in height) 
followed by a weak attraction by 
the van der Waals potential ($\sim -10$~K in depth) 
(see Fig.~\ref{fig:dimer-short} in \mbox{Appendix A.5} 
for a typical LM2M2 potential~\cite{LM2M2}). %% the $^4$He-atom dimer) 
The interaction  has
a large scattering length ($\sim100\,$\AA)  much larger than the
interaction range ($\sim5\,$\AA) and
supports a very shallow bound states ($\sim -0.001$ K).

%\vskip 0.5cm
%\noindent
\subsubsection{Difficulty in calculating $^4${\rm He}-atom tetramer}
%\vskip 0.2cm

Until the energy levels of Fig.~\ref{fig:level-cluster}
were reported~\cite{Hiyama12COLD-1}, 
a long standing problem in the study of  $^4$He-atom  
clusters was the difficulty in performing 
a reliable \mbox{4-body} calculation
of the very-weakly-bound  excited state $(\mbox{$v=1, \,$} 0^+_2)$ 
of $^4$He-tetramer in the presence of extremely strong short-range 
repulsive core; one has to
describe accurately both the short-range structure 
($\lesssim  5\!$ \AA)
and the long-range asymptotic behavior (up to $\sim 1000\,$\AA).

%%%%%%%%%%%%%%%%%%%%%%%%%%  Fig. 16  %%%%%%%%%%%%%%%%
\begin{figure}[t!]
\begin{center}
\epsfig{file=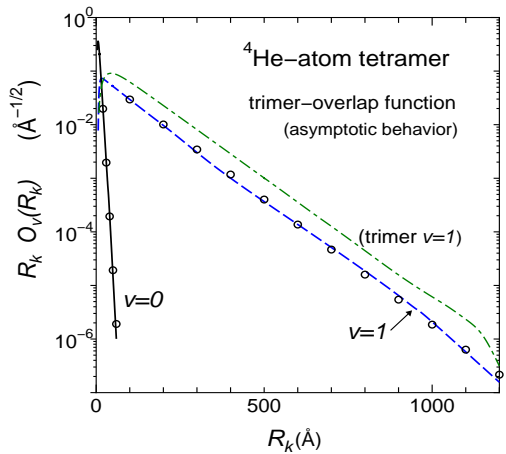,width=6.7cm,height=6.1cm}
\end{center}
\caption{
Good asymptotic behavior, up to $\sim 1000$ \AA, of the overlap function 
$O_v(R_k)\!\!=\!\!\langle \Psi_{3,v_3=0} 
| \Psi_{4,v} \rangle $, multiplied by $R_k$,
between the trimer ground state $(v_3=0)$
and the tetramer states $(v=0,1)$.
Open circles represent
the exact asymptotic behavior.
The green dash-dotted line is  the same quantity between
trimer excited state and dimer. Taken from Ref.~\cite{Hiyama12COLD-1}.
}
\label{fig:tetra-red-long}
\end{figure}
%%%%%%%%%%%%%%%%%%%%%%%%%%%%%%%%%%%%%%%%%%%

\vskip 0.1cm
The authors of Ref.~\cite{Carbonell} (2006), 
who used the \mbox{4-body} Faddeev-Yakubovsky method, said
``A direct calculation of the $^4$He-tetramer excited state
represents nowadays a hardly realizable task'';
instead, they derived  the excited-state binding energy 
by an extrapolation from 
a low-energy atom-trimer scattering $S$-matrix.

\vskip 0.1cm
However, this difficult problem was solved by the present 
authors~\cite{Hiyama12COLD-1} (2012) with 
a \mbox{4-body} GEM calculation. We employed the same set of 
all the 4-body Jacobi-coordinates of Fig.~9
as used in the 4-nucleon study in Sec.~\ref{sec:bench-mark}.
The energy of the $^4$He-tetramer excited state
was obtained as $E=-0.00093$ K with respect to the 
atom-trimer threshold (Fig.~\ref{fig:level-cluster}). 
In this calculation we took 23504 4-body basis functions
whose nonlinear parameters are all listed in a small 
table of 14 lines (Table~V of Ref.~\cite{Hiyama12COLD-1}) 
as pointed out in Sec.~\ref{sec:easy-opt}.

%%%%%%%%%%%%%%%%%%%%   Fig. 17  %%%%%%%%%%%%%%%
\begin{figure}[t!]
\begin{center}
\epsfig{file=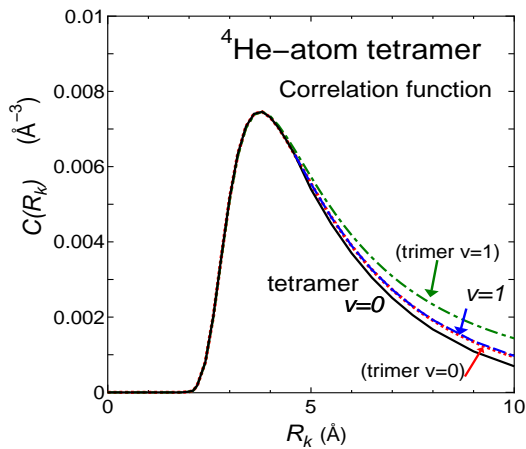,width=7.0cm,height=6.0cm}
\end{center}
\caption{
Short-range structure of the pair correlation function
$C(r)\!=\!\langle \Psi | \delta({\bf r}{\rm-}{\bf r}_{12})
 | \Psi \rangle $
of the $^4$He tetramer calculated with 
the 4-body GEM~\cite{Hiyama12COLD-1}.
The black solid line stands for the tetramer ground $(v=0)$ state
and the blue dashed line for  the excited $(v=1)$ state.  
For the sake of comparison, additionally  shown are  
the red dotted line for the trimer ground  state
and the green dash-dotted line for the trimer excited state. 
That for dimer nearly overlaps with the green line.
The lines are normalized to the peak value of the
black line.  It is striking that
the same shape of the short-range correlation ($r \lesssim 5 $ \AA $\,$)
appears in all the states. Taken from Ref.~\cite{Hiyama12COLD-1}.
}
\label{fig:tetra-den-short}
\end{figure}
%%%%%%%%%%%%%%%%%%%%%%%%%%%%%%%%%%%%%%%%%%% 
%%%%%%%%%%%%%%%%%%%%%%%%%%%%%%%%%%%%%%%%%%%

The excited-state wave function exhibits   correct
asymptotic behavior  up to \mbox{$\sim\!1000$\AA} 
as seen in Fig.~\ref{fig:tetra-red-long}
for the overlap function between 
the tetramer excited state and the trimer ground state.
In Fig.~\ref{fig:tetra-den-short}, it is interesting to see 
that behavior of the extremely-strong short-range correlations 
\mbox{($\lesssim  5\!$ \AA)} in the tetramer has almost the same 
shape as in the dimer and in the trimer. 
This justifies the assumption in some literature calculations 
that the Jastrow correlation factor 
is \mbox{{\it a priori}} employed in few-body wave functions
so as to treat the strong repulsive force 
between the interacting pair.

%\clearpage
%\vskip 0.5cm
%\noindent
\subsubsection{Efimov scenario: {\rm CAL} versus {\rm EXP}}

Here, we do not intend to enter the details of
the cold-atom physics, but our calculations mentioned below 
are closely related to the keypoint of the physics as follows:

  Surprisingly to nuclear physicists, strength 
(in other word, scattering length) of the interaction 
between some ultra-cold atoms,
such as $^{133}$Cs, $^{85}$Rb and $^7$Li at $\mu$K, can be changed/tuned  
by a magnetic field from outside
utilizing  Feshbach resonances
of the atom pair located near the threshold. 
Realization of this experimental technics (at $\sim$ 2006) 
has very much developed the cold-atom physics (Efimov physics).
One can investigate the structure change (called Efimov \mbox{scenario)}
of the atom clusters
(dimer, trimer, tetramer,...) as a function of
the scattering length of the atom-atom interaction.

%%%%%%%%%%%%%%%%%%%%%%  Fig.18   %%%%%%%%%%%%%
\begin{figure}[t!]
\begin{center} 
%\epsfig{file=efimov-tetramer-fig6.eps,width=8.5cm,height=6.5cm}
%\epsfig{file=fig7-efimov-tetramer.eps,width=8.5cm,height=6.5cm}
%##\epsfig{file=fig18-efimov-tetramer-arima.eps,width=8.5cm,height=6.5cm}
\epsfig{file=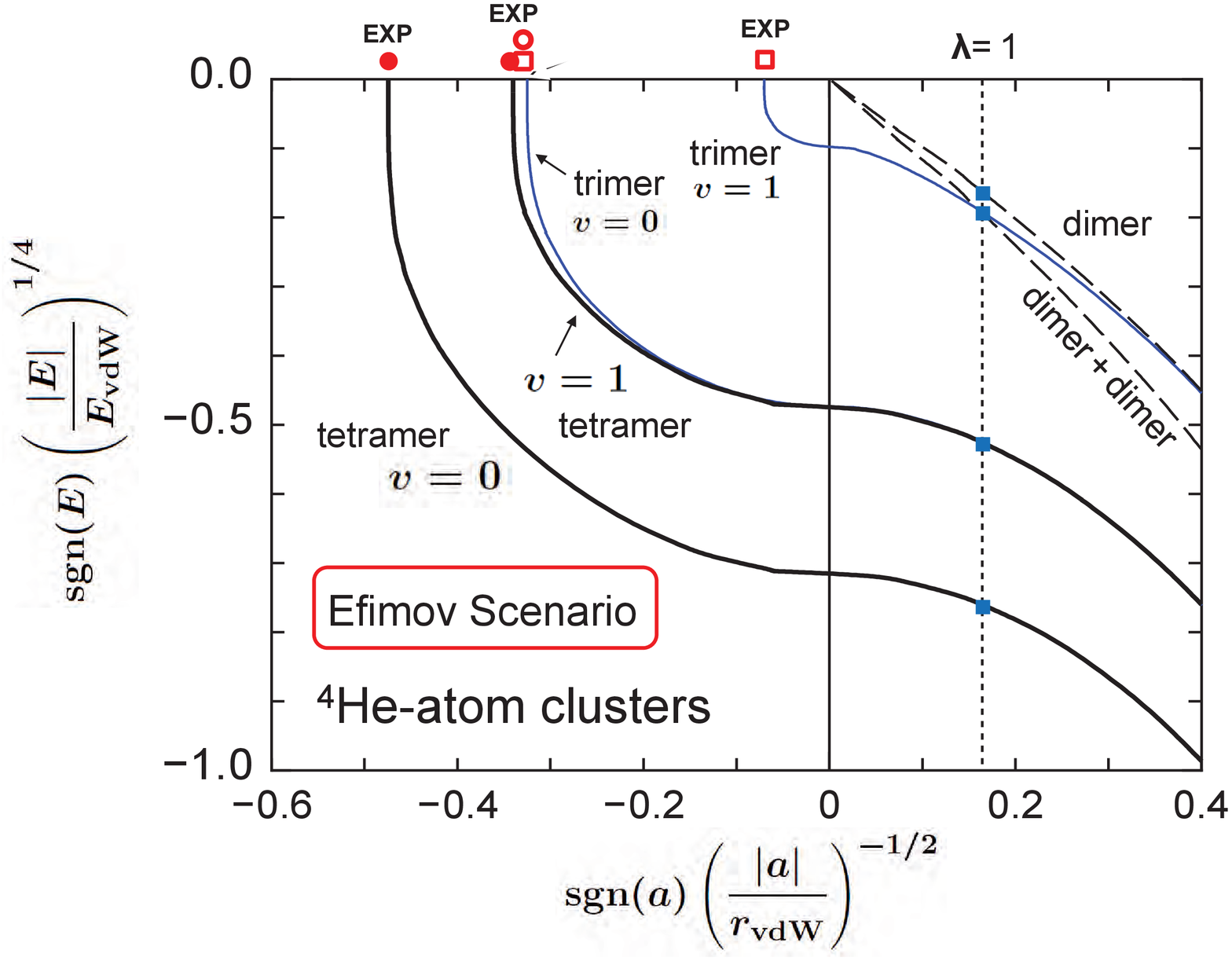,
width=8.5cm,height=6.5cm}
\caption{  
Efimov scenario (spectrum) for the $^4$He-atom clusters calculated 
in Ref.~\cite{Hiyama14COLD-3}
with the realistic $^4$He-$^4$He potentials.
The thick solid curves represent the tetramer spectrum that is 
the scaled tetramer energy $E_4^{(v)}/E_{\rm vdW}$ 
as a function of the scaled-inverse scattering length 
$(a/r_{\rm vdW})^{-1}$
for the ground $(v=0)$ and excited $(v=1)$ states.
The thin solid blue curves denote  the trimer spectrum. 
The critical scattering lengths where 
the tetramer energies $E_4^{(0)}$ and $E_4^{(1)}$ cross the 
4-atom threshold are named as $a_-^{(4,0)}$ and $a_-^{(4,1)}$, 
respectively; the corresponding observed values for the 
$^{133}$Cs, $^{85}$Rb and $^7$Li tetramers are given by red circles, 
and similarly for trimers by red boxes (see the text). 
Taken from Ref.~\cite{Hiyama14COLD-3}.}
\end{center}
\label{fig:scenario}
\end{figure}
%%%%%%%%%%%%%%%%%%%%%%%%%%%%%%%%%%%%%%%%%%%

\vskip 0.1cm
%%In Fig.~\ref{fig:scenario}, 
In Fig.~18, 
the present authors calculated~\cite{Hiyama14COLD-3}  
the \mbox{Efimov} scenario
(essentially, an energy spectrum of $E$ versus scattering length $a$)
for the first time using {\it realistic} atom-atom potential
(here, the $^4$He-atom potential).
Following the literature,
we have drawn 
$(|E|/E_{\rm vdW})^{1/4}$  versus
$(|a|/r_{\rm vdW})^{-1/2}$  so that all the curves are 
graphically represented on the same scale.
The scattering length $a$ and the energy $E$ are scaled
with the van der Waals length $r_{\rm vdW}\, 
(=5.08\, a_0$)   %~\cite{Chin10}) 
and  energy $E_{\rm vdW}=\hbar^2/m r_{\rm vdW}^2\, 
(=1.677$ K), respectively.
The dashed curve shows the dimer energy.

%\vskip 0.1cm
In Fig.~18, the scattering length $a$ are tuned by changing
the factor $\lambda$ which is multiplied
to the realistic $^4$He-$^4$He interaction:
\begin{equation}
\big[\, T +  \sum_{1=i<j}^A \lambda \, 
V(r_{ij}) - E_A \, \big] \,\Psi_A =0 ,
\end{equation}
where $T$ is the kinetic energy and $A (=2,3,4)$ is the number of 
$^4$He-atom clusters concerned.

%\vskip 0.1cm
The vertical dotted line stands for 
the physical value $\lambda=1$.
The blue circles on the line indicate the energy levels 
that are illustrated in Fig.~\ref{fig:level-cluster} with red lines;
namely, from the top, they are the energies of the dimmer, the trimer
excited state, the trimer ground state (overlapping with 
the circle for the tetramer excited state) and the tetramer ground state.

%\vskip 0.1cm
The states move to the left as $\lambda$ decreases ($a^{-1}$ decreases).
In the region  $a^{-1} < 0$, there is no 2-body bound state,
but the blue curves for the trimer show that
the 3-body system 
is bound (this is a general form of the so-called Borromine states).

\vskip 0.1cm
The critical scattering lengths where 
the tetramer energies $E_4^{(0)}$ and $E_4^{(1)}$ (black solid curves)
cross the 
4-atom threshold (the $E=0$ line)
are named as $a_-^{(4,0)}$ and $a_-^{(4,1)}$, 
respectively, and their values scaled with $r_{\rm vdW}$ are
summarized in \mbox{Table II} in Ref.~\cite{Hiyama14COLD-3}
together with the corresponding observed values (red circles
in Fig.~18 for
$^{133}$Cs, $^{85}$Rb and $^{6,7}$Li), and similarly for 
the trimers (red boxes for the corresponding observed values).

It is striking that the GEM calculation~\cite{Hiyama14COLD-3}
of the critical scattering lengths of the trimer and tetramer
using the  realistic potentials of $^4$He atoms
explains consistently the above-mentioned corresponding observed values
that are the heart of cold-atom (Efimov) physics.

%\end{document}  %arxiv

%%%%%%%%%%%%%%%%%%%%%%%%%%%%%%%%%%%%%%%%%%%%%%%%%%%%%%
%%%%%%%%%%%%%%%%%%%%%%%%%%%%%%%%%%%%%%%%%%%%%%%%%%%%%%
\section{Successful predictions by GEM calculations}
\label{sec:predict}
%%%%%%%%%%%%%%%%%%%%%%%%%%%%%%%%%%%%%%%%%%%%%%%%%%%%%%
%%%%%%%%%%%%%%%%%%%%%%%%%%%%%%%%%%%%%%%%%%%%%%%%%%%%%%

As mentioned in the previous section, 
applicability of GEM to various few-body calculations 
with high accuracy has been much improved.
Therefore, 
it became possible to make theoretical {\it prediction} 
before measurement (as long as interactions employed are reliable);
some successful examples are reviewed below.

%%%%%%%%%%%%%%%%%%%%%%%%%%%%%%%%%%%%%%%%%%%%%%%%%%%%%%%%
\subsection{Prediction of energy levels of antiprotonic He atom}
\label{antiprotonic}
%%%%%%%%%%%%%%%%%%%%%%%%%%%%%%%%%%%%%%%%%%%%%%%%%%%%%%%%%%%%%%%%%%%

As  mentioned in Sec.~\ref{sec:pbar-mass}, 
the precise 3-body GEM calculation of the antiprotonic helium atom
\mbox{(${\bar p}$He$^+$=He$^{2+}+{\bar p}+e^-$)}
contributed to the first determination of the antiproton mass
in Particle Listings 2000.
Since then, a lot of transitions between excited states of 
the atom were observed by CERN's laser experiment. 
But, due to very expensive cost of the precise 
sub-ppm laser-scan search 
of the transition energy $\Delta E$,
GEM was requested to predict $\Delta E$       
before measurements. 

\vskip 0.1cm
A typical example of the
transition frequency ($\nu$) by the
GEM prediction~\cite{Kino01}  and the experimental result~\cite{Hori01} 
is listed in Table~\ref{table:pbar-atom-energy}.
So accurate is the theoretical prediction using GEM.

%%%%%%%%%%%%%%%%%%%%%%  Table     %%%%%%%
%
\begin{table} [h!]  
\begin{center}
\small
\caption{Comparison of the prediction by GEM~\cite{Kino01}
with the CERN experiment~\cite{Hori01}
about the transition frequencies between some levels of
the antiprotonic helium atom (${\bar p}$He$^+$). 
}
\begin{tabular}{cclcl}
\noalign{\vskip 0.3 true cm} 
\hline 
\hline
\noalign{\vskip 0.1 true cm} 
 $(J, v) - (J', v') $  &&  $(32,0) - (31,0)$ &&  
 $(33,1) - (32,1)$ $\;$   \\
\noalign{\vskip  0.05 true cm} 
   &&  $\quad\nu$ (GHz) &&  $\quad\nu$ (GHz)  $\;$   \\
\noalign{\vskip 0.1 true cm} 
\hline 
\noalign{\vskip 0.15 true cm} 
  GEM && $\;$  1 012 445.559 $\;$ && $\;$   804 633.127(5)  $\;$  \\
\noalign{\vskip 0.15 true cm} 
  EXP && $\;$  1 012 445.52(17)$\;$ && $\;$ 804 633.11(11)   $\;$ \\
\noalign{\vskip 0.1 true cm} 
\hline
\noalign{\vskip 0.3 true cm} 
\label{table:pbar-atom-energy}
\end{tabular}
\end{center}
\end{table}
%%%%%%%%%%%%%%%%%%%%%%%%%%%%%%%%%%%%%%

\vskip 0.1cm
On the basis of this comparison, 
in the same way as in Sec.~\ref{sec:pbar-mass},
a relative deviation of the antiproton mass from 
the proton mass $|m_{\bar p}-m_p|/m_p < 6\times10^{-8}$ was  
reported in the 2002 edition 
of Particle Listings~\cite{Hagiwara02}.

\vskip 0.1cm
The laser spectroscopy of metastable 
antiprotonic helium atoms is a pioneering work 
toward anti-matter science.
We see that the GEM calculations was providing suggestive, 
helpful predictions for anti-matter science in a preliminary stage.

%%%%%%%%%%%%%%%%%%%%%%%%%%%%%%%%%%%%%%%%%%%%%%%%%%%%%%%%
\subsection{Prediction of shrinkage of hypernuclei}
\label{sec:shrink}

When a $\Lambda$ particle is injected into a nucleus,
how modified is structure of the nucleus?
There is no Pauli principle acting between 
$\Lambda$ and  nucleons in the nucleus.
Therefore, the $\Lambda$ particle can reach deep inside,
and attract the
surrounding nucleons towards the interior of the nucleus
(this is called "gluelike role" of $\Lambda$ particle).
However, how do we observe
the shrinkage of the nuclear size by the
$\Lambda$ participation?
In the work of Ref. \cite{Motoba83} based on
the microscopic $\alpha +x+ \Lambda$ 3-cluster
model $(x=d, t, ^3$He) for light $p$-shell
hypernuclei together with the
$\alpha +x$ \mbox{2-cluster} model for
the nuclear core, the reduction of the nuclear size was
discussed  in relation to the reduction of the
$B(E2)$ strength which is proportional
to the fourth power of the distance
between the clusters.

%
%
%%%%%%%%%%%%%%%%%%%%%%%  Fig.19  7Li-L level  %%%%%%%%%%%%%%%%%%
\begin{figure}[b!]
\begin{center}
\epsfig{file=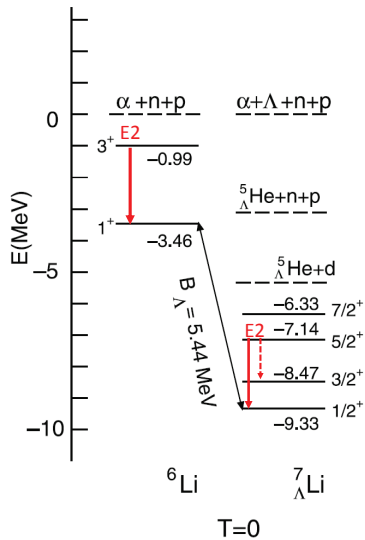,width=10.0cm}
\caption{$E2$ transitions in $^6$Li and in $^7_\Lambda$Li that are
used to discuss about the shrinkage of hypernucleus.
}
\label{fig:shrink-li7}
\end{center}
\end{figure}
%%%%%%%%%%%%%%%%%%%%%%
%%%%%%%%%%%%%%%%%%%%%%%  Fig.20   7Li-L jacobi  %%%%%%%%%%%%%%%%%%
\begin{figure}[t!]
\begin{center}
\epsfig{file=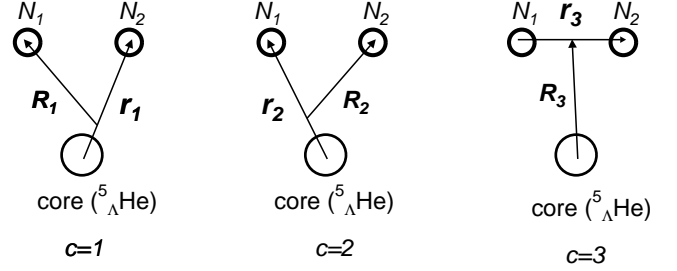,scale=0.45}
\caption{Jacobian coordinates of the 
${\rm core}+N_1+N_2$ system where
the core is hypernucleus $^5_{\Lambda}{\rm He}$ and
$N_1$($N_2$) is a nucleon.
}
\label{fig:jacobili7l}
\end{center}
\end{figure}
%%%%%%%%%%%%%%%%%%%%%%
%
\vskip 0.1cm
More precisely, in Ref.~\cite{Hiyama99},
we explicitly suggested  measurement of 
$B(E2; 5/2^+_1 \rightarrow 1/2^+_1)$ in $^7_{\Lambda}$Li 
 (Fig.~\ref{fig:shrink-li7}) and proposed a prescription to derive 
hypernuclear size %%%%% for the first time
with the aid of the
empirical values of $B(E2; 3^+_1 \rightarrow 1^+_1)$ and
the size of the ground state of $^6$Li.
We also noted that
another decay branch $B(E2; 5/2^+_1 \rightarrow 3/2^+_1)$ 
is negligibly small, measurement of the
lifetime of the $^7_{\Lambda}$Li$(5/2^+_1)$ state
can give the $B(E2; 5/2^+_1 \rightarrow 1/2^+_1)$. 
Afterwards, the experiment by Ref.~\cite{Tanida01} %%%%%%%%%%
was performed and the result was compared with our 
prediction on the size of $^7_{\Lambda}$Li. 

\vskip 0.1cm
We employed a microscopic 
$^5_{\Lambda}{\rm He} + n+p$ 3-body model
for $^7_{\Lambda}$Li~\cite{Hiyama99}.
It was examined in Ref.~\cite{Hiyama96-5L}  %%%%%%%%%%%%%%%
that the  $^5_{\Lambda}{\rm He}$ is a good cluster. 
The total 3-body wave function is constructed
on the Jacobian coordinates of
Fig.~\ref{fig:jacobili7l} %%%%%%%%%%%%
in the same manner as in the
3-body calculations in the previous sections.
Interactions employed are described in 
Ref.~\cite{Hiyama99}.
%\vskip 0.4cm

The observed energies of the $1/2^+_1$ and $5/2^+_1$
were well reproduced by the calculations,
and the value $B(E2; 5/2^+_1 \to 1/2^+_1)= 2.42\, e^2$fm${^4}$
was predicted. This is much smaller than
the observed   $B(E2; 3^+_1 \rightarrow 1^+_1)=
9.3 \pm 2.1 \,e^2$fm${^4}$
for the $^6$Li core which is well reproduced by our
$^6{\rm Li} =$ $^4{\rm He}+n+p$ 3-body model 
whose prediction is $9.26 \,e^2$fm${^4}$.
It should be noted, however, that one cannot conclude
the size-shrinkage from the
reduction of the $B(E2)$ value alone since the
$B(E2)$ operator $r^2 \,Y_{2 \mu}(\theta, \phi)$ 
includes the angle part.
Furthermore, we should note that the shrinkage of
$^7_{\Lambda}$Li can occur both along the
$n-p$ relative distance and  along the distance
between the $^5_\Lambda$He core and the c.m. of the $(np)$ pair.
%
%%%%%%%%%%%%%%%%%%%%%%%  Fig.21   7Li-L density  %%%%%%%%%%%%%%%%%%
\begin{figure}[t!]
\begin{center}
\epsfig{file=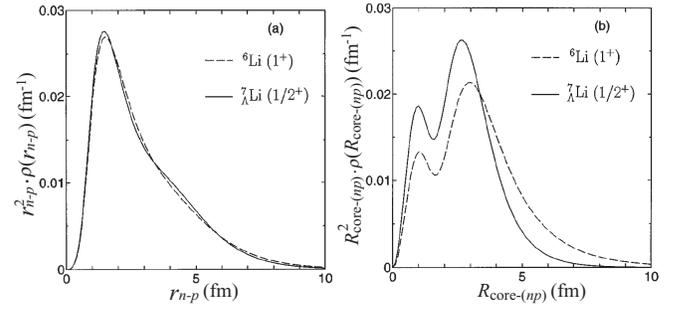,scale=0.44}
\caption{
(a) the $n-p$ relative density of $^7_{\Lambda}{\rm Li}$ 
 as a function of $r_{n-p}$ 
and (b) the $(np)$ c.m. density as a function
of $R_{{\rm core}-(np)}$
together with the corresponding densities in
$^6$Li core. This figure is taken from \cite{Hiyama99}.
}
\label{fig:li7den}
\end{center}
\end{figure}
%%%%%%%%%%%%%%%%%%%%%%
%

\vskip 0.1cm
We show  in Fig.~\ref{fig:li7den}
the $n-p$ relative density
$\rho(r_{n-p})$ 
and the $np$ c.m. density $\rho(R_{{\rm core}-(np)})$
together with the corresponding densities in
$^6$Li core.
The $n-p$ relative density
exhibits almost the same
shape for the ground state of
$^6$Li and that of $^7_{\Lambda}$Li; namely, the shrinkage
of the $n-p$ distance due to the $\Lambda$ participation
is  negligibly small.
On the other hand, 
the $n-p$ c.m. density distribution 
of $^7_{\Lambda}$Li
is remarkably different from that
of $^6$Li, showing a significant contraction
along the ${\bf R}_{{\rm core}-(np)}$
coordinate due to the $\Lambda$
addition.
In fact, the r.m.s. distance ${\bar R}_{{\rm core}-(np)}$
is estimated to be 2.94 fm for
$^7_{\Lambda}{\rm Li}(1/2^+)$ versus
3.85 fm for $^6{\rm Li}(1^+)$.

\vskip 0.1cm
Thus, we  concluded that, by the addition of the
$\Lambda$ particle to $^6{\rm Li}(1^+)$, contraction of
$^7_{\Lambda}$Li occurs between the c.m. of the
$(np)$ pair and the core whereas the $n-p$ relative
motion remains almost unchanged.
In this type change in the wave function, 
the angle operator in $B(E2)$ does not significantly
affect the magnitude of shrinkage. 
We predicted
in Ref. \cite{Hiyama99}  %%
that the size of ${\bar R}_{{\rm core}-(np)}$
in $^6$Li will shrink by 25 \% 
due to the participation of a $\Lambda$ particle.
In a later calculation
\cite{Hiyama01c}  %%%%%%%%%%%%%%5
based on more precise $^4$He$+n+p+\Lambda$ 4-body model,
we predicted it \mbox{to be 22 \%.}

%\vskip 0.4cm

The first observation of the hypernuclear $B(E2)$ strength
was made in the KEK-E419 experiment
for $B(E2; 5/2^+ \rightarrow 1/2^+)$ in
$^7_{\Lambda}$Li. The observed $B(E2)$ 
value was $3.6 \pm 0.5^{+0.5}_{-0.4}$
$e^2$fm$^4$ \cite{Tanida01}.
From this, the shrinkage of
${\bar R}_{{\rm core}-(np)}$ was estimated to be 
by $19 \pm 4$ \%,
which was consistent with our prediction.
It is to be emphasized that this interesting finding was
realized with the help of our precision few-body calculations.

Our prediction about shrinkage of the $^{13}_\Lambda{\rm C}$ states
was given in Ref.~\cite{Hiyama97,Hiyama00} though experiment 
on $^{13}_\Lambda{\rm C}$  is not yet performed.

%%%%%%%%%%%%%%%%%%%%%%%%%%%%%%%%%%%%%%%%%%%%%%%%%%%%%%%%
%%%%%%%%%%%%%%%%%%%%%%%%%%%%%%%%%%%%%%%%%%%%%%%%%%%%%%%%
\subsection{Prediction of spin-orbit splitting in hypernuclei}
\label{sec:ls-splitting}
%%%%%%%%%%%%%%%%%%%%%%%%%%%%%%%%%%%%%%%%%%%%%%%%%%%%%%%%

In this subsection, we briefly review that the present 
authors and collaborators~\cite{Hiyama00} predicted the 
spin-orbit splittings in hypernuclei $^{9}_\Lambda{\rm Be}$ and
$^{13}_\Lambda{\rm C}$ and that afterwards it was
confirmed by experiments at BNL~\cite{Akikawa02,Ajimura01}.

\vskip 0.1cm
One of the characteristic phenomena in non-strange nuclear physics is 
that there is a strong $NN$ spin-orbit
interaction which leads to magic number nuclei. 
How large is the $YN$ spin-orbit interaction in comparison
with the $NN$ spin-orbit one? 
It is known, for instance, that the antisymmetric spin-orbit ($ALS$) 
interactions are
qualitatively different between one-boson-exchange (OBE) 
models~\cite{Rijken75,Rijken99}
and quark models~\cite{Morimatsu84,Fujiwara96}.
As a typical difference, the quark models predict 
that the $ALS$ component of the $\Lambda N$ interaction is so
strong as to substantially cancel the $LS$ one, 
while the OBE models have (much) smaller $ALS$ and
various strength of $LS$.

%%%%%%%%%%%%%%%%%%%%%%%%%%  Fig. 22  %%%%%%%%%%%%%%%%
\begin{figure}[b!]
\begin{center}
\epsfig{file=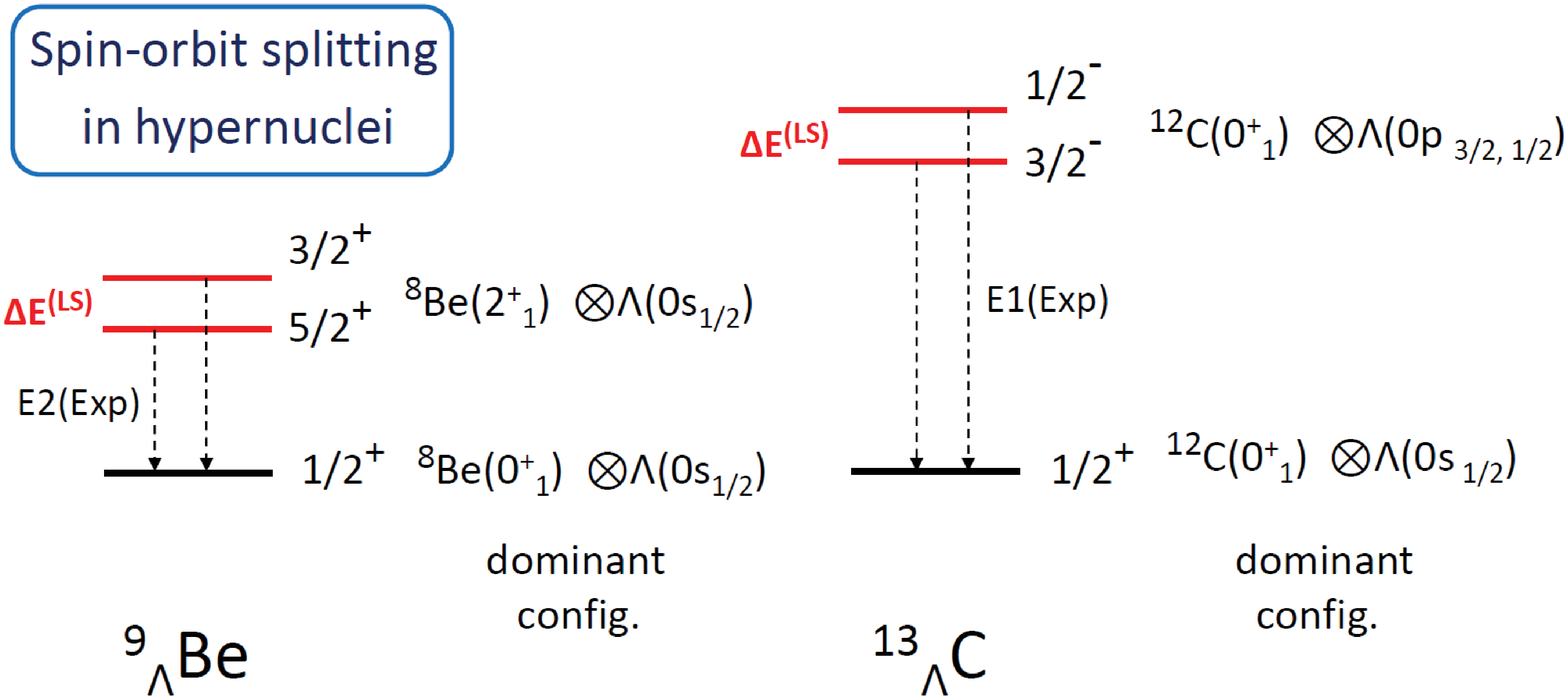,height=4.6cm,width=8.6cm}
\end{center}
\caption{Successful GEM prediction of the spin-orbit splitting
$\Delta E ^{(LS)}$ 
in the hypernuclei $^9_\Lambda$Be and $^{13}_\Lambda$C 
(see Table~\ref{table:ls-split}).
}
\label{fig:ls-split}
\end{figure}
%%%%%%%%%%%%%%%%%%%%%%%%%%%%%%%%%%%%%%%%%%%

\vskip 0.1cm
Because of no $YN$ spin-polarized scattering data, however,
we have no information on the strength of the interaction experimentally.
Therefore, in order to extract information on it, careful calculations of hypernuclear structure should be of great help because $\Lambda$
spin-orbit splittings in hypernuclei are related 
straightforwardly to the spin-orbit component of $\Lambda N$ interactions.

\begin{figure}[b!]
\begin{center}
\epsfig{file=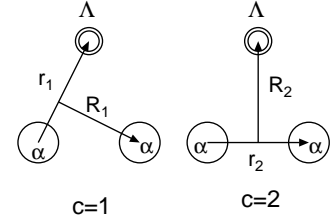,scale=0.45}
\end{center}
\caption{Jacobi coordinates for the $2\alpha +\Lambda$ model
of $^9_\Lambda$Be.
The two $\alpha$ clusters are to be symmetrized.}
\label{fig:be9jacobi}
\end{figure}

\begin{figure}[b!]
\begin{center}
\epsfig{file=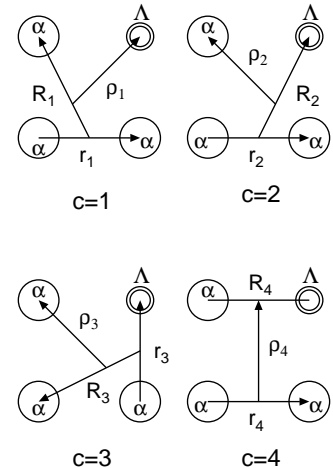,scale=0.45}
\end{center}
\caption{Jacobi coordinates for the $3\alpha +\Lambda$ model of
$^{13}_{\Lambda}$C. The three $\alpha$ clusters are to be symmetrized
(totally 18-channels).}
\label{fig:c13jacobi}
\end{figure}

\vskip 0.1cm
In $\Lambda$-hypernuclei, spin-orbit splitting energy due to the
$\Lambda N$ interaction
was first precisely calculated in Ref.~\cite{Hiyama00} (2000)
for the  $5/2^+_1-3/2^+_1$ doublet states in
$^{9}_\Lambda{\rm Be}$ and
the  $3/2^-_1-1/2^-_1$ states in
$^{13}_\Lambda{\rm C}$ (Fig.~\ref{fig:ls-split}). 
The GEM calculation  employed
the $2 \alpha+ \Lambda$ model for  
$^{9}_\Lambda{\rm Be}$ and the
$3 \alpha+ \Lambda$ model for $^{13}_\Lambda{\rm C}$~ 
(Figs.~\ref{fig:be9jacobi} and ~\ref{fig:c13jacobi}). 
The total wavefunction was described as a sum of
component functions corresponding those coordinate-channels in 
the figures, multiplied by the $\Lambda$-spin wavefunction.

\vskip 0.1cm
We note that the core nuclei $^8$Be and $^{12}$C 
in these two hypernuclei are well described by 
the $2\alpha$- and $3\alpha$-cluster models, and that the spin-spin part 
of the $\Lambda N$ interaction vanishes and 
tensor term does not work in the $\Lambda \alpha$ folding potential. 
Therefore, calculation of the spin-orbit 
level splitting in  $^{9}_\Lambda{\rm Be}$ and
$^{13}_\Lambda{\rm C}$ 
using the folded $\Lambda \alpha$ spin-orbit potential will be 
useful to examine the qualitatively different
two types of potential models, namely, 
OBE models~\cite{Rijken75,Rijken99} and 
quark models~\cite{Morimatsu84,Fujiwara96} mentioned above.
%\vskip 0.1cm
The calculated spin-orbit splitting energies~\cite{Hiyama00}
are listed in Table~\ref{table:ls-split}.

%%%%%%%%%%%%%%%%  Fig.23  %%%%%%%%%%%%%%%%%
%%%%%%%%%%%%%%%%%%%%%%%%%%%%%%%%%%%%%%
%%%%%%%%%%%%%%%%%%%%  Fig.24  %%%%%%%%%%%%%%%%%%
%%%%%%%%%%%%%%%%%%%%%%%%%%%%%%%%%

\vskip 0.1cm
Afterwards, experimental values were reported as
$\Delta E_{\rm EXP}(5/2^+_1 {\rm -} 3/2^+_1)=  31.4^{+2.5}_{-3.6}$ keV
in $^9_{\Lambda}{\rm Be}$ by BNL-E930 \cite{Akikawa02} in 2002
and $\Delta E_{\rm CAL}(3/2^-_1 {\rm -} 1/2^-_1)=150 \pm 54 \pm 36$ keV
in $^{13}_{\Lambda}$C  by BNL-E929
\cite{Ajimura01} in 2001, which is consistent with our prediction
using the quark-based $\Lambda N$ spin-orbit force.
The very weak spin-orbit component of the $\Lambda N$ interaction
compared with that of the $N N$ interaction was
confirmed.

%%%%%%%%%%%%%%%%%%%%%  Table 4   LS splitting %%%%%%%%%%%%%%
\begin{table}[h!]
\caption{
Spin-orbit splitting energies in 
$^9_{\Lambda}$Be and $^{13}_{\Lambda}$C.
Calculated values by GEM are given in Ref.~\cite{Hiyama00}
using the OBE-model-based~\cite{Rijken75,Rijken99} 
and quark-model-based~\cite{Morimatsu84,Fujiwara96}
$\Lambda N$ spin-orbit forces.
Experimental values are taken from 
\cite{Akikawa02} for $^9_{\Lambda}$Be and
from \cite{Ajimura01} for $^{13}_{\Lambda}$C.
The theoretical  prediction using the quark-based 
$\Lambda N$ spin-orbit force was confirmed by the
experiments. 
}
\begin{center}
%\noalign{\vskip -0.25 true cm}  
 \small
\begin{tabular}{c c c c c } 
\hline \hline
\noalign{\vskip 0.15 true cm} 
  &  & CAL & CAL & EXP \\ 
\noalign{\vskip -0.1 true cm} 
  &           &    (OBE)  & (quark)      &  \\
\noalign{\vskip 0.1 true cm} 
  & splitting          &    (keV)  &      (keV)  & (keV) \\
\noalign{\vskip 0.1 true cm} 
\hline
\noalign{\vskip 0.2 true cm} 
$^9_\Lambda$Be $\;$ & $E(5/2^+_1 {\rm -} 3/2^+_1) $ & 
 $\!\!$ $80 - 200$$\!\!$ & $\!\!$$35 - 40$$\!\!$ & 
$\!\!$$ 31.4^{+2.5}_{-3.6}$$\!\!$ \\
\noalign{\vskip 0.2 true cm} 
$^{13}_\Lambda$C $\;$ & $E(3/2^-_1 {\rm -} 1/2^-_1) $ &
 $\!\!$ $390 - 960$ & $\;150 - 200\;$ &
 $ 150 \pm 54 \pm 36$$\!\!$ \\
\noalign{\vskip 0.1 true cm} 
%\noalign{\vskip 0.1 true cm} 
\hline \hline
\end{tabular}
\label{table:ls-split}
\end{center}
\end{table}
%%%%%%%%%%%%%%%%%%%%%%%%%%%%%%%%%%%%%%

%%%%%%%%%%%%%%%%%%%%%%%%%%%%%%%%%%%%%%%%%%%%%%%%%%%%%%%%
\subsection{Prediction for neutron-rich hypernuclei}
\label{sec:neutron-rich}
%%%%%%%%%%%%%%%%%%%%%%%%%%%%%%%%%%%%%%%%%%%%%%%%%%%%%%%%

It is of importance to produce {\it neutron-rich} 
$\Lambda$ hypernuclei
for the fundamental study of hyperon-nucleon ($YN$) interaction.
It is quite helpful to the newly developing 
experiments to predict energy levels
of  these $\Lambda$ hypernuclei before measurement.

\vskip 0.1cm
In 2009, the present authors and collaborators~\cite{Hiyama09He7L} 
predicted energies of
the ground and excited states of a neutron-rich hypernucleus
$^7_{\Lambda}$He together with $^7_\Lambda{\rm Li}(T=1)$ 
and $^7_\Lambda{\rm Be}$ 
using an $\alpha +\Lambda +N+N$ 4-body 
cluster model. A part of the aim of this work was 
to help the new $^7{\rm Li}(e,e'K^+)^7_\Lambda{\rm He}$ 
experiment scheduled at JLAB.

\begin{figure}[b!]
\begin{center}
\epsfig{file=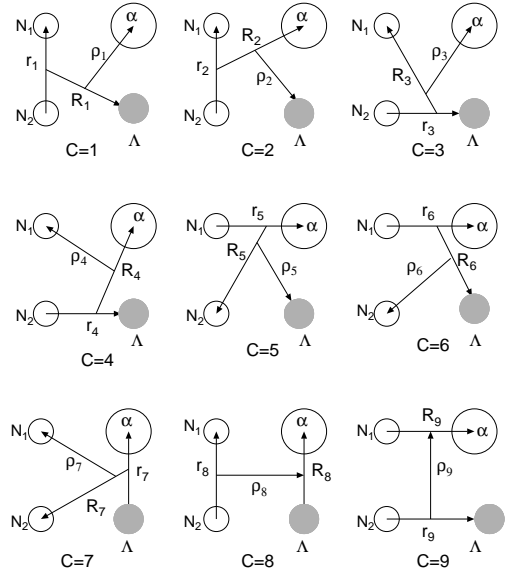,width=6.7cm}
\end{center}
\caption{Jacobi coordinates for all the rearrangement channels
($c=1,...,9)$ of the $\alpha+ \Lambda+ N_1+ N_2$ 
4-body model for $\Lambda$-hypernuclei
$^7_\Lambda{\rm He}$, $^7_\Lambda{\rm Li}$ and 
$^7_\Lambda{\rm Be}$~\cite{Hiyama09He7L}.
Two nucleons are to be antisymmetrized.
}
\label{fig:jacobi-Li7lam}
\end{figure}

\vskip 0.1cm
We constructed 4-body Gaussian basis functions
on all the Jacobi coordinates in Fig.~\ref{fig:jacobi-Li7lam}
in order to take 
account of the full correlations among all the constituent
particles. It is to be stressed that 
2-body interactions among those particles
were chosen so as to reproduce  satisfactorily the observed 
low-energy properties 
of the subsystems ($NN$, $N\Lambda$, $N\alpha$, $\Lambda\alpha$,
$NN\alpha$ and $N\Lambda \alpha$),
at least all the existing binding energies of the 
subsystems~\cite{Hiyama09He7L}.

\vskip 0.1cm
This condition for interactions is 
important in the analysis of the energy levels of these
hypernuclei. Our analysis is performed systematically for both
ground and excited states of $\alpha \Lambda NN$ systems with no more
adjustable parameters in the stage of
full 4-body calculation.  Therefore, these predictions can
offer an important guidance to the interpretation of upcoming
hypernucleus experiments,
$^7{\rm Li}(e,e'K^+)^7_\Lambda{\rm He}$ 
reaction at JLAB.

%%%%%%%%%%%%%%%%%%%%%%%%%%  Fig. 25  %%%%%%%%%%%%%%%%
%%%%%%%%%%%%%%%%%%%%%%%%%%%%%%%%%%%%%%%%%%%
%%%%%%%%%%%%%%%%%%%%%%%%%%  Fig. 26  %%%%%%%%%%%%%%%%
\begin{figure}[b!]
\begin{center}
\epsfig{file=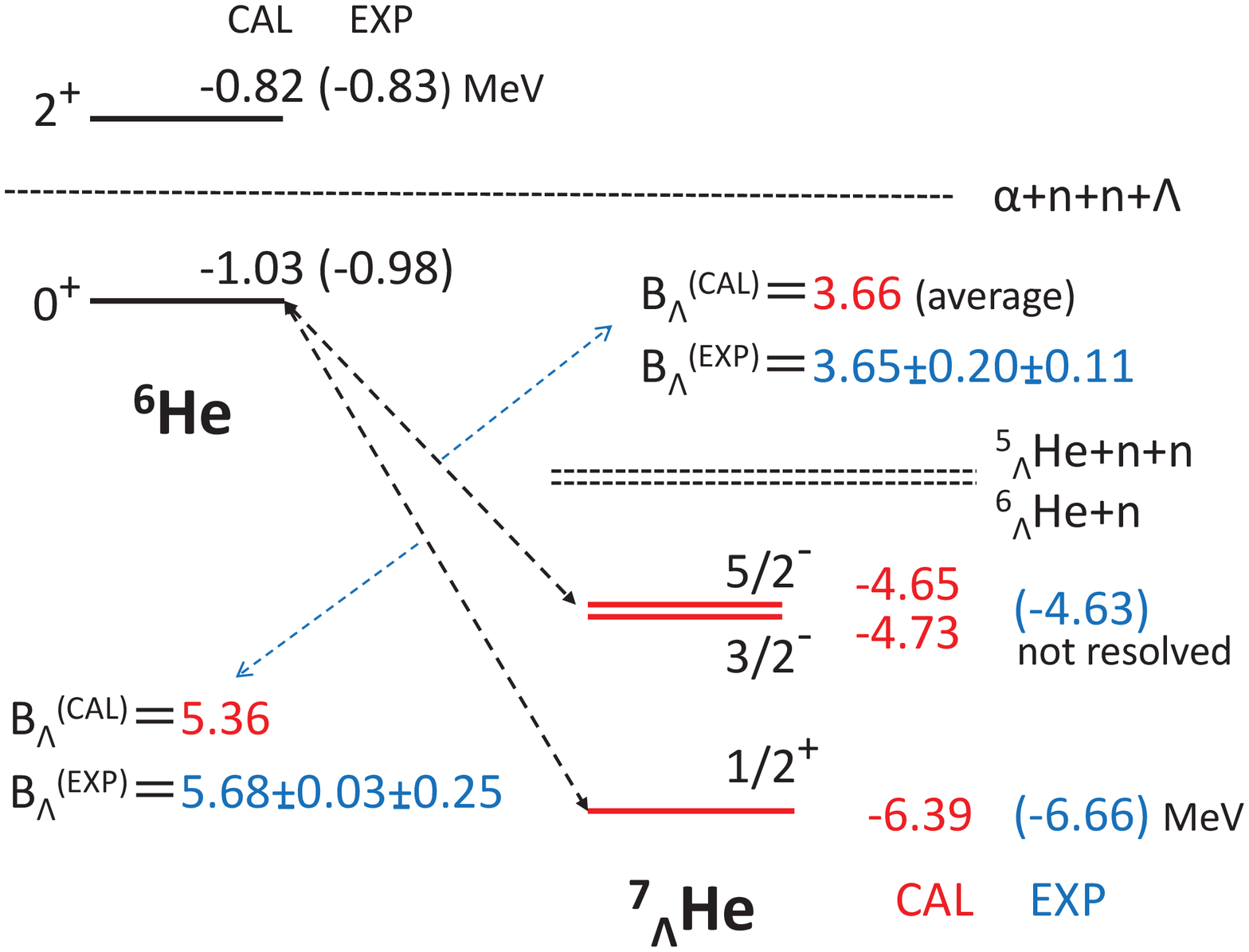,height=7.0cm,width=9.0cm}
\vskip -0.5cm
\end{center}
\caption{
Calculated energy levels of $^6$He and $^7_\Lambda$He
by Ref.~\cite{Hiyama09He7L}.
The predicted $\Lambda$ binding energy $B_\Lambda^{({\rm CAL})}=3.66$ MeV
for the excited states was afterwards reproduced by the 
experiment~\cite{Gogami}.
This figure is taken from  Ref.~\cite{Hiyama09He7L}.
}
\label{fig:he7-lam}
\end{figure}
%%%%%%%%%%%%%%%%%%%%%%%%%%%%%%%%%%%%%%%%%%%

\vskip 0.1cm
As shown in Fig.~\ref{fig:he7-lam},
the $\Lambda$ binding (separation) energy $B_{\Lambda}$ 
of the $1/2^+$ ground  state
(namely, the binding energy measured from the 
$^6{\rm He(g.s.)}+\Lambda$ threshold) is calculated as 
$B_{\Lambda}^{\rm cal}=5.36$ MeV, while
the  $3/2^+$ and $5/2^+$ excited states are
given at 1.66 and 1.74 MeV above the $1/2^+$ ground state, respectively.

\vskip 0.1cm
In 2013, this hypernucleus $^7_{\Lambda}$He 
was observed by the JLAB E01-011 experiment with the 
$^7{\rm Li}(e,e'K^+)^7_\Lambda{\rm He}$ reaction 
and the $\Lambda$ separation energy was reported~\cite{Nakamura13} 
as $B_{\Lambda}^{\rm exp}=
5.68 \pm 0.03 ({\rm stat.}) \pm 0.25 ({\rm sys.})$ MeV,
which is consistent with the theoretical  prediction.
Observation of the first excited-state peak 
($3/2^+_1$ and $5/2^+_1$ unresolved) 
by the JLAB E01-015 experiment was reported~\cite{Gogami} with
$B_{\Lambda}^{\rm exp} =3.65 \pm 0.20 ({\rm stat.}) \pm 0.11 ({\rm sys.})$ MeV,
which agrees with the theoretical prediction
$B_{\Lambda}^{\rm cal} =3.66$ MeV (average for the two excited states).

\vskip 0.1cm
Those theoretical and experimental studies of the energies
of $^7_{\Lambda}$He states are newly attracting strong attentions
from the viewpoints of  CSB (charge symmetry breaking) of
the $YN$ interactions.
For more details, see Ref.~\cite{Hiya14He7L}.

%%%%%%%%%%%%%%%%%%%%%%%%%%%%%%%%%%%%%%%%%%%%%%%%%%%%%%%%
\subsection{Prediction of hypernuclear 
states with strangeness $S=-2$}
\label{sec:S=-2}

Study of $\Lambda \Lambda$ interaction and $\Xi N$ interaction 
(both $S=-2$) is important.
However, since hyperon-hyperon ($YY$) 
scattering experiment is difficult to perform,
it is essential to extract information on these interactions
from the structure study of $S=-2$ hypernuclei 
such as double $\Lambda$ hypernuclei and $\Xi$ hypernuclei.

\vskip 0.1cm
For this aim,  KEK-E373 emulsion  experiment was performed and 
the $^6_{\Lambda \Lambda}$He was observed without ambiguity for the first time.
The reported $\Lambda \Lambda$ 
binding energy (binding energy of $^6_{\Lambda \Lambda}$He
measured from the $^4{\rm He (g.s.)}+\Lambda+\Lambda$ threshold)
is $B_{\Lambda \Lambda}=6.91 \pm 0.16$ MeV;
analysis of the emulsion data to find new hypernuclei 
is still in progress. 
Besides, it is planned to perform, in 2017, 
new emulsion experiment at J-PARC (J-PARC-E07).
However, since it is difficult to determine spins and parities 
of observed states, theoretical analysis is 
important for the identification of those states.
The present authors and collaborators
have succesfull experiences in interpreting 
the states of the following two double $\Lambda$ hypernuclei.

%\vskip 0.5cm
%\noindent
\subsubsection{Double $\Lambda$ hypernucleus 
$^{10}_{\Lambda \Lambda}${\rm Be}}

%\vskip 0.1cm
The KEK-E373 experiment observed a
double $\Lambda$ hypernucleus, $^{10}_{\Lambda \Lambda}$Be, 
which is called  Demachi-Yanagi event \cite{Demachi-1,Demachi-2,Hida}.
The reported $\Lambda \Lambda$ binding energy was 
$B_{\Lambda \Lambda}^{\rm exp}=12.33_{-0.21}^{+0.35}$ MeV.
However, it was not determined whether this event was observation of
the ground state or any excited state in $^{10}_{\Lambda \Lambda}$Be.

%%%%%%%%%%%%%%%%%%%%%%%%%%  Fig. 27  %%%%%%%%%%%%%%%%
\begin{figure}[b!]
\begin{center}
\epsfig{file=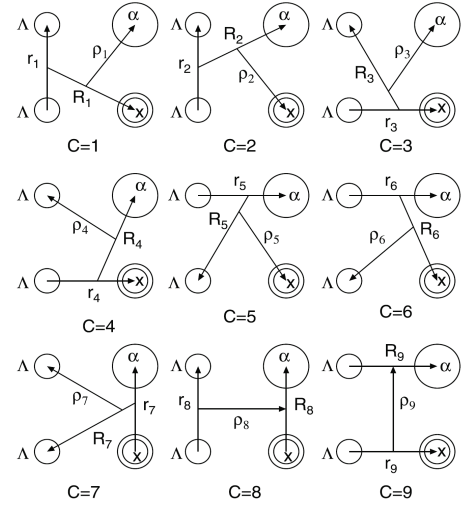,width=6.27cm}
\end{center}
\caption{Jacobi coordinates for all the rearrangement channels
($c=1,...,9)$ of the $\alpha+ X + \Lambda+ \Lambda$ 
4-body model.  For the double $\Lambda$ hypernuclei
$^{10}_{\Lambda \Lambda}$Be, we take \mbox{$X=\alpha$.}
The two $\alpha$'s are to be symmetrized and the two $\Lambda$'s are
to be antisymmetrized.  Taken from Ref.~\cite{Hiyama02doubleL}.
}
\label{fig:jacobi-be10L}
\end{figure}
%%%%%%%%%%%%%%%%%%%%%%%%%%%%%%%%%%%%%%%%%%%
%%%%%%%%%%%%%%%%%%%%%%%%%%  Fig. 28  %%%%%%%%%%%%%%%%
\begin{figure}[b!]
\begin{center}
\epsfig{file=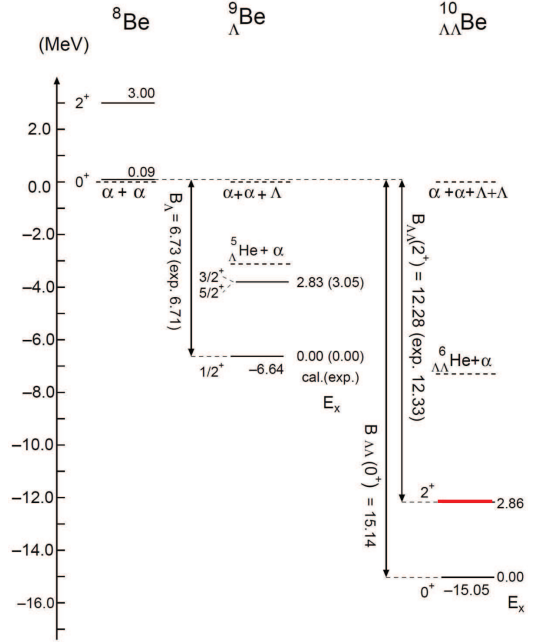,width=11.5cm}
%\hskip -5.5cm
\end{center}
\caption{
Calculated energy levels of 
$^{8}$Be, $^{9}_{\Lambda}$Be and
$^{10}_{\Lambda \Lambda}$Be on
the basis of the $\alpha \alpha$, $\alpha \alpha \Lambda$ and 
$\alpha \alpha \Lambda \Lambda$  models, respectively.
The level energies are measured from the particle
breakup thresholds or are given by excitation energies $E_{\rm x}$.
This figure is taken from Ref.~\cite{Hiyama02doubleL}.
}
\label{fig:be10LL}
\end{figure}
%%%%%%%%%%%%%%%%%%%%%%%%%%%%%%%%%%%%%%%%%%%

We studied $^{10}_{\Lambda \Lambda}$Be with the framework of
$\alpha +\alpha +\Lambda +\Lambda$ 4-body 
model~\cite{Hiyama02doubleL}, constructing 
4-body Gaussian basis functions
on all the Jacobi coordinates in Fig.~\ref{fig:jacobi-be10L}
in order to take 
account of the full correlations among all the constituent
particles.  
Two-body interactions among those particles
were chosen so as to reproduce  satisfactorily the observed 
low-energy properties 
of the subsystems ($\alpha \Lambda$, $\alpha \alpha$
and  $\alpha \Lambda\Lambda$, $\alpha \alpha \Lambda$).
We then predicted,
with no more adjustable parameters, the energy level of
$^{10}_{\Lambda \Lambda}$Be. 

\vskip 0.1cm
As seen in Fig.~\ref{fig:be10LL}, the calculated ${\Lambda \Lambda}$  
binding energy of the $2^+$ state is  
$B_{\Lambda \Lambda}^{\rm cal}=12.28$ MeV,  %11.88$ MeV, 
which is in good agreement with the experimental data.
The Demachi-Yanagi event was then interpreted as the
observation of the $2^+$ excited state of $^{10}_{\Lambda\Lambda}$Be
(the ground state is located 2.86 MeV below). 
For more details, see Ref.~\cite{Hiyama02doubleL}
in which more energy levels of %% more double $\Lambda$ hypernuclei
$^{7}_{\Lambda \Lambda}$He, $^{7}_{\Lambda \Lambda}$Li, 
$^{8}_{\Lambda \Lambda}$Li, $^{9}_{\Lambda \Lambda}$Li and
$^{9}_{\Lambda \Lambda}$Be are predicted though no experiment on them
is done  yet.

%\vskip 0.5cm
%\noindent
\subsubsection{Double $\Lambda$ hypernucleus 
$^{11}_{\Lambda \Lambda}${\rm Be}}
%\vskip 0.1cm

The KEK-E373 experiment observed another new double $\Lambda$
hypernucleus, called  Hida event~\cite{Hida}.
This event had two possible interpretations: 
one is $^{11}_{\Lambda \Lambda}$Be 
with $B_{\Lambda \Lambda}=20.83 \pm 1.27$ MeV, and the other is
$^{12}_{\Lambda \Lambda}$Be 
with 
and $B_{\Lambda \Lambda}=22.48 \pm 1.21$ MeV.
It is uncertain whether this is observation
of a ground state or an excited~state.

\vskip 0.1cm
Assuming this event to be  $^{11}_{\Lambda \Lambda}$Be,
we calculated the energy spectra of this hypernucleus within the framework of
$\alpha +\alpha +n+\Lambda +\Lambda$ 5-body cluster 
model~\cite{Hiyama10-Hida}.
All the interactions are tuned to reproduce the binding energies
of  possible subsystems (cf. Ref.~\cite{Hiyama10-Hida} for
the details). 
There is no adjustable parameter when entering the 
5-body calculation of $^{11}_{\Lambda \Lambda}$Be.
The calculated $\Lambda \Lambda$ binding energy 
was $B_{\Lambda \Lambda}=18.23$ MeV, which
does not contradict the interpretation that
the Hida event is  observation of the ground state of
$^{11}_{\Lambda \Lambda}$Be.
%%%For more details, see Ref.~\cite{Hiyama10-Hida}.

\vskip 0.1cm
As for $\Xi^-$ hypernuclei,  there are a few experimental data at present.
Among them,  
the observed spectrum of the
$(K^-,K^+)$ reaction on a $^{12}$C target
seems to indicate that the $\Xi$-nucleus interactions are attractive
with a depth of $\sim 14$ MeV when a Woods-Saxon shape is assumed. 
Taking this information into consideration, we performed 
$\alpha+ n+n+ \Xi^-$ and $\alpha+ \alpha+n+ \Xi^-$ 
four-body cluster-model
calculations, and predicted bound states for these hypernuclei.
It is expected to perform search experiments for these $\Xi^-$ hypernuclei 
at J-PARC in the future.
For more details, see Ref.~\cite{Hiyama10Suppl-2}.

%%%%%%%%%%%%%%%%%%%%%%%%%%%%%%%%%%%%%%%%%%%%%%%%%%%%%%%%
\subsection{Strategy of studying hypernuclei and
$YN$ and $YY$ interactions}
\label{sec:hyper-strategy}
%%%%%%%%%%%%%%%%%%%%%%%%%%%%%%%%%%%%%%%%%%%%%%%%%%%%%%%%%%%%%%%%%%%

\begin{figure}[h!]
\begin{center}
\epsfig{file=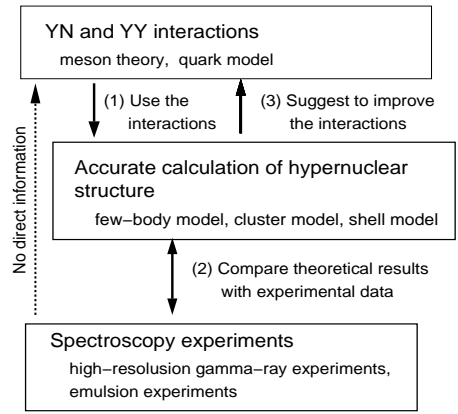,width=6.0 cm,height=5.5cm}
\caption{Strategy for extracting information about $YN$ and $YY$
interactions from the study of the structure of light hypernuclei.
Taken from Ref.~\cite{Hiyama12PTEP}.
}
\label{fig:strategy-hyper}
\end{center}
\end{figure}

In the previous Secs.~\ref{sec:shrink} --~\ref{sec:S=-2}, 
we have reviewed some of 
our GEM studies of hypernuclei and $YN$ and $YY$ interactions.
Here, we emphasize that 
one can obtain useful information on the $YN$ and $YY$ interaction combining
few-body calculations of the hypernuclear structure and the related spectroscopy experiments on
the basis of the following strategy (cf.~Fig.~\ref{fig:strategy-hyper}):

\vskip 0.1cm
(i) Firstly, we begin with candidates of $YN$ and $YY$ interactions 
that are based on the meson theory and/or the constituent quark model.

\vskip 0.1cm
(ii) We then utilize spectroscopy experiments of hypernuclei. Generally, 
    the experiments do not directly give any information about 
the $YN$ and $YY$ interactions.

\vskip 0.1cm
(iii) Using the interactions in (i), accurate calculations of 
hypernuclear structures are performed.
The calculated results are compared with the experimental data.

\vskip 0.1cm
(iv) On the basis of this comparison, improvements for the underlying 
interaction models are proposed.

\vskip 0.1cm
Following this strategy, we have succeeded in extracting information on the 
$YN$ and $YY$ interactions proposed so far with the use of GEM. 
These efforts are summerized in review 
papers~\cite{Hiyama09PPNP,Hiyama10Suppl-1,Hiyama10Suppl-2,
Hiyama12FEW,Hiyama12PTEP}
on the physics of hypernuclei and  $YN$ and $YY$ interactions.

%%%%%%%%%%%%%%%%%%%%%%%  Fig.29  Hyper strategy  %%%%%%%%%%%%%%%%%%
%%%%%%%%%%%%%%%%%%%%%%

%\newpage
%%%%%%%%%%%%%%%%%%%%%%%%%%%%%%%%%%%%%
\section{Extension of GEM}
\label{sec:extension}
%%%%%%%%%%%%%%%%%%%%%%%%%%%%%%%%%%%%%%%%%%%%%%%%%%
\subsection{Few-body  resonances with the Complex-scaling method}
\label{sec:CSM}
%%%%%%%%%%%%%%%%%%%%%%%%%%%%%%%%%%%%%%%%%%%%%%%%%%

We extended  GEM to the case of 
calculating the energy and width of few-body resonances,
employing the complex scaling method
(CSM)~\cite{CSM-ref1,CSM-ref2,CSM-ref3,Ho,Moiseyev} whose
applications to nuclear physics problems are
reviewed, for example, in Refs.~\cite{Aoyama2006}.
We applied GEM+CSM to the study of
i) possibility of narrow 4-neutron resonance~\cite{Hiyama16CSM}
using real-range Gaussian basis functions, and
ii) new broad $0^+_3$ resonance in $^{12}$C~\cite{Ohtsubo2013}
using complex-range Gaussian basis functions.

\vskip 0.1cm
The resonance energy (its position
and width) is obtained as a stable complex eigenvalue  of the
complex scaled Schr\"{o}dinger equation:
\begin{equation}
[H(\theta) -E(\theta)] \Psi_{JM, TT_z}(\theta)=0 \;
\label{eq:sccsm},
\end{equation}
where $H(\theta)$ is obtained by making the complex radial scaling
with an  angle $\theta$
\begin{equation}
r_{\rm c} \to r_{\rm c} \,e^{i \theta}, \;
R_{\rm c} \to R_{\rm c} \,e^{i \theta}, \;
{\rho}_{\rm c} \to    \rho_{\rm c} \,e^{i \theta},
\;\;% ({\rm c}={\rm K}, {\rm H})  ,
\end{equation}
for example, in the case of 4-body system of Fig.~9.   %\ref{fig:jacobi-4body}.
According to the ABC theorem~\cite{CSM-ref1,CSM-ref2},
the eigenvalues of
Eq.~(\ref{eq:sccsm}) may be separated into three groups:

%\noindent
%\begin{enumerate}
i) The bound state poles, remain unchanged under the complex
scaling transformation and remain on the negative real axis.

%\noindent
ii) The
cuts, associated with discretized continuum states, are rotated
downward making an angle of $2\theta $ with respect to the real axis.

%\noindent
iii) The resonant poles are independent of parameter $\theta$ and
are isolated from the discretized non-resonant continuum spectrum
lying along the $2\theta$-rotated line when the relation tan\,2$\theta
> -{\rm Im}(E_{\rm res})/{\rm Re}(E_{\rm res})$ is satisfied.
The resonance width is defined by \mbox{$\Gamma=-2\,{\rm Im}(E_{\rm res})$.}
%\end{enumerate}

%%%%%%%%%%%%%%%%%%%%%%%%%%%%%%%
\subsubsection{Tetraneutron $(^4n)$ resonances}
%%%%%%%%%%%%%%%%%%%%%%%%%%%%%%%%%555

\begin{figure}[b!]
\begin{center}
\begin{minipage}[b!]{7.0 cm}
\epsfig{file=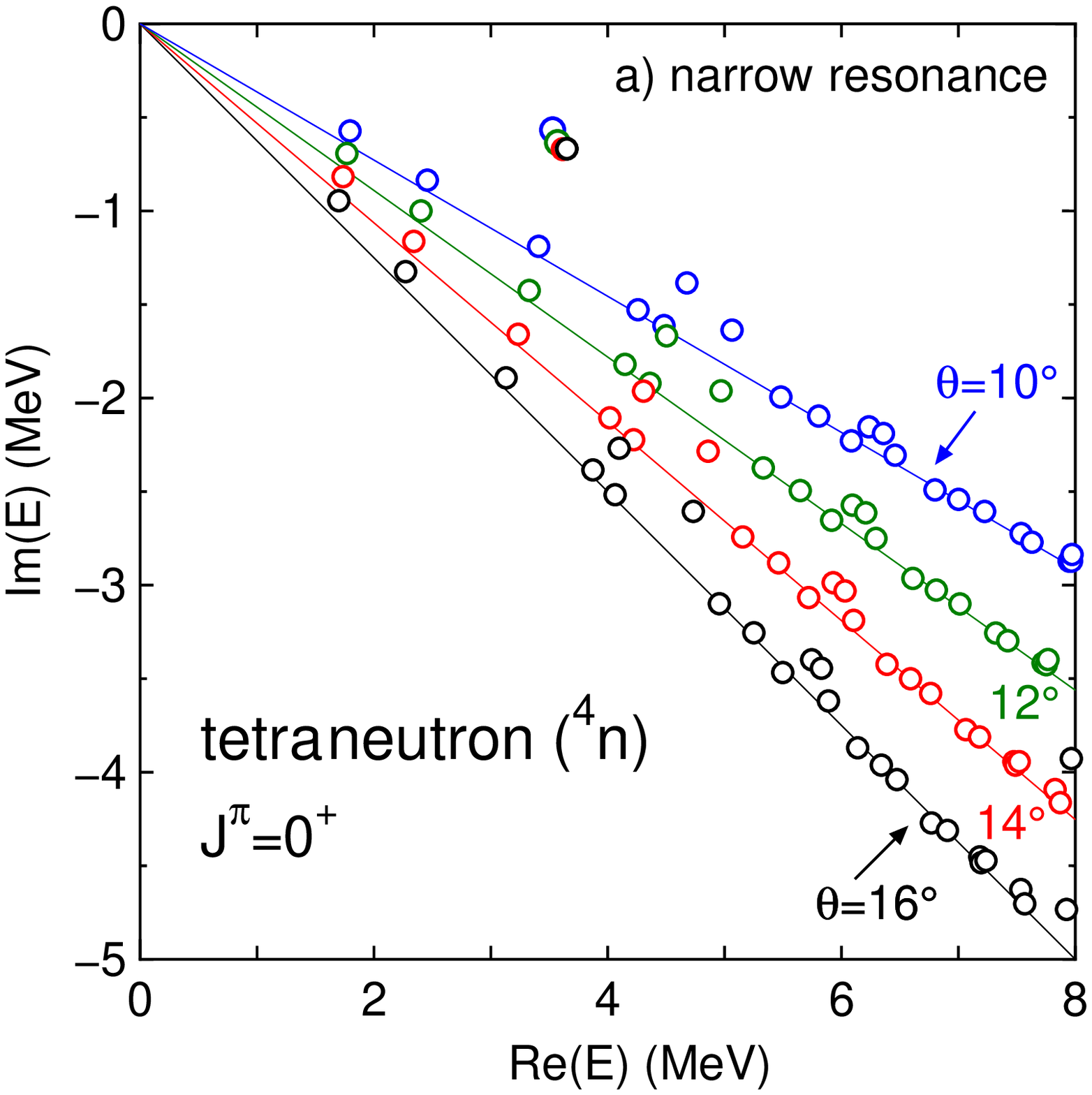,scale=0.37}
\end{minipage}
\vskip 0.4cm 
%\hspace{\fill}
\begin{minipage}[b!]{7.0 cm}
\epsfig{file=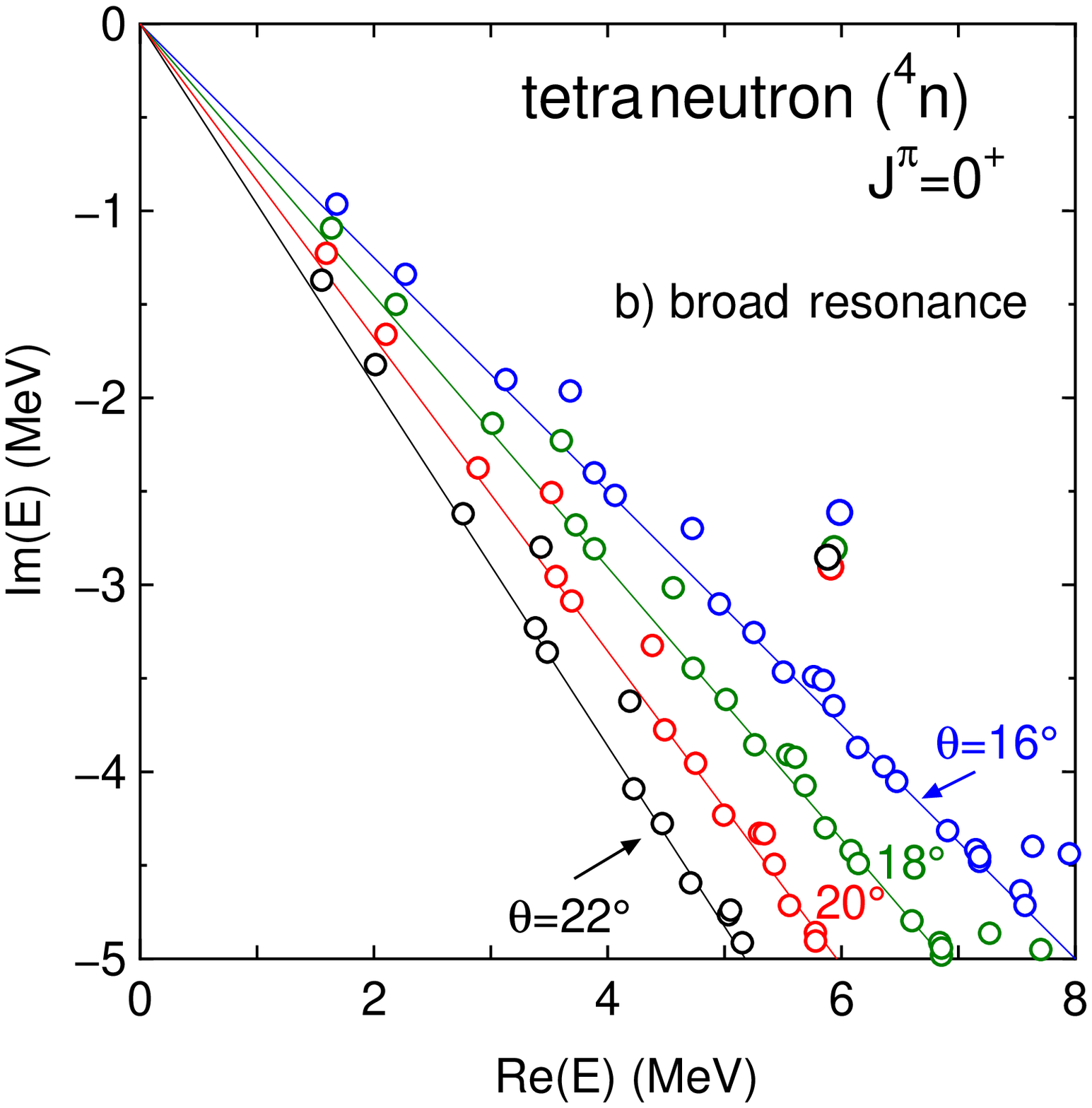,scale=0.37}
\end{minipage}
\end{center}
\caption{Dependence of the eigenenergy distribution
on the complex scaling angle $\theta$ for the  $^4n$ system with
$J^\pi=0^+$. Two different cases are considered a) presence of a
narrow resonance at $E_{\rm res}=3.65-0.66i$ MeV for
$W_1(T=3/2)=-28$ MeV and b) presence of a broad resonance at
$E_{\rm res}=5.88-2.85i$ MeV for $W_1(T=3/2)=-21$ MeV. 
Taken from Ref.~\cite{Hiyama16CSM}. }
\label{fig:pole-example}
\end{figure}

As a beautiful example 
that satisfies the above properties i)-iii), we show, in
Fig.~\ref{fig:pole-example},  narrow and
broad resonances as well as the non-resonant continuum spectrum
of the 4-neutron system (tetraneutron, $^4n$)~\cite{Hiyama16CSM};
they are  rotated
in the complex energy plane from $\theta=10^\circ-22^\circ$.

%%%%%%%%%%%%%%%%%%%%%%%%  Fig. 30 %%%%%%%%%%%%%%%%%%%%
%%%%%%%%%%%%%%%%%%%%%%%%%%%%%%%%%%%%%%%%%%%%%%

\vskip 0.1cm
In Ref.~\cite{Hiyama16CSM}, we discussed about 
the theoretical possibility to generate a narrow resonance
in the 4-neutron system as suggested by a recent experimental 
result ($E_{\rm res}=0.83 \pm 0.65 \pm 1.25$ MeV and 
$\Gamma \leq 2.6$ MeV)~\cite{Kisamori-PRL}. This experiment provides 
a good chance to investigate the isospin $T=3/2$ component of the
3-nucleon ($3N$) force since the $T=1/2$ component does not work 
in this system;  the $T=3/2$ component has been
considered to be  smaller than the $T=1/2$ one in the literature.

\vskip 0.1cm
To investigate this problem, we introduced a phenomenological $3N$ force  
for $T=3/2$ (in the same functional form of
the $T=1/2$ one; cf. Eq.~(2.2) of Ref.~\cite{Hiyama16CSM}) 
in addition to a realistic $NN$ interaction (AV8$'$). 
We inquired what should be  the strength of the $T=3/2$ $3N$ force
(compare with the $T=1/2$ one)  in order 
to generate such a resonance; 
we performed this by changing the strength parameter 
$W_1(T\!=\!3/2)$ of the $T=3/2$ $3N$ force.  
As for the $T=1/2$ $3N$ force, 
$W_1(T\!=\!1/2)=-2.04$ MeV is known from our study of the
ground and second $0^+$ states of $^4$He (cf.~Sec.~\ref{sec:bench-mark}).
% and Ref.~\cite{Hiya04SECOND}). 

\vskip 0.1cm
The reliability  of the $3N$ force in the  $T=3/2$ channel was examined 
by analyzing
its consistency with the low-lying $T=1$ states of $^4$H, $^4$He 
and $^4$Li and the  $^3{\rm H} + n$ scattering.
 The {\it ab initio} solution of the $4n$ Schr\"{o}dinger equation 
was obtained using the complex scaling method   
with boundary conditions appropriate to the 4-body resonances.
We found that, in order to generate narrow $4n$ resonant states,   
unrealistically strong attractive $3N$ force 
is required as is explained  below.

%%%%%%%%%%%%%%%%%%%%%%%%  Fig. 30  %%%%%%%%%%%%%%%%%%%%
\begin{figure}[b!]
\begin{center}
\epsfig{file=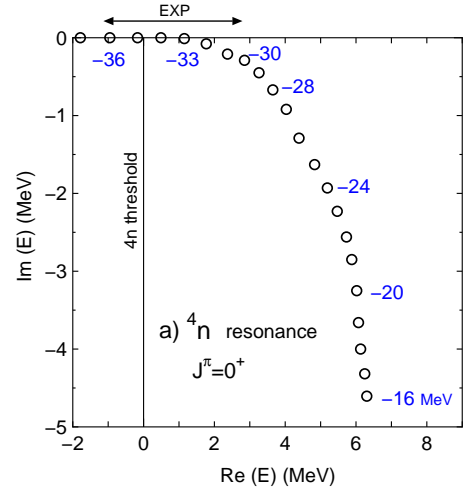,scale=0.37}
%\hspace{\fill}
\end{center}
\caption{Tetraneutron ($^4$n) resonance trajectory for the $J^\pi=0^+$ state. 
The circles correspond to resonance positions calculated 
in Ref.~\cite{Hiyama16CSM}. The strength parameter of the $T=3/2$ $3N$ force, 
$W_1(T=3/2)$,  is changed from $-37$ to $-16$ MeV 
in steps of 1 MeV.
To guide the eye, the resonance region suggested by the 
measurement~\cite{Kisamori-PRL} is
indicated by the arrow at the top. Very strong attructive force of
$W_1(T=3/2)=-36$ to $-30$ is required to generate a resonance
in the energy region.  Taken from Ref.~\cite{Hiyama16CSM}.
}
\label{fig:nnnn-trajectry-j0p}
\end{figure}
%%%%%%%%%%%%%%%%%%%%%%%%%%%%%%%%%%%%%%%%%%%%%%

\vskip 0.1cm
In Fig.~\ref{fig:nnnn-trajectry-j0p}, we display the trajectory of
the $^4n$ S-matrix pole (resonance) with $J=0^+$ state by reducing the
$3N$-force strength parameter 
from $W_1(T\!=\!3/2)=-37$ to $-16$ MeV in step of $1$ MeV.
We were unable to continue the resonance trajectory beyond $W_1
(T=3/2) =-16$ MeV with the CSM, the resonance becoming too broad to
be separated from the non-resonant continuum.
To guide the eye, at the top of the same figure, we present an arrow
to indicate the $^4n$ energy range 
($E_{\rm res}=0.83 \pm 0.65 \pm 1.25$ MeV)
suggested by the recent measurement~\cite{Kisamori-PRL}.
In order to generate a $^4n$ resonance in our calculation,
we need the strength of the $3N$ force in
the $T=3/2$ channel so large as $W_1(T\!=\!3/2)= -36 \:{\rm to} \: -30$ MeV.
%In that range the maximum value of the calculated
%decay width $\Gamma$ is 0.6 MeV, which is to be compared
%with the observed upper limit width $\Gamma=2.6$ MeV.

\vskip 0.1cm
In Ref.~\cite{Hiyama16CSM}, showing many reasons,
we concluded that we find no physical justification for the issue that
the $T =3/2$ term should be one order of magnitude more
attractive than the $T = 1/2$ one, as is required to generate
tetraneutron states compatible with the ones claimed in the
recent experimental data~\cite{Kisamori-PRL}.
We therefore requested the authors of the experiment paper
to re-examine their result. They say that additional experiment has 
been performed and  analysis is under way.

%%%%%%%%%%%%%%%%%%%%%%%%%%%%%%%%%%%%%%%%%%%%%%%%%%
\subsubsection{3-body resonances in $^{12}${\rm C} studied with 
complex-range Gaussians}
\label{sec:12C-resonance}
%%%%%%%%%%%%%%%%%%%%%%%%%%%%%%%%%%%%%%%%%%%%%%%%%%

Use of the complex-range Gaussian basis functions, 
introduced in Sec.~\ref{sec:complex-range}, is 
powerful in CSM calculations
since the CSM resonace wave function becomes very
oscillatory when the rotation angle $\theta$ becomes large (though
the wave function is still $L^2$ integrable).

\vskip 0.1cm
In Sec.~\ref{sec:12C-resonance}, 
we show a typical example in order to demonstrate 
that the use of the complex-range Gaussians gives rise to
much more  precise result than that  of  the real-range Gaussians.
In Ref.~\cite{Ohtsubo2013} the present authors and collaborators
studied the $3\alpha$-cluster resonances performing 
the 3-body GEM calculation
with the complex-range Gaussian basis functions
in the $3\alpha$ OCM (orthogonality condition model).
The main purpose of the work was to discuss about the newly observed  broad
$0^+_3$ resonance, but here we do not enter it.
Instead,  we show a comparison of the two results by the use of two different
types of Gaussian basis functions; both calculations took the same 
$3\alpha$-cluster model and the same interactions.

\begin{figure}[t!]
\centering
\epsfig{file=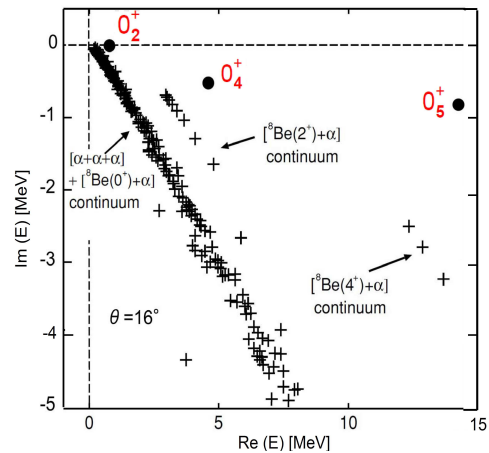,width=6.3cm,height=5.94cm}
\caption{
The $0^+$ eigenvalue distribution of the complex scaled Hamiltonian
for the $3 \alpha$ cluster OCM model 
obtained by Kurokawa and Kat\={o}~\cite{Kurokawa}
%The CSM was applied to the $3 \alpha$ OCM 
using the {\it real-range} Gaussian basis functions. 
The scaling angle is $\theta=16^\circ$.
This figure is to be compared with Fig.~\ref{fig:Ohtsubo}.
Taken from Ref.~\cite{Kurokawa}.
}
\label{fig:Kurokawa}%
\end{figure}

\begin{figure}[t!]
\begin{minipage}{0.48\textwidth}
\centering
\epsfig{file=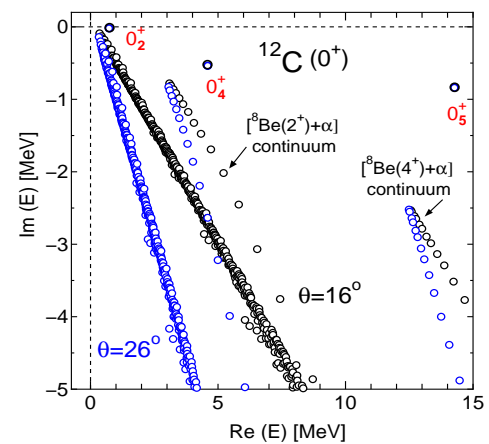,width=6.3cm,height=5.85cm}
\caption{
The $0^+$ eigenvalue distribution of the complex scaled Hamiltonian
for the $3 \alpha$ cluster OCM model 
with the use of the {\it complex-range} Gaussian basis functions.
The scaling angles are $\theta=16^\circ$ (black) and $26^\circ$ (blue).
This figure is to be compared with Fig.~\ref{fig:Kurokawa}.
Taken from Ref.~\cite{Ohtsubo2013}.
}
\label{fig:Ohtsubo}%
\end{minipage}%
\end{figure}

\vskip 0.1cm
Figure~\ref{fig:Kurokawa} illustrates
the $0^+$ eigenvalue distribution of the complex scaled Hamiltonian
$H(\theta)$ for the $3 \alpha$-cluster OCM model 
obtained by Kurokawa and Kat\={o}~\cite{Kurokawa} (2005)
using the real-range Gaussian basis functions. 
The scaling angle is $\theta=16^\circ$.
On the other hand,
Fig.~\ref{fig:Ohtsubo} by our calculation~\cite{Ohtsubo2013} (2013)
shows the same quantity as 
in Fig.~\ref{fig:Kurokawa},  but using 
the complex-range Gaussian basis functions for
$\theta=16^\circ$ (black) and $26^\circ$ (blue).

\vskip 0.1cm
One sees that Fig.~\ref{fig:Ohtsubo} gave much more precise result than 
that in Fig.~\ref{fig:Kurokawa}; especially,  the non-resonant continuum
spectra are almost on straight lines even at $\theta=26^\circ$.

%%%%%%%%%%%%%%%%%%%%%%%%%%%  Fig.31  %%%%%%%%%%%%%%%%%%%
%%%%%%%%%%%%%%%%%%%%%%%%%%%%%%%%%%%%%%%%%%%%%%

%%%%%%%%%%%%%%%%%%%%%%%%%%%  Fig. 5 , 6  %%%%%%%%%%%%%%%%%%
%%%%%%%%%%%%%%%%%%%%%%%%%%%%%%%%%%%%%%%

\vskip 0.1cm
In order to investigate the new broad $0^+_3$ resonance
that was predicted in Ref.~\cite{Kurokawa}, we performed
the CSM calculation for scaling angles from $\theta= 22^\circ$ up to 
$36^\circ$. These large angles are
required to reveal explicitly such a low-lying broad resonance separated 
from the 3-body continuum spectra.
In our calculation~\cite{Ohtsubo2013}, it was really possible to have 
the $0^+_3$ state at $E_{\rm res}=0.79 - i\,0.84$ MeV
as a clearly  isolated and stable resonance pole  
against so large $\theta$ as $30^\circ-36^\circ$ 
(cf. Fig.~7 of Ref.~\cite{Ohtsubo2013}).  
See the paper for more about the $0^+_3$ state.

%\clearpage
%%%%%%%%%%%%%%%%%%%%%%%%%%%%%%%%%%%%%%%%%%%%%%%%%%
\subsection{Few-body  reactions with the Kohn-type variational 
principle to  $S$-matrix} 
\label{sec:reaction}
%%%%%%%%%%%%%%%%%%%%%%%%%%%%%%%%%%%%%%

GEM is applicable to few-body
reactions.  In Sec.~\ref{sec:reaction}, we review briefly three examples:

\noindent
i)  Muon transfer reaction in the cycle of 
muon \mbox{catalyzed}

\hskip 0.2cm fusion ($\mu$CF) (cf.~Sec.~\ref{sec:MCF}), 

\noindent
ii)  {\it Catalyzed} big-bang nucleosynthesis (CBBN) \mbox{reactions}

\hskip 0.2cm (for review, see Ref.~\cite{Kusakabe-review} 
and Sec.~9.2 of Ref.~\cite{Iocco}). 

\noindent
iii) Scattering calculation of
5-quark $(uudd{\bar s})$ systems.

\vskip 0.1cm
The  subjects \mbox{i) and ii)} give good 
tests to 3-body reaction theories
for elastic and transfer processes 
in the presence of strong \mbox{3-body} distortions 
(virtual excitations) in the intermediate stage of reaction.

%\vskip 0.4cm
%%%%%%%%%%%%%%%%%%%%%%%%%%%%%%%%%%%%%%%%%%%%%%%%
\subsubsection{Muon transfer reaction in $\mu$CF cycle} 
\label{sec:muon-transfer}
%%%%%%%%%%%%%%%%%%%%%%%%%%%%%%%%%%%%%%%%%%%%%%%%

In the $\mu$CF cycle (cf. Fig.~22 of~\cite{Hiyama03}), 
muons injected into the $D_2/T_2$ mixture
form finally $(d\mu)_{1s}$ and $(t\mu)_{1s}$,
and then $(d\mu)_{1s}$ is changed to $(t\mu)_{1s}$ by the
muon transfer reaction due to the difference in their binding energies:
%
%\vskip -0.8cm
\begin{equation}
(d\mu)_{1s}+t \to d+(t\mu)_{1s} + 48\, {\rm eV}.  
\label{eq:muon-transfer}
\end{equation}

%%%%%%%%%%%%%%%%%  Fig. 33 (Jacobi) %%%%%%%%%%%
\begin{figure}[h!]
\begin{center}
\epsfig{file=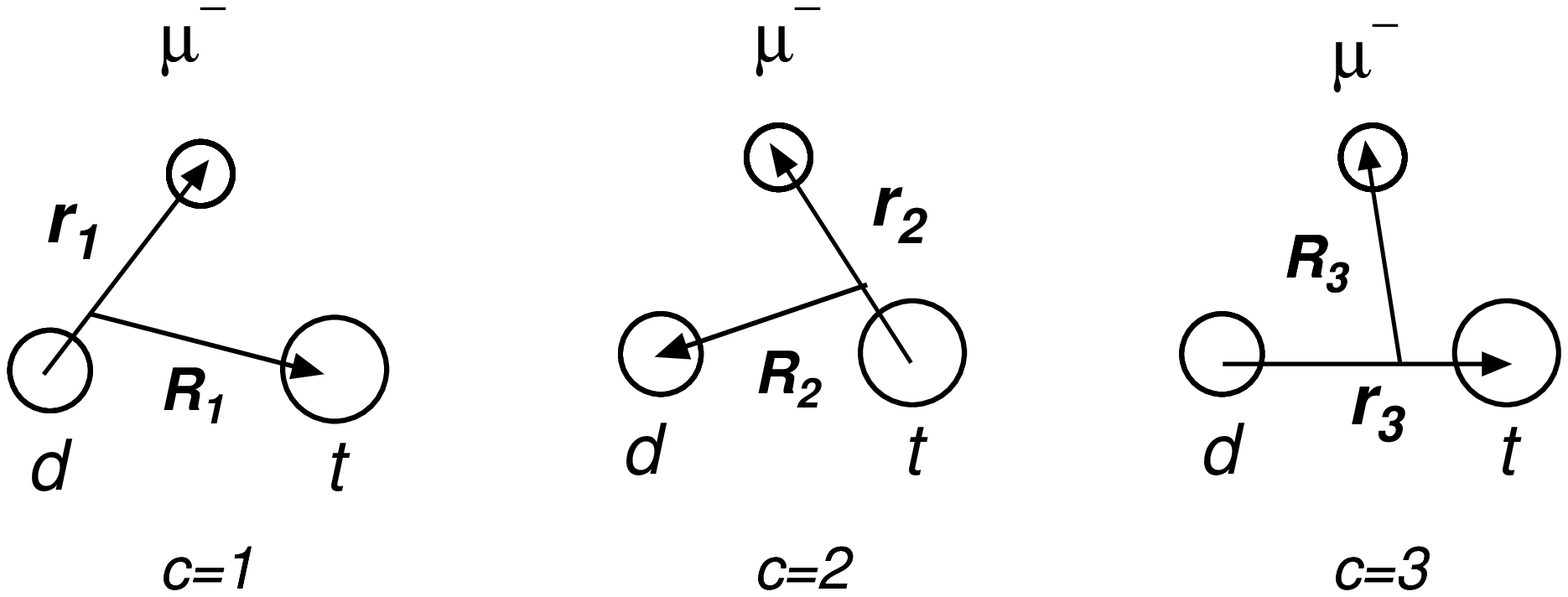,scale=0.45}
\end{center}
\vskip -0.3cm
\caption{
Three Jacobi coordinates of the $d + t + \mu^-$ system.
}
\label{fig:Jacobi-transfer}
\end{figure}
%%%%%%%%%%%%%%%%%%%%%%%%%%%%%%%%%%%%%%%%%%%%%%%%%
%
%\vskip -0.2cm
This reaction (cf. Fig.~\ref{fig:Jacobi-transfer})
was extensively studied theoretically in
1980's and 1990's as
an important doorway process to the $\mu$CF
and also by the following reason:
Calculation of the cross section of this reaction
at $E_{\rm cm}=0.001 - 100$ eV has been a stringent
benchmark test for the calculation methods of 
Coulomb 3-body reactions. Since the muon mass 
is 207 times the electron mass,
fully non-adiabatic treatment is necessary. 
The GEM calculaion~\cite{Kamimura88b,Kino93a} gave one of the
most precise results so far (cf. a brief review in Sec.~8.1
of Ref.~\cite{Hiyama03}).%(cf.~Fig.~\ref{fig:Jacobi-transfer}).

\vskip 0.1cm
We consider the reaction (\ref{eq:muon-transfer}) 
at incident c.m. energies $0.001 - 100$ eV which are much
less than the excitation energy of
the $n=2$ state of $(t\mu)$ and $(d\mu)$, $\sim~\!2$ keV.
The formulation below follows Sec.~8.1 of Ref.~\cite{Hiyama03}:

The wave function which describes the transfer reaction
(\ref{eq:muon-transfer}) as well as the diagonal
$(t\mu)_{1s}-d$ and $(d\mu)_{1s}-t$  processes
with the total energy $E$ may be written as 
\begin{eqnarray}
&&\Psi_{JM}(E)=\phi^{(d\mu)}_{1s,\varepsilon_1}({\bf r}_{1})\,
               \chi^{(d\mu-t)}_{JM}(k_1,{\bf R}_{1})  \nonumber \\
&&  \qquad  \quad  \quad  
          + \; \phi^{(t\mu)}_{1s,\varepsilon_2}({\bf r}_{2})\,\, 
               \chi^{(t\mu-d)}_{JM}(k_2,{\bf R}_{2})  \nonumber \\
&&  \qquad  \quad  \quad  
               + \; \sum_{\nu=1}^{\nu_{\rm max}} b_\nu(E)  
                  \Phi^{(\nu)}_{J M}(E_\nu).
%    + \Psi^{({\rm closed})}_{JM}\;.
\label{eq:psi12}
\end{eqnarray}
%  
%
%%%%%%%%%%%%%%%%%%%%%%%%  Fig.34  dtmu sigma21  %%%%%%%%%%%%%%%
\begin{figure}[b!]
\begin{center}
\epsfig{file=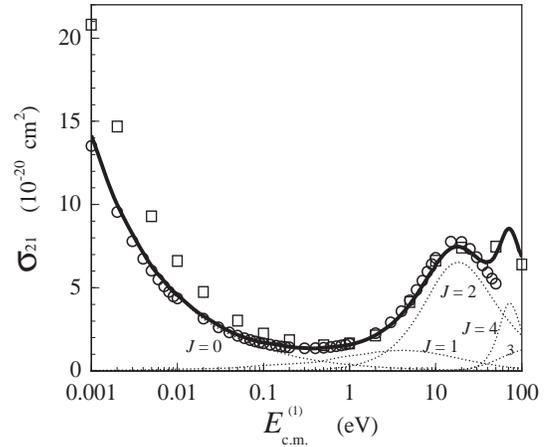,width=7.2cm,height=6.0cm}
\caption{
Calculated transfer cross sections $\sigma_{21}$ of
$(d\mu)_{1s}+t \to d+(t\mu)_{1s} + 48\ {\rm eV}$.
$E_{\rm c.m.}^{(1)}
=E-\varepsilon_{1s}^{(1)}$ is the
collision c.m. energy in the incident channel.
The results are given 
by GEM \cite{Kino93a} (solid line),
by Ref.~\cite{Cohen91} (open boxes) and
by Ref.~\cite{Chiccoli92} (open circles).
Dotted lines are partial-wave cross sections for each $J$
by GEM. 
Precise numbers of the cross sections are seen partially  
in Table~18 of Ref.~\cite{Hiyama03}.
This figure is taken from Ref.~\cite{Hiyama03}.
}
\label{fig:transfer}
\end{center}
\end{figure}
%%%%%%%%%%%%%%%%%%%%%%%%%%%%%%%%%%%%%%
%
\vskip -0.1cm
The first and second terms describe 
the open channels \mbox{$(d\mu)_{1s}-t$} and 
$(t\mu)_{1s}-d$, respectively.
Here, $k_1$ is the wave number of the channel $c=1$
and is given 
as $\hbar^2 k_1^2/(2\mu_1)= E - \varepsilon_1$
with the intrinsic energy $\varepsilon_1$
; and similarly for the channel $c=2$.

The third term 
is responsible, in the interaction region,
for the 3-body degrees of freedom that are not included 
in the first and second terms. 
The third term is expanded by a set of 
$L^2$-integrable 3-body eigenfunctions (should nearly be
a complete set in the restricted region).
As such eigenfunctions, we employ $\{\Phi^{(\nu)}_{J M}(E_\nu); 
\nu=1,...,\nu_{\rm max} \}$ with the eigenenergy $E_\nu$
that are obtained by
diagonalizing the total Hamiltonian with the use of the 
3-body Gaussian
basis functions, Eqs.~(\ref{eq:3-particles}), 
 whose total number is $\nu_{\rm max}$.

\vskip 0.1cm
The authors of Refs.~\cite{Kamimura88b,Kino93a}
solved the unknown functions
$\chi^{(d\mu-t)}_{JM}(k_1,{\bf R}_{1})$ and 
$\chi^{(t\mu-d)}_{JM}(k_2,{\bf R}_{2})$ as well as
the unknown coefficients $\{ b_\nu(E); \nu=1,...,\nu_{\rm max} \}$
by using the Kohn-type variational principle
to $S$-matrix (see Sec.~4 of Ref.~\cite{Kamimura77} for the general 
formulation and Secs.~2.5 and~8.1 of Ref.~\cite{Hiyama03}). 

\vskip 0.1cm
Figure~\ref{fig:transfer} illustrates the calculated cross sections
$\sigma_{21}$ of the reaction~(\ref{eq:muon-transfer})
%$(d\mu)_{1s}+t \to d+(t\mu)_{1s} + 48\ {\rm eV}$
by GEM \cite{Kino93a} (solid line),
by Ref.~\cite{Cohen91} (open boxes) and
by Ref.~\cite{Chiccoli92} (open circles).
As reviewed in Ref.~\cite{Tolstikhin}, the GEM 
calculations %~\cite{Kino93a}
provides a standard \mbox{result} for the
benchmark test calculations of this Coulomb 3-body reaction. 

\vskip 0.1cm
Here, we emphasize an important role of the 
third term   of the 
total wave function~(\ref{eq:psi12}); the term is responsible
for the 3-body degrees of freedom in the interaction region.
If we omit the term, the  cross section
$\sigma_{21}$ of the transfer reaction becomes more than ten times 
larger than $\sigma_{12}$ obtained above with
the third term included.

%\pagebreak
 %%%%%%%%%%%%%%%%%%%%  Table III (CBBN)  %%%%%%%%%%%% 
\begin{table*}[t!] 
\caption{Summary of the calculated reaction rates 
of  catalyzed big-bang nucleosynthesis (CBBN) reactions obtained by
the 3-body GEM calculations~\cite{Kamimura09}.   
 The first three are for  
 $T_9 \lesssim 0.2$ 
 and the others are for $T_9 \lesssim 0.5$. 
Taken from Ref.~\cite{Kamimura09}.
%The rate of the   resonant reaction g)  depends on the 
% mass of $X^-$ particle. 
 } 
 \begin{center} 
 %\noalign{\vskip -0.2 true cm}  
 \small
 \begin{tabular}{ll}  
 \hline \hline 
 \noalign{\vskip 0.2 true cm}  
 $\qquad\qquad$  CBBN Reaction  &  $\qquad \qquad$ 
 Reaction rate (${\rm cm}^3 \,{\rm s}^{-1}\, {\rm mol}^{-1}$) by 
 GEM~\cite{Kamimura09} \\  
 \noalign{\vskip 0.2 true cm} 
 \hline  
 \noalign{\vskip 0.2 true cm} 
 $\qquad $ {\it non-resonant} reaction \\ 
 \noalign{\vskip -0.0 true cm} 
 a) $(\alpha X^-) + d \to~^6{\rm Li} + X^- $ & 
 $\quad 
 2.78  \times 10^8 \, T_9^{-\frac{2}{3}} \,  
 {\rm exp}\,(- 5.33 \,T_9^{-\frac{1}{3}})  
 ( 1 - 0.62\, T_9^\frac{2}{3} - 0.29\, T_9 ) $\\ 
 % -------------- 
 b) $(\alpha X^-) + t \to~^7{\rm Li} + X^- $ & 
 $\quad 
 1.4  \times 10^7 \, T_9^{-\frac{2}{3}} \,  
 {\rm exp}\,(- 6.08 \,T_9^{-\frac{1}{3}}) 
 (  1 + 1.3\, T_9^\frac{2}{3} + 0.55\, T_9  ) $\\ 
 % ------------------------- 
 c) $(\alpha X^-) +~\!\!^3{\rm He} \to~^7{\rm Be} + X^- $ & 
 $\quad 
 9.4  \times 10^7 \, T_9^{-\frac{2}{3}} \,  
 {\rm exp}\,(- 9.66 \,T_9^{-\frac{1}{3}})  
 \, (  1 + 0.20\, T_9^\frac{2}{3} + 0.05\, T_9   ) $\\ 
 % ----------------------- 
 d) $(^6{\rm Li} X^-) + p \to~\alpha +~\!\!^3{\rm He} + X^- $ & 
 $\quad 
 2.6  \times 10^{10} \,T_9^{-\frac{2}{3}} \,  
 {\rm exp}\,(- 6.74 \,T_9^{-\frac{1}{3}}) $\\ 
 % ----------------- 
 e) $(^7{\rm Li} X^-) + p \to~\alpha +~\!\!\alpha + X^- $ & 
 $\quad  
 3.5  \times 10^8 \, 
 \, T_9^{-\frac{2}{3}} \,  
 {\rm exp}\,(- 6.74 \,T_9^{-\frac{1}{3}}) 
 \,(1 + 0.81 \, T_9^\frac{2}{3} + 0.30\, T_9 )$\\ 
 % ------------------ 
 f) $(^7{\rm Be} X^-) + p \to~(^8{\rm B}X^-)+ \gamma $ & 
 $\quad  
 2.3  \times 10^5 \,  T_9^{-\frac{2}{3}} \,  
 {\rm exp}\,(- 8.83 \,T_9^{-\frac{1}{3}})  
 \,(  1 + 1.9\, T_9^\frac{2}{3} + 0.54\, T_9   ) $\\ 
 % ---------------------- 
 \noalign{\vskip 0.20 true cm}  
 \hline  
 % ---------------------------------------- 
 \noalign{\vskip 0.1 true cm} 
 $\qquad $ {\it resonant} reaction \\ 
 \noalign{\vskip -0.00 true cm} 
 g) $(^7{\rm Be} X^-) + p \to (^8{\rm B}X^-)_{2p}^{\rm res.}$ & 
 $\quad$ $1.44  \times 10^6 \, T_9^{-\frac{3}{2}} \,  
 {\rm exp}\,(- 2.15  T_9^{\,-1}) \qquad   m_X=100 {\rm GeV}$ \\ 
  $ \quad \: \to (^8{\rm B}X^-)+\gamma $ & \\
%  & 
 \noalign{\vskip 0.30 true cm} 
 \hline  
 \hline\\ 
 \end{tabular} 
 \end{center} 
\label{table:CBBN-rate}
 \end{table*} 
 %%%%%%%%%%%%%%%%%%%%%%%%%%%%%%%%%%%%%% 

This is due to the fact that, 
in such a low-energy reaction,
the effect of the 3-body distortion (virtual excitation) induces 
a strong {\it attractive} force in the interaction region and
causes severe mismatching of the wave length
between the interaction region and the outside region,
which results in  the strong reduction of the transfer cross section.

%%%%%%%%%%%%%%%%%%%%%%%%%%%%%%%%%%%%%%%%%%%%%%%%
\subsubsection{Catalyzed big-bang nucleosynthesis {\rm (CBBN)} reactions} 
\label{sec:CBBN}
%%%%%%%%%%%%%%%%%%%%%%%%%%%%%%%%%%%%%%%%%%%%%%%%

The present authors and collaborators~\cite{Hamaguchi,Kamimura09}
applied the 3-body reaction-calculation method 
in Sec.~\ref{sec:muon-transfer} 
to the calculation of the reaction rates
in the {\it catalyzed} big-bang nucleosynthesis (CBBN) reactions
a) to g) in Table~\ref{table:CBBN-rate} and several more
reactions (so-called rate-time CBBN reactions)
in Table~II of Ref.~\cite{Kamimura09}.
Those CBBN reaction rates were  incoorpolated in the
BBN network calculation in the literature
and have been used for the study of 
the $^6$Li-$^7$Li abundance problem, etc.

\vskip 0.1cm
In the CBBN reactions a) to g),
the particle  $X^-$ stands for
a hypothetical long-lived 
negatively-charged, massive ($\gtrsim 100$ GeV) leptonic particle 
such as a supersymmetric (SUSY) particle {\it stau}, 
a scalar partner of the tau lepton.
%\vskip 0.1cm
It is known that if  the $X^-$ particle has a lifetime  of
$\tau_X \gtrsim 10^3$ s, \mbox{it would} capture 
a light element previously synthesized in the standard BBN and
forms a Coulombic bound state,  for example,
$(^7{\rm Be} X^-)$ at temperature $T_9$  $\lesssim 0.4$ 
(in units of $10^9$ K),
$(\alpha X^-)$ at $T_9 \lesssim 0.1$ and 
$(p X^-)$ at  $T_9 \lesssim 0.01$.
Those exotic-atom bound states are expected to induce
the reactions a) to g) in which $X^-$ works as a catalysis. 

\vskip 0.1cm
Recent literature papers have claimed that 
some of these $X^-$-catalyzed reactions 
have significantly large cross sections so that 
inclusion of the reactions into the BBN network calculation
can change drastically abundances of some elements; 
this can give not only a solution to  the  
$^6$Li-$^7$Li problem (calculated underproduction of 
$^6$Li by $\sim 1000$ times  and  
overproduction of $^7$Li$+$$^7$Be by $\sim 3$~times)
but also a constraint on the lifetime and abundance
of the elementary particle $X^-$.

\vskip 0.1cm
However, most of these literature calculations of
the reaction cross sections were made assuming 
too naive models or approximations that are not suitable
for those complicated low-energy nuclear reactions.
We performed a fully quantum three-body calculation
of the cross sections of 
the above types of $X^-$-catalyzed reactions~\cite{Hamaguchi,Kamimura09},
and provided their reaction rates 
to the BBN network calculations. 
Our reaction rates are cited in 
recent review papers of BBN ~\cite{Iocco}
and CBBN~\cite{Kusakabe-review} and have been actually used, 
for example, in Refs.~\cite{Bailly2009CBBN,Kusakabe-use,Kubo-CBBN-use}. 

\vskip 0.1cm
We note that  GEM is responsible
for such BBN network calculations using our CBBN reaction rates since 
{\it absolute} values of the cross sections were predicted
(usually, such a prediction is  difficult for nuclear reactions).

%\end{document}
%\pagebreak
%\vskip 0.2cm
%%%%%%%%%%%%%%%%%%%%%%%%%%%%%%%%%%%%%%%%%%%%%%%%
\subsubsection{Scattering calculation of
5-quark $(uudd{\bar s})$ systems
}
\label{sec:penta} 
%%%%%%%%%%%%%%%%%%%%%%%%%%%%%%%%%%%%%%%%%%%%%%%%

\vskip 0.1cm

In Ref.~\cite{Hiyama06-penta}, the present authors and collaborators
performed a 5-body ($uudd{\bar s}$) {\it scattering} calculation,
for the first time, about the penta-quark resonance 
$\Theta^+ (1540)$ (experiment by Ref.~\cite{penta}).

%%%%%%%%%%%%%%%%%%%%%%%%%%%  Fig.36  %%%%%%%%%%%%%%%%%%%%%%
\begin{figure}[h!]
\centering
\epsfig{file=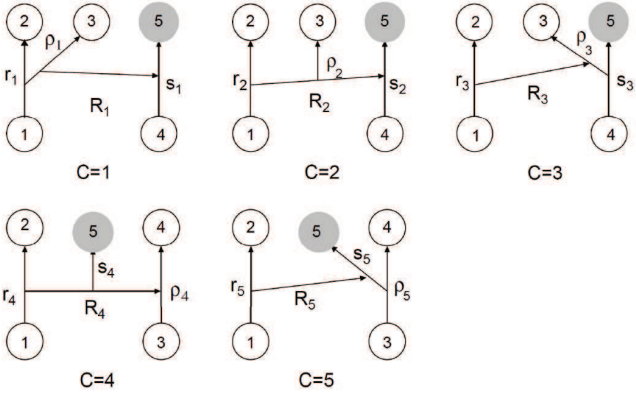,scale=0.35}
\caption{Five sets of Jacobi coordinates for the $uudd{\bar s}$ systems. 
Four $u, d$ quarks, labeled by particle $1-4$, are to be 
antisymmetrized,
while particle 5 stands for ${\bar s}$ quark. 
Sets $c=4, 5$ contain  two $qq$ correlations,  
while sets $c = 1-3$ do both $qq$ and $q{\bar q}$ correlations. 
Sets $c=1-3$ describe molecular configurations and 
sets $c=4, 5$ does connected ones.  
The $NK$ scattering channel is treated with $c=1$.
Taken from Ref.~\cite{Hiyama06-penta}} 
\label{fig:pen5-jacobi}
\end{figure}
%%%%%%%%%%%%%%%%%%%%%%%%%%%%%%%%%%%%%%%%%%%%%%%%%

\vskip 0.1cm
We took the five sets of Jacobi coordinates (Fig.~\ref{fig:pen5-jacobi}) 
%in which  particles $1-4$ are to be antisymmetrized.
and employed the same framework of the previous Secs.~\ref{sec:muon-transfer} 
and \ref{sec:CBBN}.
The $NK$ scattering channel is treated with $c=1$,
described similarly as the first term of Eq.~(\ref{eq:psi12}) 
(note that no second term in the present case).
The channels $c=2-4$ stand for the 5-body degrees of freedom in the
interaction region,  described similarly as
the third term of Eq.~(\ref{eq:psi12}).
We prepared a very large set of 5-body GEM basis functions 
%for $J = \frac{1}{2}^\pm$ and $\frac{3}{2}^\pm$
and generated, by the bound-state approximation 
(diagonalization of the total Hamiltonian),
the 5-body eigenstates $\{\Phi^{(\nu)}_{J M}(E_\nu);
\nu=1,...,\nu_{\rm max} \}$ with $\nu_{\rm max} \simeq 15,000$.

\vskip 0.1cm
There is no bound state below the $NK$ threshold at $E=1.4$ GeV.
Therefore, all the eigenstates $\Phi^{(\nu)}_{J M}(E_\nu)$ are so-called
pseudo-states, namely, discretized continuum states. It is not  
{\it a priori} known  whether the pseudo-states become real resonances or 
non-resonant continuum states when the Schr\"{o}dinger equation
is fully solved under the $NK$-scattering boundary condition imposed.

\vskip 0.1cm
Although a lot of pseudo-states $\Phi^{(\nu)}_{J M}(E_\nu)$
with  $J^\pi = \frac{1}{2}^\pm$ and $\frac{3}{2}^\pm$
were obtained within the bound-state approximation,
all the pseeudo-states  in $1.4 - 1.85$ GeV~in~mass
around $\Theta^+ (1540)$ 
{\it melt into} non-resonant continuum states 
when the coupling with the $NK$ scattering state 
is switched on (see the phase shifts in Fig.~\ref{fig:penta-phase}).

%%\end{document}  %arxiv

\vskip 0.1cm
We then concluded, at  the early stage of various discussions 
on $\Theta^+ (1540)$, that
there appears no 5-quark ($uudd{\bar s}$) resonance below 1.85 GeV in mass. 
%For more details, see Ref.~\cite{Hiyama06-penta}.

%%%%%%%%%%%%%%%%%  Penta-phase Fig.37 %%%%%%%%%%%
\begin{figure}[t!]
\begin{center}
\epsfig{file=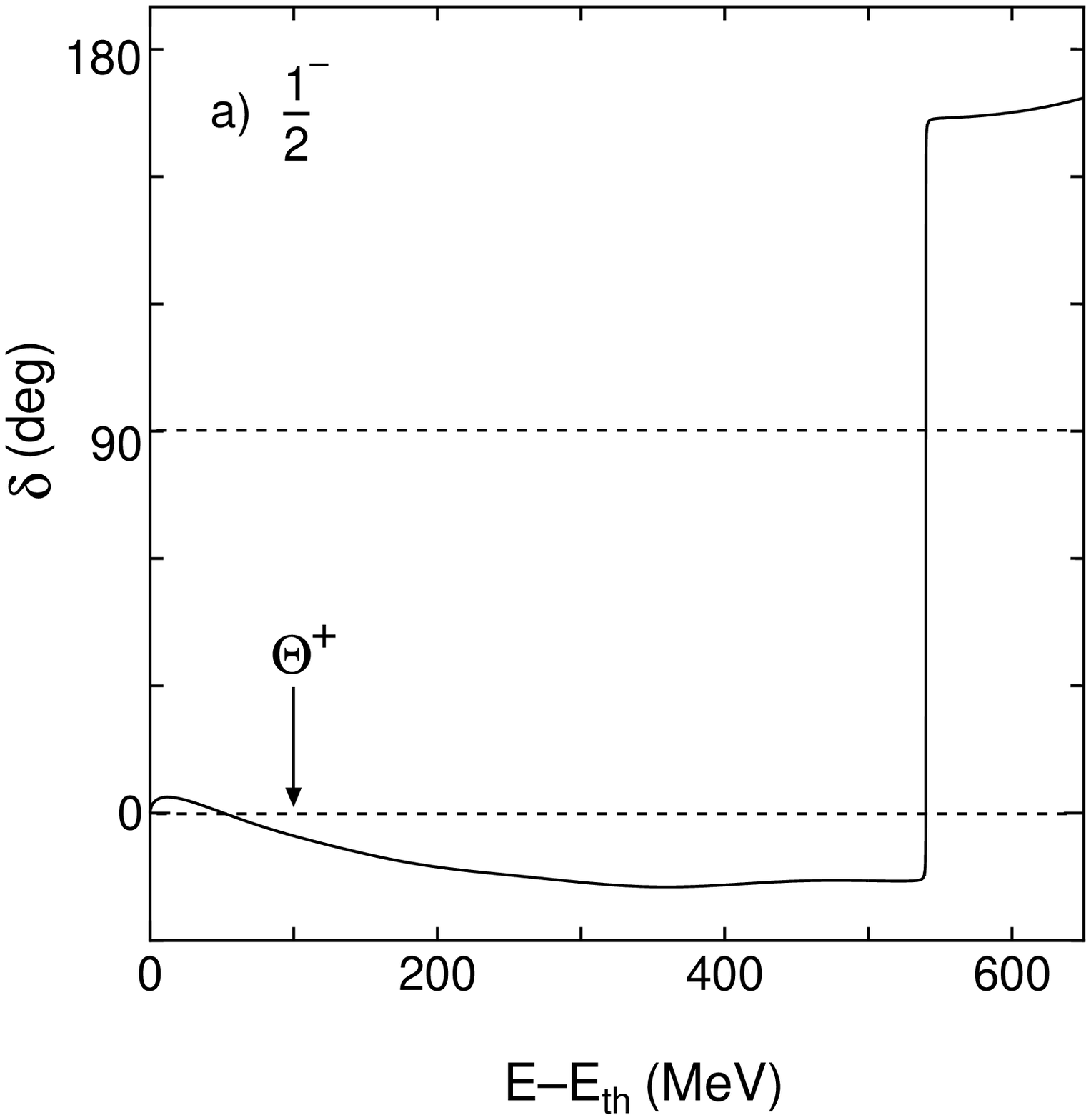,width=6.0cm,height=5.2cm}
\end{center}
\end{figure}
\begin{figure}[t!]
\begin{center}
%\vskip -1.2cm
%\epsfig{file=transition-he4.eps,width=6.0cm,height=4.6cm}
\epsfig{file=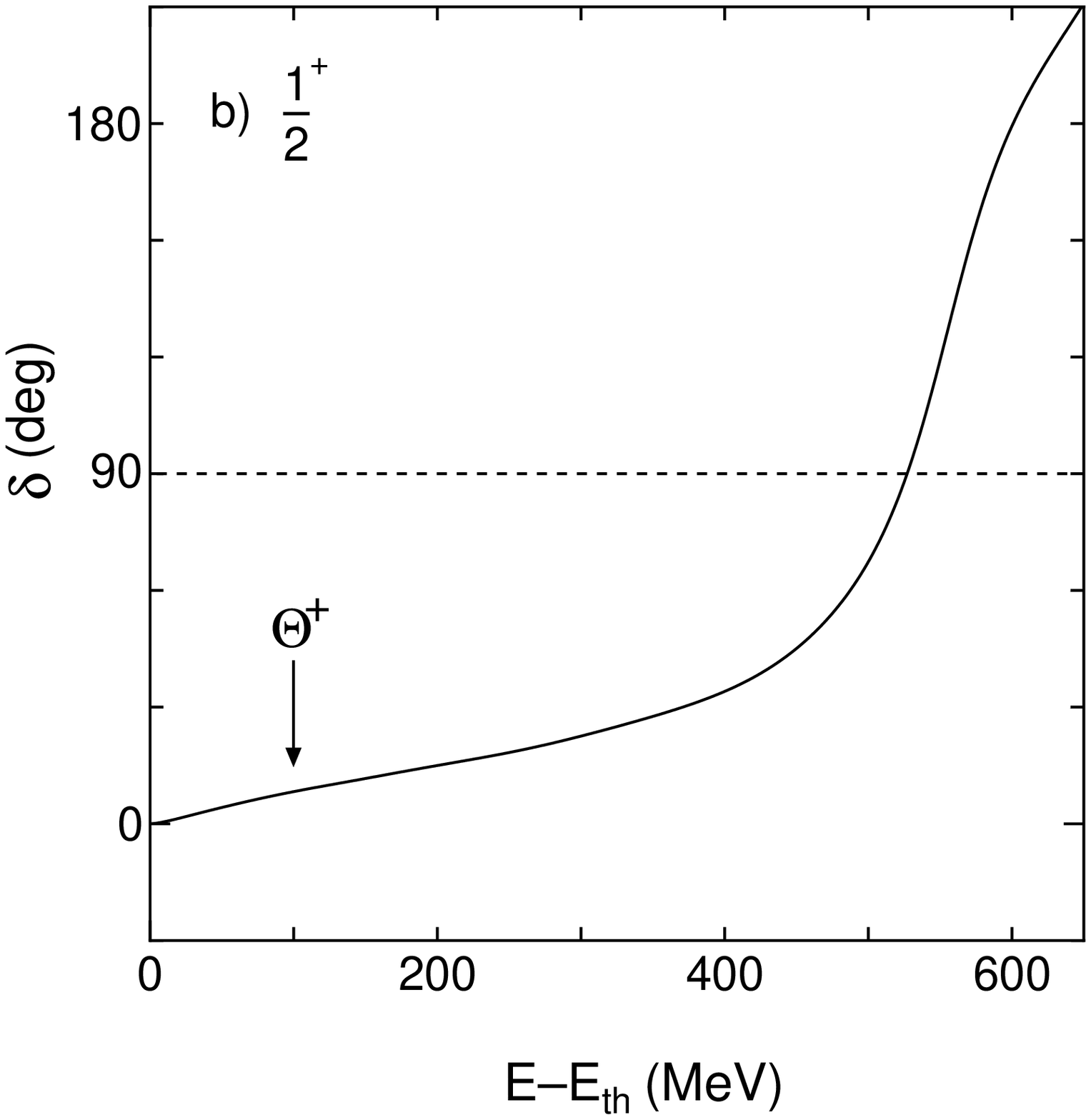,width=6.0cm,height=5.2cm}
\end{center}
\vskip -0.4cm
\caption{
Calculated phase shifts for (a) $J^\pi = \frac{1}{2}^-$
and (b) $J^\pi = \frac{1}{2}^+$ states. The
solid curves are given by the full-fledged calculation, 
while the dash-dotted curves
are by the calculation with the elastic $NK$ channel alone.
Energies are measured from the $NK$ threshold $(E_{\rm th})$. 
The arrow indicates the energy of $\Theta^+ (1540)$ in  $E-E_{\rm th}$.
Taken from Ref.~\cite{Hiyama06-penta}.
}
\label{fig:penta-phase}
\end{figure}
%%%%%%%%%%%%%%%%%%%%%%%%%%%%%%%%%%%%%%%%%%%%%%%%%
%

%\pagebreak
%%%%%%%%%%%%%%%%%%%%%%%%%%%%%%%%
\section{Summary}
\label{sec:summary}

We have reviewed our calculation method, Gaussian expansion method
(GEM)~\cite{Kamimura88,Kameyama89,Hiyama03,Hiyama12FEW,Hiyama12PTEP}
for few-body systems, and its applications to various subjects.
Those applications have been performed under our research strategy
illustrated in Fig.~\ref{fig:strategy}.
We studied few-body problems on

\vskip 0.1cm
a) bound states using the Rayleigh-Ritz variational \\
\hspace*{8mm}method, 

\mbox{b)~resonant} states using the complex-scaling method\\
\hspace*{8mm}(Sec.~\ref{sec:CSM})  and 

c) reaction processes using the Kohn-type variational \\
\hspace*{8mm}principle to $S$-matrix  (Sec.~\ref{sec:reaction}). 

\vskip 0.1cm
\noindent
We have explained 

\mbox{$\:$1)  high} accuracy of GEM calculations (Sec.~\ref{sec:AccuracyGEM}),

\mbox{$\:$2) successfull} predictions by GEM calculations before \\
\hspace*{7mm} measurements (Sec.~\ref{sec:predict}) and

\mbox{$\:$3) wide} applicability of GEM to few-body problems 
\hspace*{7mm} in various reseach fields.

\vskip 0.1cm
\noindent
%In order to perform those calculations,
We introduced three types of Gaussian basis functions: 

$\;\,$i)~real-range Gaussians (Sec.~\ref{sec:geomet}), 

%\vskip 0.1cm
$\;$ii)~complex-range Gaussians (Sec.~\ref{sec:complex-range}) and 

%\vskip 0.1cm
\mbox{iii)~infinitesimally-shifted} Gaussian lobe functions 
\hspace*{8mm} (Sec.~\ref{sec:ISGL}). 

%%%%%%%%%%%%%%%%%%%%%%%%%%%%%%%%%%%%%
 
\vskip 0.2cm
All of the Gaussians  have range parameters  
chosen to form  {\it geometric progression} which is
dense at short distances so that the description of the dynamics 
mediated by short-range interactions can be properly treated. 
Moreover approprite superposition of many Gaussians can decay accurately
(exponentially) up to a sufficiently large distance 
(cf. Figs.~\ref{fig:tetra-den-short} and \ref{fig:tetra-red-long} for a
4-body case).

\vskip 0.2cm
The function space spanned by the basis functions of 
the \mbox{second} type ii) is much wider than that of the first type i),
and is particularly good at describing 
highly oscillatory wave functions (cf. Figs.~\ref{fig:howf-complex} 
and \ref{fig:sincos-coul}).

\vskip 0.2cm
Use of the third type iii),
mathematically equivalent to the first two, makes  
the calculation of few-body Hamiltonian matrix
elements quite \mbox{easier} (with no tedius angular-momentum algebra) 
since the \mbox{basis} functions do not require any spherical harmonics
function $Y_{lm}(\theta, \phi)$ to describe the angular part.

\vskip 0.2cm
One of the advantages of taking the Gaussian ranges in geometric progression
is that the number of variational 
parameters are so small that optimization of them can  easily be performed.
The GEM calculation is quite transparent in the sense that
{\it all} the nonlinear variational parameters can be explicitly
reported in a small table even in 4-body 
calculations (Sec.~\ref{sec:easy-opt}).

\vskip 0.2cm
The total wave function of bound (resonant) state is 
expanded in terms of few-body  Gaussian basis functions 
of the Jacobi coordinates for {\it all} the rearrangement channels 
(Sec.~\ref{sec:3body-jacobi} for 3-body and Sec.~\ref{sec:bench-mark} 
for 4-body systems).
This multi-channel representation makes the function
space much wider than that spanned by single-channel basis functions. 
Therefore, those basis functions are particularly suitable 
for describing both the short-range behavior and long-range behavior (or
weak binding) along any Jacobi coordinate of the system.

\vskip 0.1cm
We are careful about all the pair interactions in order to
reproduce the binding energies of all the subsystems.
Therefore, there is no adjustable parameters
when entering the full few-body calculation;
the calculated result is 'predicted' 
in this sense (Sec.~\ref{sec:predict}).

\vskip 0.1cm
We are interested in applying GEM to few-body problems in any fields that
we have not enter yet (cf. Fig.~\ref{fig:strategy} of 
our research strategy); collaboration for it is welcome. 

\vskip 0.5cm

%\pagebreak
%%%%%%%%%%%%%%%%%%%%%%%%%%%%%%%%%%%%%%%%%%%%%%%%%%%%%%%
\section*{ACKNOWLEDGEMENTS}

It is our great pleasure to submit
this invited review paper to the International Symposium 
in honor of  Professor Akito Arima
for the celebration of his 88th birthday. 
%for his 88th birthday. 
We are very grateful for his continuous encouragement on our work.
\mbox{We would} like to thank Professor Y.~Kino for valuable discussions
on GEM and its applications.
The writing of this review was partially supported by the Japan Society for
the Promotion of Science under grants 16H03995 and 16H02180 and by
the RIKEN Interdisciplinary Theoretical Science Research Group project.

%\clearpage
%%%%%%%%%%%%%%%%%%%%%%%%%%%%%%%%
%%%%%%%%%%%%%%%%%%%%%%%%%%%%%%%%
\section*{APPENDIX} 
%\label{appendix}

\vskip -0.2cm
%\begin{center}
--- {\it Examples of  accurate 2-body GEM calculations} --- 
%\end{center}
%%%%%%%%%%%%%%%%%%%%%%%%%%%%%%%%

%\vskip 0.3cm
%%%%%%%%%%%%%%%%%%%%%%%%%%%%%%%%%%%%%%%%%%%%%%%%%%
\subsection*{A.1  Harmonic oscillator potential}
\label{HOpot}
%%%%%%%%%%%%%%%%%%%%%%%%%%%%%%%%%%%%%%%%%%%%%%%%%%

It is a good test to solve a problem whose
exact analytical solution is known.
We consider nucleon motion in a 3-dimensional 
harmonic oscillator (HO) potential:
\begin{eqnarray}
\big[\, -\frac{\hbar^2}{2m_N} \nabla^2 + \frac{1}{2}m_N \omega^2 r^2
 - E\: \big] \,\phi_{lm}({\bf r}) =0    \nonumber
\label{eq:HO-schr}
\end{eqnarray}
with $\hbar^2/m_N=41.47$ MeV fm$^2$ and $\hbar \omega= 15$ MeV. 
Radial part of the wave function is expanded
in terms of the Gaussian basis functions of 
Eq.~(\ref{eq:gauss-r}).
The Hamiltonian and norm-overlap  matrix elements can be
calculated with Eqs.~(12)-(15) in Ref.~\cite{Hiyama03}. 
We take $l=0$.

%%%%%%%%%%%%%%%%%%%%%%  Table I (H.O. potential)    %%%%%%%
%
\begin{table} [b!]  
\begin{center}
\caption{Test of the accuracy of GEM calculation for 
a nucleon in a harmonic oscillator potential with $\hbar \omega=15$ MeV 
using a set
\{$n_{\rm max}=10, \,r_1=1.5$ fm, \,$r_{n_{\rm max}}=4.0$ fm\}
for $l=0$. 
The calculated and exact
eigenenergies $(E^{(k)}; k=1,...,7)$ 
are listed in terms of the number of quanta,
$\varepsilon^{(k)}=E^{(k)}/{\hbar \omega} - 3/2$.  
}
\begin{tabular}{clrrc} 
\noalign{\vskip 0.3 true cm} 
\hline 
\hline
\noalign{\vskip 0.1 true cm} 
k  & $\qquad$ &$\varepsilon^{(k)}$ (GEM) & $\quad$&  
$\varepsilon^{(k)}$ (Exact) \\
%\noalign{\vskip -0.1 true cm} 
\hline 
\noalign{\vskip 0.1 true cm} 
  1  & $(0s)$ & $ 0.000000$ & & $ 0 \quad$  \\
%\noalign{\vskip -0.2 true cm} 

  2 &$(1s)$  &$  2.000000$ & & $ 2 \quad$  \\
%\noalign{\vskip -0.2 true cm} 

  3 &$(2s)$  &$  4.000000$ &&  $ 4 \quad$  \\
%\noalign{\vskip -0.2 true cm} 
  4 &$(3s)$  &$  6.000005$ &&  $ 6 \quad$  \\
%\noalign{\vskip -0.2 true cm} 
  5 & $(4s)$ &$  8.000064$ &&  $ 8 \quad$  \\
%\noalign{\vskip -0.2 true cm} 
  6 &$(5s)$  &$  10.002508$ &&  $10 \quad$  \\
%\noalign{\vskip -0.2 true cm} 
  7 & $(6s)$ &$  12.015534$ &&  $12 \quad$  \\
\noalign{\vskip 0.1 true cm} 
\hline
\noalign{\vskip 0.3 true cm} 
\label{table:HO-energy}
\end{tabular}
\end{center}
\end{table}
%%%%%%%%%%%%%%%%%%%%%%%%%%%%%%%%%%%%%%
%\vskip -2.1cm

The Gaussian range parameters are chosen  as 
\{$n_{\rm max}=10$, $r_1=1.5$~fm, $r_{n_{\rm max}}=4.0$ fm\}
after a little try-and-error effort about $r_1$ and $r_{n_{\rm max}}$.
More precise optimization is not necessary for practical use since the
result is satisfactorily good as follows:

%\clearpage
In Table~\ref{table:HO-energy}, calculated energy $E^{(k)}$
of the $k$-th eigenstate $(k=1,...,7)$ is compared 
with the exact value;
here $\varepsilon^{(k)}=E^{(k)}/{\hbar \omega} - 3/2$ is presented.
Wave function of the $5s\,(k=6)$ state %% and $6s\, (k=7)$ states 
is illustrated in Fig.~38.
We obtain precise energies and wave functions
for the lowest 6 states using 10 Gaussians. 
It can be said that the GEM well describes  oscillating functions 
with 4 or 5 nodes (except for the origin); this will be enough 
in actual nuclear-potential problems.

As will be shown in Appendix A.6, 
use of the complex-range Gaussians
can much more accurately describe, 
for example, the excited state with 
19 oscillations in terms of 28 Gaussians for the
same Schr\"{o}dinger equation.

%%%%%%%%%%%%%%%%%%  Fig. 38  (H.O. wave function)  %%%%%%%%%%%%%
\begin{figure}[t!]
\begin{center}
\epsfig{file=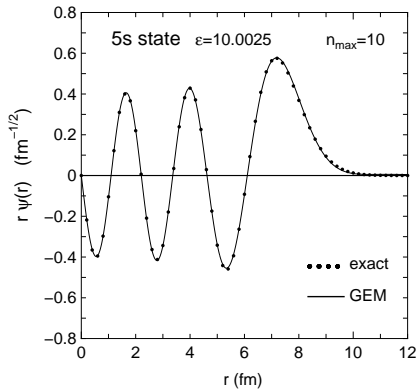,scale=0.31}
\caption{ 
Wave function of a nucleon in a harmonic oscillator potential of 
$\hbar \omega=15$ MeV. The 5s~state %% (upper) and the 6s~state (lower)
is illustrated.
The solid line shows the GEM result using a set
\{$n_{\rm max}=10, \,r_1=1.5$ fm, \,$r_{n_{\rm max}}=4.0$ fm\}, 
while the closed circles
denote the exact one.
}
\end{center}
\label{fig:ho-wf}
\end{figure}
%%%%%%%%%%%%%%%%%%%%%%%%%%%%%%%%%%%%%%%%%%%%%%
%\vskip -0.5cm

%%%%%%%%%%%%%%%%%%%%%%%%%%%%%%%%%%%%%%%%%%%%%%%%%%%%%%%%%%%%
\subsection*{A.2  Coulomb potential : hydrogen atom}
%%%%%%%%%%%%%%%%%%%%%%%%%%%%%%%%%%%%%%%%%%%%%%%%%%%%%%%%%%%%%

Here, we consider the eigenstates of the hydrogen atom 
$(p+e^-)$ as solution of the Schr\"{o}dinger equation
\begin{eqnarray}
\big[\, -\frac{1}{2} \nabla^2 - \frac{1}{r}
 - E\: \big] \,\psi_{lm}({\bf r}) =0 \:,  \nonumber
\label{eq:pe-schr-au}
\end{eqnarray}
where radius $r$ and energy $E$ are given
in the atomic units of $\hbar^2/(m_{\rm e} e^2) =0.5291$\AA~and
$m_e e^4/\hbar^2= 27.21$ eV.  %, respectively. 

%%%%%%%%%%%%%%%%%%%%%%  Table II    %%%%%%%
%
\begin{table} [b!]  
\begin{center}
\caption{Calculated eigenenergies $E^{(k)}$ (in atomic unit)
of the hydrogen atom with $l=0$
are compared with the exact values for the lowest 7 states.
We took real-range Gaussians of  
\{$n_{\rm max}=20$, $r_1=0.1$ a.u., $r_{n_{\rm max}}=80$ a.u.\}.
}
\begin{tabular}{ccccc} 
\noalign{\vskip 0.3 true cm} 
\hline 
\hline
\noalign{\vskip 0.1 true cm} 
k  & $\qquad$ &$E^{(k)}$ (GEM) & $\qquad$&  $E^{(k)}$ (Exact) \\
%\noalign{\vskip -0.1 true cm} 
\hline 
\noalign{\vskip 0.1 true cm} 
  1 && $-0.499982$  &&   $-0.500000$  \\
%\noalign{\vskip -0.2 true cm} 

  2 & &$  -0.124998$  &&  $ -0.125000$  \\
%\noalign{\vskip -0.2 true cm} 

  3 & &$  -0.055555 $ & & $  -0.055556$ \\
%\noalign{\vskip -0.2 true cm} 
  4 & &$ -0.031249 $& & $   -0.031250 $\\
%\noalign{\vskip -0.2 true cm} 
  5 & &$ -0.019998 $& & $  -0.020000 $\\
%\noalign{\vskip -0.2 true cm} 
  6 & &$ -0.013883 $ & & $  -0.013889 $ \\
  7 & &$ -0.010203 $ & & $  -0.010204 $\\
%\noalign{\vskip -0.2 true cm} 
\noalign{\vskip 0.1 true cm} 
\hline
\noalign{\vskip 0.3 true cm} 
\label{table:Pe-energy}
\end{tabular}
\end{center}
\end{table}
%%%%%%%%%%%%%%%%%%%%%%%%%%%%%%%%%%%%%%

In Table~\ref{table:Pe-energy}, calculated eigenenergies 
$E^{(k)}(k=1,...,7)$ are compared with the exact values, $-1/(2 k^2)$, 
for $l=0$. We took the Gaussian range parameters as 
\{$n_{\rm max}=20$, $r_1=0.1$ a.u., $r_{n_{\rm max}}=80$ a.u.\},
which might be nearly the best set for $n_{\rm max}=20$;
since $n_{\rm max}=20$ is sufficiently large for the lowest-lying 7 states,
a little effort was necessary to optimize $r_1$ and $r_{n_{\rm max}}$
taking round numbers with the accuracy of  0.00001 a.u. in energy.
Of course, we can obtain better solutions if we employ a larger basis set,
but here we do not enter the problem.
Much more accurate solution will be presented in Appendix A.6
with {\it complex-range} Gaussian basis functions.

%%%%%%%%%%%%%%%%%%%%%%%%%%%%%%%%%%%%%%%%%%%%%%%%%%%%%%%%%%%%%
\subsection*{A.3  Woods-Saxon potential}
%%%%%%%%%%%%%%%%%%%%%%%%%%%%%%%%%%%%%%%%%%%%%%%%%%%%%%%%%%%%%

We solve $0s, 1s$ and $0d$ bound states of neutron
in a Woods-Saxon potential;
namely, in the Schr\"{o}dinger equation of Sec. A.1,
we replace the H.O. potential by
\begin{eqnarray}
V(r)=\frac{V_0}{1+{\rm e}^{(r-R_0)/a}}         \nonumber
\label{eq:woods}
\end{eqnarray}
with $V_0=-55$ MeV, $R_0=3.0$ fm, $a=0.6$ fm and  
$\hbar^2/m=41.47$ MeV. 
The energy  by the direct numerical calculation 
is listed in the first column of Table~\ref{table:WS-energy}.
Use of GEM calculation with a Gaussian basis set
$\{n_{\rm max}=8, r_1=1.0\, {\rm fm}, r_{n_{\rm max}}=6.0$ fm\}
gives the result in the second
column of Table~\ref{table:WS-energy}.

%%%%%%%%%%%%%%%%%%%%%%  Table II    %%%%%%%
%
\begin{table} [b!]  
\begin{center}
\caption{Binding energies of 
the $0s$, $1s$ and $0d$ states of a neutron
in the Woods-Saxon potential (see text) 
by the direct numerical calculation and the GEM calculations.
The Gaussian basis set is 
$\{n_{\rm max}=8, r_1=1.0\, {\rm fm}, 
r_{\rm max}=6.0\, {\rm fm} \}$.
}
\begin{tabular}{ccrcr}
\noalign{\vskip 0.3 true cm} 
\hline 
\hline
\noalign{\vskip 0.1 true cm} 
  & & Exact $\;\;$ &&  GEM $\;\;$   \\
\noalign{\vskip -0.1 true cm} 
  &&   &&  ($n_{\rm max}=8$)  \\
\noalign{\vskip 0.0 true cm} 
\hline 
\noalign{\vskip 0.1 true cm} 
  $E_{0s}$ (MeV) && $  -33.2531$ &&  $ -33.2528 $  \\

  $E_{1s}$ (MeV) && $  -3.2221$ &&  $ -3.2208 $  \\

  $E_{0d}$ (MeV) && $  -2.1897$ &&  $ -2.1893 $  \\
\noalign{\vskip 0.1 true cm} 
\hline
\noalign{\vskip 0.3 true cm} 
\label{table:WS-energy}
\end{tabular}
\end{center}
\end{table}
%%%%%%%%%%%%%%%%%%%%%%%%%%%%%%%%%%%%%%
%%%%%%%%%%%%%%%%%%%%  Fig. 39  %%%%%%%%%%%%%%%%%%%%%%%%%%
\begin{figure} [b!]
\begin{center}
\epsfig{file=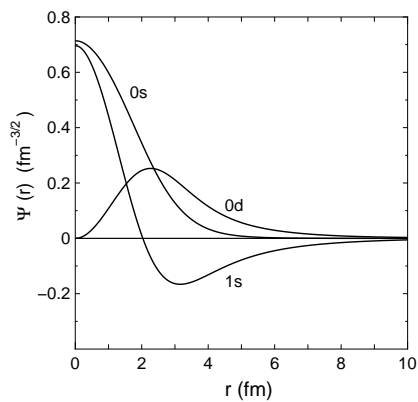,width=5.5cm}
\caption{ 
Wave functions of the $0s$, $1s$ and $0d$ states of a neutron
in a Woods-Saxon potential (see text).
The solid curve denotes the GEM result
with eight Gaussians, 
whereas the dotted curve is the direct numerical one,
but both curves are almost overlap in the whole region.
%This figure is taken from Ref.~\cite{Hiyama12PTEP}.
}
\label{fig:WS-wf}
\end{center}	
\end{figure}
%%%%%%%%%%%%%%%%%%%%%%%%%%%%%%%%%%%%%%%%%%%%%%
%\end{document}

A satisfactorily accurate result is obtained by GEM.
In Fig.~\ref{fig:WS-wf}, the wave functions given by the
8 Gaussian basis functions agree with those by the
direct calculation.

%%%%%%%%%%%%%%%%%%%%%%%%%%%%%%%%%%%%%%%%%%%%%%%%%%%%%%%%%%%%%
\subsection*{A.4  Realistic $NN$ potential : deuteron}
%%%%%%%%%%%%%%%%%%%%%%%%%%%%%%%%%%%%%%%%%%%%%%%%%%%%%%%%%%%%%

\begin{figure}[h!]
\begin{center}
%\begin{minipage}[t]{7.0cm}
%\epsfig{file=av8-potential.eps,width=6.0cm}
%\epsfig{file=av8-potential.eps,width=5.5cm}
\epsfig{file=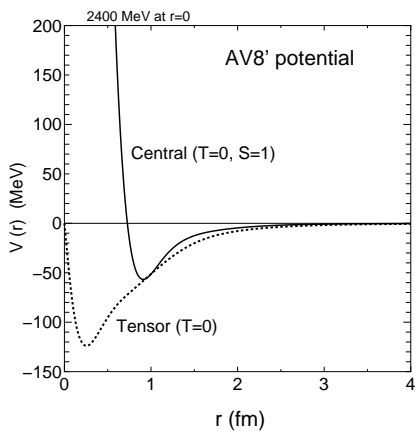,width=5.5cm}
\caption{
The AV8$'$ $NN$ potential.  
The central part $(T=0, S=1)$ and 
tensor part $(T=0)$ are shown.
}
\label{fig:AV8-pot}
%\end{minipage}
\end{center}
\end{figure}

As a realistic $NN$ potential for solving deuteron, we employ the AV8$'$
potential~\cite{AV8P97} which is often used in few-body calculations such as
the benchmark test calculation of $^4$He ground state~\cite{Kamada01}
which is mentioned in Sec.~III C.
The AV8$'$ potential is expressed as a sum of central, 
spin-orbit and tensor forces; 
Fig.~\ref{fig:AV8-pot} shows its central part $(T=0, S=1)$ having a strong
repulsive core and tensor part $(T=0)$.

\vskip 0.1cm
Purpose of the GEM calculation of this system is
to describe simultaneously both the strong short-range 
correlation and the asymptotic behavior accurately.

%%%%%%%%%%%%%%%%%%%%%%%%%%%%  Fig. 40  %%%%%%%
%
%%%%%%%%%%%%%%%%%%%%%%%%% Fig. 41  %%%%%%%%%%%%%%%%%%%%%%%%%%%%%
\begin{figure}[b!]
\begin{center}
%\begin{minipage}[h!]{7.0cm}
%\epsfig{file=deuteron.eps,width=6.0cm}
%\epsfig{file=deuteron.eps,width=5.5cm}
\epsfig{file=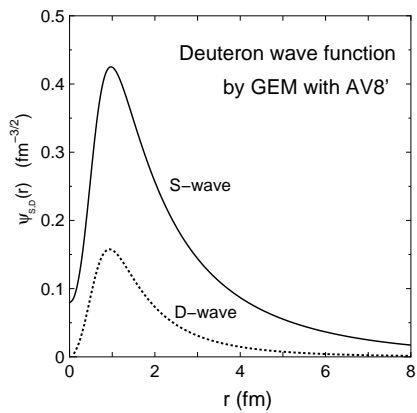,width=5.5cm}
\caption{
The $S$-wave and $D$-wave components of the deuteron wave function 
calculated by GEM with AV8$'$.
%The $S$-wave one
%is to be compared with
%Fig. 43 on the right about the strength of the short-range correlation.
}
\label{fig:deuteron-av8-short}
%\end{minipage}
\end{center}
\end{figure}

We employ a Gaussian parameter set: \\
$\quad\{n_{\rm max}=15,\: r_1=0.2 \,{\rm fm},\: 
r_{n_{\rm max}}=20\,$fm\} for $S$-wave,\\
$\quad\{n_{\rm max}=20,\: r_1=0.2 \,{\rm fm},\: 
r_{n_{\rm max}}=25\,$fm\} for $D$-wave, 
namely, 35 basis functions totally. 

\begin{figure}[t!]
\begin{center}
%\begin{minipage}[h!]{7.0cm}
%\epsfig{file=deuteron-log.eps,width=6.0cm}
%\epsfig{file=deuteron-log.eps,width=5.5cm}
\epsfig{file=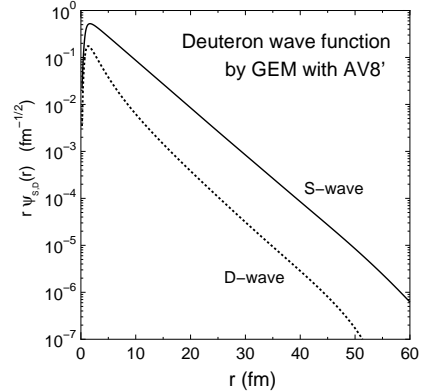,width=5.5cm}
\caption{
Asymptotic behavior of the deuteron $S$- and $D$-wave components 
(multiplied by $r$) by GEM with AV8$'$. 
%This is to be compared with
%Fig. 44 on the right. 
}
\label{fig:deuteron-av8-long}
%\end{minipage}  
\end{center}
\end{figure}

\vskip 0.1cm
Calculated wave function in the interaction region and 
that in the asymptotic region are illustrated respectively
in Fig.~\ref{fig:deuteron-av8-short} and Fig.~\ref{fig:deuteron-av8-long}.
Strong reduction and steep increase of the wave-function magnitude
due to the repulsive core is well derived. 
The correct asymptotic behavior
(exponential decaying) of the wave function (multiplied by $r$) is
demonstrated 
up to $r \sim 50$ fm where the amplitude is reduced by five-order of 
magnitude from the maximum value at $r \sim 1$ fm. 

%%%%%%%%%%%%%%%%%%%%%%  Fig. 42  %%%%%%%%%%%%%%%%%%%%%%%%%%%%%%%%
%
%%%%%%%%%%%%%%%%%%%%%%%%%%%%%%%%%%%%%%%% 

%\hspace{\fill}
%\vskip 0.3cm
%
%\vskip 2.3cm
%\clearpage
%\newpage
%%%%%%%%%%%%%%%%%%%%%%%%%%%%%%%%%%%%%%%%%%%%%%%%%%%%%%%%%%%%%
\subsection*{A.5  Very strong short-range correlation and very long tail :
$^4${\rm He}-atom dimer}
%%%%%%%%%%%%%%%%%%%%%%%%%%%%%%%%%%%%%%%%%%%%%%%%%%%%%%%%%%%%%

To the authors' knowledge, the most weakly bound 2-body state
in nature is the ground state of $^4$He-atom  dimer, and 
the most difficult problem to solve 2-body bound state
with a central potential is this dimer state.

\begin{figure}[b!]
\begin{center}
\epsfig{file=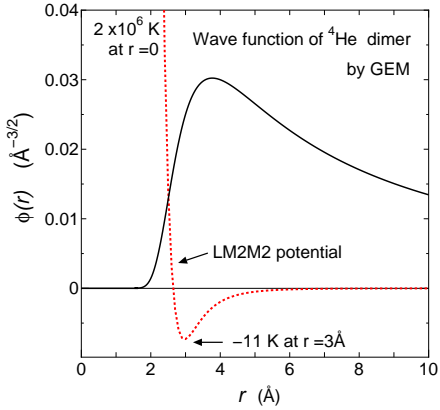,scale=0.33}
\caption{
Wave function of the $^4$He-atom dimer 
(short-range region) calculated by GEM.
The LM2M2 potential  between $^4$He atoms is illustrated (in red curve)
in arbitrary units. The only bound state
is located at $E=-0.00130$ K, so shallow.
}
\label{fig:dimer-short}
\end{center}
\end{figure}

An example of the $^4$He-$^4$He potential is
the one called  LM2M2 potential
\cite{LM2M2} illustrated in Fig.~\ref{fig:dimer-short} in red curve: 
this potential has a very strong repulsive core ($\sim\! 10^6$~K at
$r=0$) accompanied by
shallow attractive tail ($\sim \!-10$~K pocket at $r=3$ \AA)
which results in a very weak bound state at $E=-0.00130 $ K
according to a precision direct numerical calculation 
by the step-by-step method.
If we  roughly scale this problem into a nuclear system,
we would have
a potential core height of $\sim 10^6$ MeV and an attractive pocket 
of $-10$ MeV at $r \sim 2$ fm, resulting in
an extremely shallow bound state at $\sim -0.001$ MeV.
%%%%%%%%%%%%%%%%%%%%%%%%%%%%%%%%%%%%%%%%%
%%%%%%%%%%%  Fig. 43 (He4-dimer)  %%%%%%%%%
%%%%%%%%%%%%%%%%%%%%%%%%%%%%%%%%%%%%%%%%%%%%%%%
%
%%%%%%%%%%%%%%%%%%%%   Fig. 44  %%%%%%%%%%%%%%%%%%%%%%%%%%%
\begin{figure}[b!]
\begin{center}
\epsfig{file=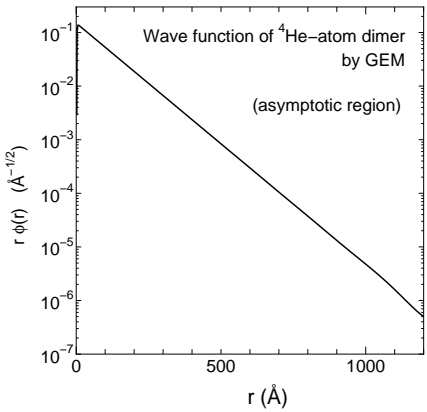,scale=0.33}
\caption{
Wave function of the $^4$He-atom dimer 
(asymptotic region) calculated by GEM
with the set $ \{ n_{\rm max}=60,\:
r_1=0.25 \,{\rm \AA},\: r_{n_{\rm max}}=700\, \AA \}$.
The asymptotic behavior is correctly reproduced up to
$\sim\!1000$ \AA}. 
\label{fig:dimer-long}
\end{center}
\end{figure}
%%%%%%%%%%%%%%%%%%%%%%%%%%%%%%%%%%%%%%%%%%%%%%

\vskip 0.1cm
Therefore, one might think 
that it would be almost impossible  
for any variational approach
to solve this problem accurately, particularly the wave function
having strong short-range correlations and a long-range asymptotic tail.
But, it is possible to solve it using GEM.
Diagonalization of Hamiltonian using
our basis functions with the set $ \{ n_{\rm max}=60,\:
r_1=0.25 \,{\rm \AA},\: r_{n_{\rm max}}=700\, $ \AA \} gives
the same energy ($E=-0.00130$ K) and wave function as those with direct 
numerical method; in Figs.~\ref{fig:dimer-short} and  \ref{fig:dimer-long}
small difference between the results of the two method is not visible.
%In this GEM calculation, the matrix elements of the 
%potential energy, Eq.~(\ref{eq:Vmat}), was obtained by
%the Gauss-quadrature integration up to $\sim\! 1000$\AA$\,$ 
%with $\sim\! 3700$ mesh points (cf. Sec.2.5). 
 
\vskip 0.3cm
It is striking that both the short-range correlations 
and the exponentially-damped tail are simultaneously 
reproduced very accurately.
This owes to the geometric-progression
Gaussian ranges which have a dense distribution  
in the short-range region
and a coherent superposition of 
long-range Gaussians in the asymptotic region. 
It will be  difficult to reach 
this degree of agreement
if other types of Gaussian-range set are chosen.
This short-range correlations in the $^4$He dimer
is relatively very much stronger than that in
the realistic nucleon-nucleon interaction (AV8$'$);
notice the large difference in the degree of amplitude-attenuation 
in the short-range region in Fig.~\ref{fig:dimer-short}
for the $^4$He dimer and that in Fig.~\ref{fig:deuteron-av8-short} for
the deuteron $S$-wave.
 
\vskip 0.1cm
In the cases of 3- and 4-body systems,
the authors presented similar figures as 
Figs.~\ref{fig:dimer-short} and  \ref{fig:dimer-long}
in Ref.~\cite{Hiyama12COLD-1} (Figs.~3, 4, 6, 8, 10 and 11) 
and in Ref.~\cite{Hiyama14COLD-3} (Figs.~4, 5, 8, 9 and 10)
for $^4$He-atom \mbox{clusters} in the cold-atom physics.

%%%%%%%%%%%%%%%%%%%%%%%%%%%%%%%%%%%%%%%%%%%%%%%%%%%%%%%%%%%%%
\subsection*{A.6  Complex-range Gaussians basis functions}
%\vskip 0.2cm
\subsubsection*{A.6.1. Highly excited states in HO potential}
%%%%%%%%%%%%%%%%%%%%%%%%%%%%%%%%%%%%%%%%%%%%%%%%%%%%%%%%%%%%%

A good test of the use of complex-range 
Gaussian basis functions
is to calculate the wave functions of 
highly excited states in a harmonic
oscillator (HO) potential.  We take the case of a nucleon 
with  angular momentum $l=0$ in
a potential having $\hbar \omega=15.0 $ MeV.
We expand the $s$-state wave function, $\Psi_0$, 
using Eq.~(\ref{eq:sin-cos}) as
\begin{eqnarray}
\Psi_0(r) = \sum_{n=1}^{n_{\rm max}}
\big[ c_n^{\rm (cos)} \phi^{\rm (cos)}_{n 0} (r) 
      + c_n^{\rm (sin)} \phi^{\rm (sin)}_{n 0} (r) \big].  \nonumber
\label{eq:new-base-swave}
\end{eqnarray}
Parameters of the complex-range
Gaussians are 
$\{\,2 n_{\rm max} =28 , r_1=1.4 \, {\rm fm}, 
\,r_{n_{\rm max}}=5.8 \,{\rm fm},
\: \omega=\frac{\pi}{2} \frac{1}{1.2^2}=1.09 \,\} $.
For the sake of comparison,
we also tested
the real-range Gaussian basis functions with the parameters
$ \{ n_{\rm max}=28,\:
r_1=0.5 \,{\rm fm},\: r_{n_{\rm max}}=11.3\, $fm\}.
Optimized $r_1$ and $r_{n_{\rm max}}$ are 
different between the two types of bases though their 
total numbers are the same.
In Table~\ref{table:accuracy}, 
we compare the calculated energy eigenvalues 
with the exact ones.  
It is evident that the complex-range Gaussians
can reproduce up to much more highly excited states 
than the real-range Gaussians do.

%%%%%%%%%%%%%%%%%%%%%%%%%%%%%%%%%%%%%%
\begin{table}[h!]
\caption{Test of  
accuracy of real-range Gaussian and complex-range Gaussian
basis functions
for highly excited states with $l=0$ of a HO
potential. The number of basis functions is 28
for both cases.
Energies are listed in terms of the number of quanta,
$E/\hbar\omega-\frac{3}{2}$. Reproduced from Ref.~\cite{Hiyama03}.
}
\label{table:accuracy}
\begin{center}
\begin{tabular}{rrrrr}
\hline 
\hline 
\vspace{-3 mm} \\
  &Exact  & $\;\;$$\;\;$Real-range   & $\;\;$$\;\;$$\;\;$Complex-range & \\
\hline 
\vspace{-3 mm} \\
&$\;0$   & $0.0000$ & $\:0.0000$ &   \\
&$\;4$  & $\:4.0000$ & $\:4.0000$ &    \\
&$\;8$ & $\:8.0000$ & $\:8.0000$ &  \\
&12   & 12.0000 & 12.0000 &    \\
&16   & $16.002\;\,$ & 16.0000 & \\
&20 & $20.01\;\;\;$ & 20.0000 &  \\
&  24 & $24.1\;\;\;\;\;$ & 24.0001 &\\
&  28 & $29.5\;\;\;\;\;$ & 28.0003 &\\
& 32 & $37.3\;\;\;\;\;$ & $32.002\;\,$ &\\
 &  {\bf 38} & $53.8\;\;\;\;\;$ & $\;\;{\bf 38.003}\;$ &\\
&  46 & $91.6\;\;\;\;\;$ & $46.3\;\;\;\;\;$ &\\
\hline 
\hline \\
\end{tabular}
\end{center}
\end{table}

%%%%%%%%%%%%%%%%%%%   fig.45   Ho wf %%%%%%%%%%%%%
%\begin{wrapfigure}{l}{\halftext}
\begin{figure}[h!]
\begin{center}
\epsfig{file=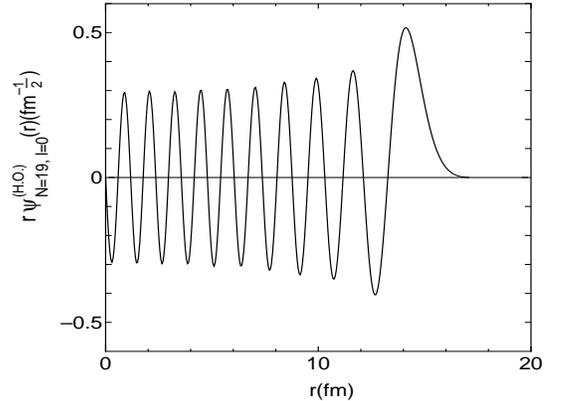,width=7.2cm,height=5.4cm}
\caption{
Wave function of the $l=0$, N=19 (38-quanta) state  
obtained by diagonalizing the HO-potential
Hamiltonian using 28 complex-range Gaussian basis functions.
It is compared with the exact wave 
function but the difference is
invisible  since the error is less than a few \%
everywhere. See text for the Gaussian parameters.
Reproduced from Ref.~\cite{Hiyama03}.}
\label{fig:howf-complex}
\end{center}
\end{figure}
%\end{wrapfigure}

\vskip -0.8cm
%\vskip 0.1cm
Figure~\ref{fig:howf-complex} shows 
good accuracy of the wave function
of 19-th excited 
state having 38 quanta. Error is within a few \%,
much smaller than the thickness of the line.

\vskip 0.1cm
Owing to the advantage mentioned above, the complex-range Gaussian
basis functions can be used to describe discretized continuum-energy  states
in various calculations 
(for example, cf. Refs.~\cite{Matsumoto03PS,Matsumoto03CDCC}).

%%%%%%%%%%%%%%%%%%%%%%%%%%%%%%%%%%%%%%%%%%%%%%%%%%%%%%%%%%%%%
\subsubsection*{A.6.2 Highly excited states of hydrogen atom}
%%%%%%%%%%%%%%%%%%%%%%%%%%%%%%%%%%%%%%%%%%%%%%%%%%%%%%%%%%%%%

\begin{figure}[h!]
\vskip 0.0cm
\begin{center}
\epsfig{file=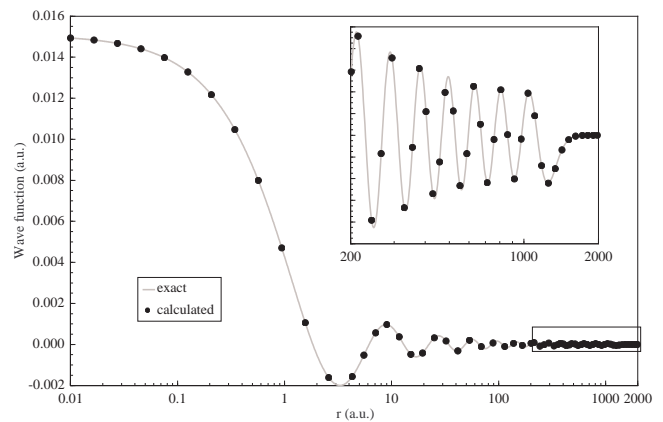,width=8.5cm}
\end{center}
\caption[]{
Wave function of the $l=0,\, n=26$ state of
the hydrogen atom. The solid line
is the exact one, and the dots are given by
the complex-range Gaussian basis functions with the 
same parameters as in Table 2. Relative error of
the latter is $10^{-7}-10^{-5}$ up to $r=1500$ a.u. at which
absolute value of the wave function is 
four-order  of magnitude smaller than that at $r=0$.
This figure is taken from Ref.~\cite{Hiyama03}.
}
\label{fig:sincos-coul}
\end{figure}

We explore another typical example in which 
the complex-range Gaussian basis functions 
reproduce 
highly oscillatory functions with high accuracy.
Table~\ref{table:b}
lists the calculated 
energy eigenvalues of the hydrogen atom with $l=0, n=1-40$ 
compared with the exact values.  Parameters of the complex-range
Gaussian basis functions are 
$\{\, 2 n_{\rm max} =160 , r_1=0.015 \,{\rm a.u.} \, , 
\,r_{n_{\rm max}}=2000 \,{\rm a.u.},
\: \omega=1.5\} $.
The energy is reproduced within a relative error of
$5\times 10^{-8}$
up to the state with $n=30$.
The wave function of the state with $n=26$ is illustrated 
in Fig.~\ref{fig:sincos-coul}, both for 
the exact solution 
and the calculated one. 
The relative error of the calculated \mbox{wave function is} 
$10^{-7}-10^{-5}$ up to $r=1500$ a.u..

\vskip 0.5cm
%
%%%%%%%%%%%%%%%%%%%%%%%%%%%%%
\begin{table}[h!]
\caption{Calculated energy eigenvalues of the 
hydrogen atom with $l=0, n=1-40$ 
compared with the exact values.  Parameters 
of the complex-range
Gaussian basis functions are taken to 
be $\{\,2n_{\rm max}=160, \,
r_1=0.015$\,a.u.$,\,r_{n_{\rm max}}=2000\,
$a.u., $\: \omega=1.5 \, $\}. 
This table is taken from Ref.~\cite{Hiyama03}.
}
\vskip -0.2cm
\label{table:b}
\begin{center}
\small
\begin{tabular}{cccc}
\hline 
\hline 
\vspace{-3 mm} \\
$\!\!$$\!\!$  $n$  & $E_{\rm cal}\;{\rm (a.u.)}$  
& $E_{\rm exact}\;{\rm (a.u.)}$ & 
rel. error  $\!\!$$\!\!$\\
\hline 
\vspace{-3 mm} \\
   1  &  -4.999999845$\times 10^{-1}$ &   
 -5.000000000$\times 10^{-1}$ &
           3.1$\times 10^{-8}$ $\!\!$$\!\!$\\
   3  &  -5.555555494$\times 10^{-2}$ & 
 -5.555555556$\times 10^{-2}$ &
          1.1$\times 10^{-8}$ $\!\!$$\!\!$\\
$\!\!$$\!\!$  10  &  -4.999999983$\times 10^{-3}$ & 
 -5.000000000$\times 10^{-3}$ &
           3.5$\times 10^{-9}$ $\!\!$$\!\!$\\
$\!\!$$\!\!$  26  &  -7.396449686$\times 10^{-4}$ & 
 -7.396449704$\times 10^{-4}$ &
          2.4$\times 10^{-9}$ $\!\!$$\!\!$\\
$\!\!$$\!\!$  30 &   -5.555555323$\times 10^{-4}$ &
   -5.555555556$\times 10^{-4}$ &
          4.2$\times 10^{-8}$ $\!\!$$\!\!$\\
$\!\!$$\!\!$  36 &   -3.856834714$\times 10^{-4}$  &
  -3.858024691$\times 10^{-4}$ &
          3.1$\times 10^{-4}$ $\!\!$$\!\!$\\
$\!\!$$\!\!$  40  &  -3.106429115$\times 10^{-4}$ & 
 -3.125000000$\times 10^{-4}$ &
         5.9$\times 10^{-3}$ $\!\!$$\!\!$\\
\hline
\hline\\
\end{tabular}
\end{center}
\end{table}
%%%%%%%%%%%%%%%%%%%%%%%%%%%%%%%%%%%%
%%%%%%%%%%%%%%%%%%   Fig. 46 (sincos)  %%%%%%%%%%%%%%%%%%%%%%%%%
%%%%%%%%%%%%%%%%%%%%%%%%%%%%%%%%%%%%%%%%%

\pagebreak

%%%%%%%%%%%%%%%%%%%%%%%%%%%%%%%%%%%%%%%%%%%%%%%%%%%%%%%%%%%%%%%%%%%%%%
%%%%%%%%%%%%%%%%%%%%%%%%%%%%%%%%%%%%%%%%%%%%%%%%%%%%%%%%%%%%
%%%%%%%%%%%%%%%%%%%%% Reference  %%%%%%%%%%%%%%%%%%%%%%%%%%%
%%%%%%%%%%%%%%%%%%%%%%%%%%%%%%%%%%%%%%%%%%%%%%%%%%%%%%%%%%%%

\end{document}